\documentclass[usenatbib, useAMS]{mn2e}
\usepackage{natbib, graphics, graphicx, epsfig, color, times, amssymb, amsmath, verbatim,amsbsy}
\usepackage{soul}

\usepackage{times} 
\usepackage{amsmath}

\title[Stability of disks]
    {The stability of stellar disks in Milky-Way sized dark matter halos}

\author[D. Yurin and V. Springel]{\parbox{18.5cm}{
Denis Yurin$^{1,2}$ and
Volker Springel$^{1,2}$}\vspace{0.2cm}\\
$^1$Heidelberg Institute for Theoretical Studies, Schloss-Wolfsbrunnenweg 35, 69118 Heidelberg, Germany\\
$^2$Zentrum f\"ur Astronomie der Universit\"at Heidelberg, ARI, M\"onchhofstr. 12-14, 69120 Heidelberg, Germany\\
}
\begin{document}

\maketitle
\begin{abstract}
  We employ an improved methodology to insert live stellar disks into
  high-resolution dark matter simulations of Milky Way sized halos,
  allowing us to investigate the fate of thin stellar disks in the
  tumultuous environment of cold dark matter structures. We study a
  set of eight different halos, drawn from the Aquarius simulation
  project, in which stellar disks are adiabatically grown with a
  prescribed structure, and then allowed to self-consistently
  evolve. The initial velocity distribution is set-up in very good
  equilibrium with the help of the {\small GALIC} code. We find that
  the residual triaxiality of the halos leads to significant disk
  tumbling, qualitatively confirming earlier work. We show that the
  disk turning motion is unaffected by structural properties of the
  galaxies such as the presence or absence of a bulge or bar. In
  typical Milky Way sized dark matter halos, we expect an average
  turning of the disks by about 40 degrees between $z=1$ and $z=0$,
  over the coarse of 7.6 Gyr.  We also investigate the impact of the
  disks on substructures, and conversely, the disk heating rate caused
  by the dark matter halo substructures. The presence of disks reduces
  the central subhalo abundance by a about a factor of two, due to
  an increased evaporation rate by gravitational shocks from disk
  passages. We find that substructures are important for heating the
  outer parts of stellar disks but do not appear to significantly affect
  their inner parts.
 \end{abstract}


\section{Introduction}
\label{sec:intro}

The $\Lambda$ cold dark matter ($\Lambda$CDM) cosmogony represents the leading theoretical model
for structure formation and predicts a hierarchical growth of galaxies
in which dark matter halos have prominent triaxial shapes
\citep{Frenk1988} and are full of substructures \citep{Moore1999}. It
is not well understood how grand design spiral galaxies can form and
survive in this violent environment, which seems at the outset quite
hostile towards the long-term survival of thin, cold stellar
disks. Such disks may easily become distorted and tilted by triaxial
potentials, they may suffer from bar instabilities or excessive
heating from dark matter substructures and merger events.

Until recently, full hydrodynamical cosmological simulations have not
been particularly successful in making realistic thin disk galaxies
without dominating bulges, even though some remarkable progress has
lately been achieved on this long-standing problem
\citep{Heller2007, Governato2010, Agertz2011, Guedes2011, Aumer2013, Stinson2013,
  Marinacci2014}. But even in the most recent simulation works, the
disks have generally been found to be too thick
\citep[e.g.][]{Marinacci2014}. This could plausibly be related to
inaccurate modelling of the disk formation physics, to a lack of
numerical resolution, or to a combination thereof. However, another
possibility is that the hierarchical nature of dark matter halo growth
in cold dark matter models, combined with the strong triaxiality of CDM
halos, quite generally causes excessive disk heating. In this case,
CDM scenarios may pose restrictive intrinsic limits on the possible
abundance of thin stellar disks. It is therefore important to shed
more light on the question under which conditions thin stellar disks
can comfortably survive in CDM halos.

\begin{table*}
\centering
\begin{tabular}[t]{llcrccrcl}
\hline
Series & Simulation names & \multicolumn{2}{c}{Disk parameters} &
\protect\hspace*{-0.1cm}Orientation\protect\hspace*{-0.1cm} & \multicolumn{2}{c}{Bulge parameters} & $\epsilon_{\rm grav}$ & Notes \\
\#           &                  & \protect\hspace*{-0.7cm}$M_d\;[10^{10}{\rm M}_\odot]$\protect\hspace*{-0.7cm} & $N_{\rm
            disk}$ &   & $M_b\;[10^{10}{\rm M}_\odot]$\protect\hspace*{-0.5cm} & $N_{\rm
            bulge}$ & $[{\rm kpc}]$ &   \\
\hline
1 & `A5..H5-minor' & $5.0$ & $2\times 10^5$ & minor & & &
0.68 \\
  & `A4-minor'                & $5.0$ & $1.6\times 10^6$ & minor  & & &
0.34&\\
  & `A3-minor'                & $5.0$ & $1.28\times 10^7$ & minor  & & &
0.17&\\
2 & `A5..H5-major' & $5.0$ & $2\times 10^5$ & major & & &
0.68&\\
3 & `A5..H5-with-bulge-minor' & $3.33$ & $2\times 10^5$ & minor &  
$1.67$ & $1\times 10^5$& 0.68 \\
  & `A4-with-bulge-minor' & $3.33$ & $1.6\times 10^6$ & minor &  
$1.67$ & $8\times 10^5$& 0.34 \\
  & `A3-with-bulge-minor' & $3.33$ & $1.28\times 10^7$ & minor &  
$1.67$ & $3.2\times 10^6$& 0.17 \\
4 & `A5..H5-lighter-disk-minor' & $3.33$ & $2\times 10^5$ & minor & & &
0.68 \\
5 & `A5..H5-massive-bulge-minor' & $1.67$ & $2\times 10^5$ & minor &  
$3.33$ & $2\times 10^5$& 0.68 \\
6 & `A5..H5-rounded-minor' & $5.0$ & $2\times 10^5$ & minor & & &
0.68 & rounded dark halo (like \#1)\\
7 & `A5..H5-withbulge-rounded' & $3.33$ & $2\times 10^5$ & minor &  
$1.67$ & $1\times 10^5$& 0.68 & rounded dark halo (like \#3)\\
8 & `A5..H5-subs-wiped-minor' & $5.0$ & $2\times 10^5$ & minor & & &
0.68 & rounded dark halo (like \#1)\\
9 & `A5..H5-withbulge-subs-wiped' & $3.33$ & $2\times 10^5$ & minor &  
$1.67$ & $1\times 10^5$& 0.68 & rounded dark halo (like \#3)\\
10 & `A5..H5-reorient-minor' & $5.0$ & $2\times 10^5$ & minor & & &
0.68 & growth phase reorientation\\
11 & `A5..H5-reorient-major' & $5.0$ & $2\times 10^5$ & major & & &
0.68 & growth phase reorientation\\
12 & `A5..H5-late-insert-minor' & $5.0$ & $2\times 10^5$ & minor & & &
0.68 & $z_{\rm ins}= 0.5$, $z_{\rm live}= 0.364$\\
\hline 
\end{tabular}
\caption{Overview of our simulation sets and their basic numerical
  parameters. We have organized the runs in different series, as
  illustrated in the table. The first series consists of our default
  pure disk runs, inserted along the minor axis. Here we also carried
  out runs for the A-halo at 8 times and 64 times higher resolution,
  respectively. Series \#2 repeats the level-5 runs with a major
  orientation of the disk. In series \#3 we replace the disk with a
  bulge+disk system in which one third of the mass is moved to a
  disk. In series \#4 this lighter disk is kept but the bulge is
  omitted, while in series \#5 we swap the masses of disk and bulge
  so that we end up with a relatively massive bulge and a disk of
  half the mass of the bulge. The remaining series represent special
  simulations to test various aspects of our procedures. Series \#6 and
  \#7 repeat the runs of series \#1 and \#3, respectively, but this time
  the dark matter halo is ``rounded''when the disk is inserted, as
  described in the text. Series \#8 and \#9 restrict the rounding to
  particles bound in substructures, so that smooth dark matter halos
  are produced. In series \#10 and \#11 we have tested for minor and
  major orientations whether a
  continuous reorientation of the disk during the growth phase between
  $z_{\rm insert}=1.0$ and $z_{\rm live}=1.3$ helps in reducing disk
  tumbling. Finally, series \#12 delays the disk insertion to much
  later time, where the dark matter halos have relaxed more and
  therefore may potentially make it easier for disks to survive unaffected.
}
\label{tab:sims}
\end{table*}

It is difficult to study this issue systematically with full
cosmological hydrodynamical simulations, due to their very high
computational cost and the lack of freedom to prescribe the structural
disk properties of the galaxies.  An alternative approach is to study
the stability of well-resolved collisionless stellar disks that are
inserted into dark matter halos in a suitable fashion. This has been
explored in numerous previous works, but most of these studies
employed isolated toy models of halos and disks, and not the full
cosmological context \citep[e.g.][]{Walker1996, Sellwood1998,
  Velazquez1999, Athanassoula2002, Debattista2006, Gauthier2006,
  Kazantzidis2008, Read2008, Machado2010}.  Only a few works have
tried to achieve a consistent cosmological embedding, where the disk
is somehow inserted by hand in a suitable fashion in a growing dark
matter halo. Pioneering work in this direction was done by
  \citet{Berentzen2006A} for isolated halos and by
  \citet{Berentzen2006B} in a cosmological context where the ``seed''
  stellar disk was inserted and quasi-adiabatically grown in
  an assembling dark matter halo. These studies revealed the mutual
  influence of halo and disk on each other and allowed to gain
  valuable insights on bar formation in the local
  Universe. Later \citet{D'Onghia2010} applied a similar approach to
  study the influence of the disk on dark matter substructures and
  found that the disk can be responsible for a partial depletion of
  substructures.

The most sophisticated variant of this approach has recently been
presented by \citet{DeBuhr2012}, who introduced disks in a subset of
the halos studied in the ``Aquarius'' project \citep{Springel2008},
which consists of high-resolution simulations of Milky Way-sized halos
in a $\Lambda$CDM universe. The Aquarius halos have been particularly
well studied, with a subset of them also being followed up
hydrodynamically \citep{Scannapieco2009, Scannapieco2012, Aumer2013,
  Aumer2014, Marinacci2014, Marinacci2014b, Pakmor2014,
  Okamoto2015}. The corresponding initial conditions exist at
different resolutions and are of high quality, allowing converged
results even including the density profiles of individual dark matter
subhalos \citep{Springel2008}.

The study of \citet{DeBuhr2012} has examined halos A, B, C, and D of
the original Aquarius set. A general result of their work was that
disks in CDM halos of Milky-Way sized halos appear to be rather
brittle, and can be expected to substantially change over the coarse of
a few Gyrs. In particular, \citet{DeBuhr2012} have found that their
disks tumble substantially, and universally grow bars, unless very
light disks are used. A substantial fraction of the initial disk
material reached large heights above and below the disk plane, and
significant warps in the disks where detected as well.

However, \citet{DeBuhr2012} used a comparatively simple method to
initialize the disk velocity distribution function, and only one
family of initial galaxy structures was considered.  In the present
work we try to improve on this earlier work in several respects, in
particular by extending the study to a larger halo sample that
encompasses eight Aquarius halos, by using a more sophisticated and
flexible method to initialize the initial disk models based on our
{\small GALIC} code \citep{Yurin2014}, by including also models with
stellar bulges, and by checking the robustness of our results with a
convergence study at much higher resolution than used in previous
work. We also consider additional lines of analysis, for example by
examining the mutual impact of the disk and the substructures on to
each other. Substructures may induce heating of the disk, but they may
also themselves be depleted through gravitational shocks experienced
during disk transition or pericentre passage \citep{D'Onghia2010}.

In this work we are especially interested in the question how
universal the disk tumbling phenomenon is, and to what extent it is
affected by the initial orientation and structure of the disk galaxy
relative to its hosting dark matter halo. We would also like to better
understand under which conditions strong bars can be avoided in live
dark matter halos, and whether the large number of dark matter
substructures poses a significant problem for disk stability.

This work is structured as follows. In Section~\ref{sec:methods} we
review the methodology we apply, while in Section~\ref{sec:simset} we
describe the simulation set we have carried
out. Section~\ref{sec:default} is then devoted to an analysis of our
results for pure disk models, whereas in Section~\ref{sec:bulges} we
turn to models that also include a central bulge. In
Section~\ref{sec:subs} we briefly analyze the impact of disks on
substructures and vice versa, and in Section~\ref{sec:resolution} we
examine the robustness of our results with respect to numerical
resolution. Finally, we conclude with a discussion and summary in
Section~\ref{sec:discussion}.

\begin{figure*}
\begin{center}
\resizebox{14cm}{!}{\includegraphics{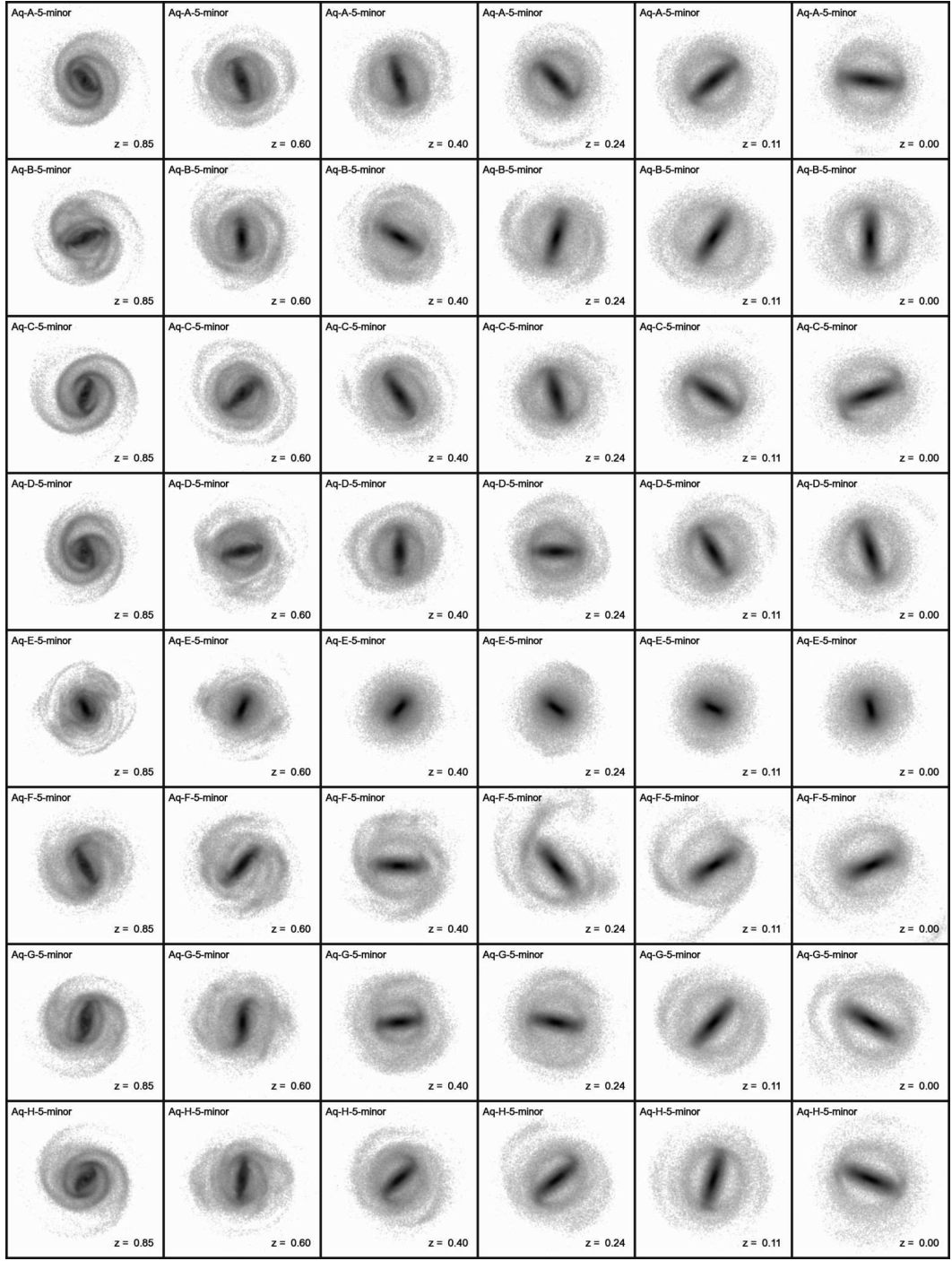}}
\caption{Face-on projections of the stellar mass of the disks in our
default runs (series \#1). Each row shows a different Aquarius halo, as labelled,
with the columns from left to right showing different times ranging
from $z=0.85$ to $z=0$. Each panel has a fixed physical size of $54\,{\rm
kpc}$ on a side, and uses the same logarithmic grey scale (covering a
dynamic range of 1000 in surface density) for
visualizing the adaptively smoothed surface density of star particles.
In each panel, the stellar particles have been turned independently 
into a face-on orientation as determined by the angular momentum vector
of the stars within the central $5\,{\rm kpc}$ of the disk.
\label{fig:defaultruns_faceon}
}
\end{center}
\end{figure*}

\begin{figure*}
\begin{center}
\resizebox{14cm}{!}{\includegraphics{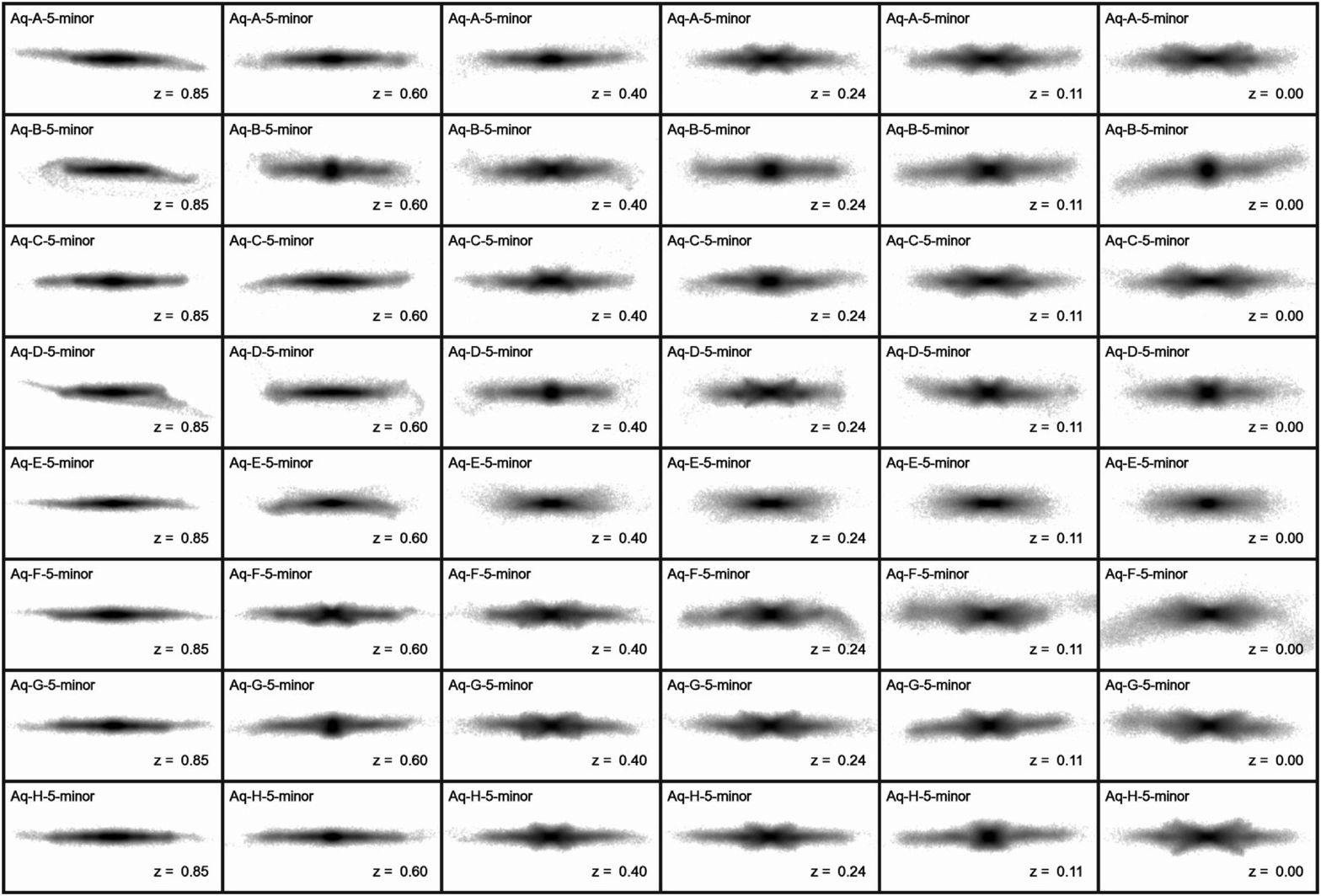}}%
\caption{Edge-on projections of the stellar disks in our default runs with
  pure disks. These images correspond to side-views of the
  corresponding images shown in Figure~\ref{fig:defaultruns_faceon},
  with an unchanged physical size of each panel in the horizontal
  direction ($54\,{\rm kpc}$), and an identical colour-scale.
\label{fig:defaultruns_edgeon}
}
\end{center}
\end{figure*}

\begin{figure*}
\begin{center}
\setlength{\unitlength}{1cm}
\resizebox{8cm}{!}{\includegraphics{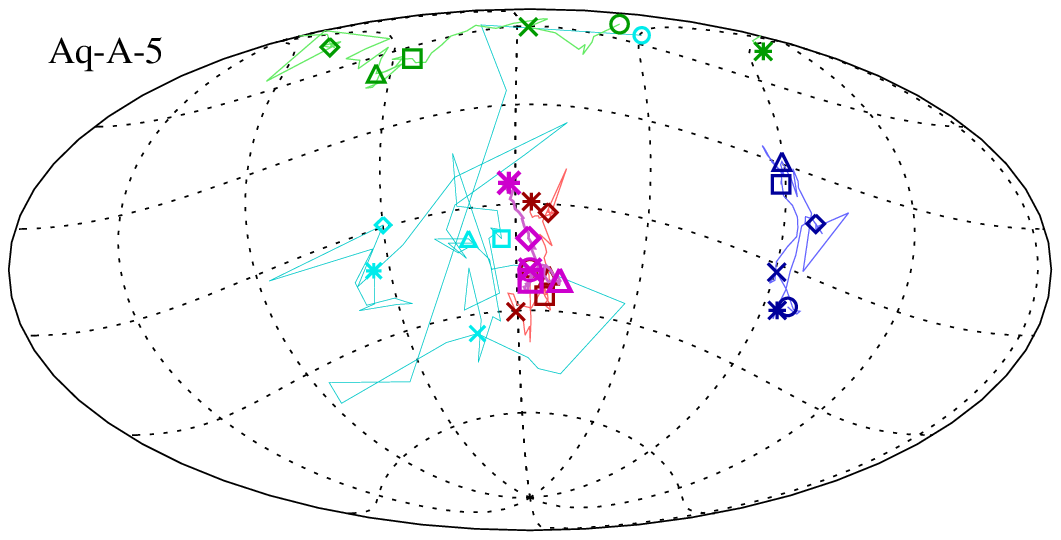}}%
\resizebox{8cm}{!}{\includegraphics{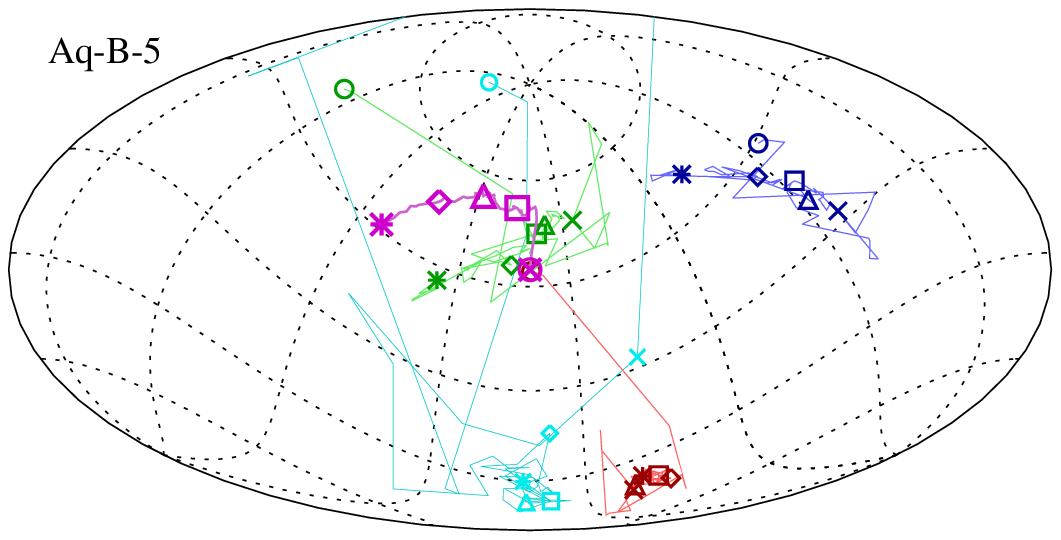}}\\%
\resizebox{8cm}{!}{\includegraphics{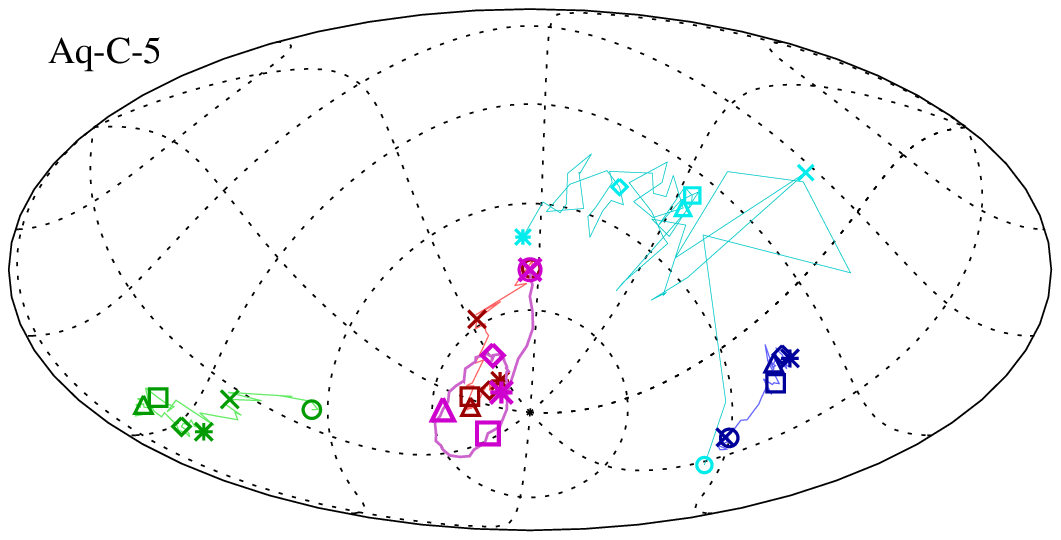}}%
\resizebox{8cm}{!}{\includegraphics{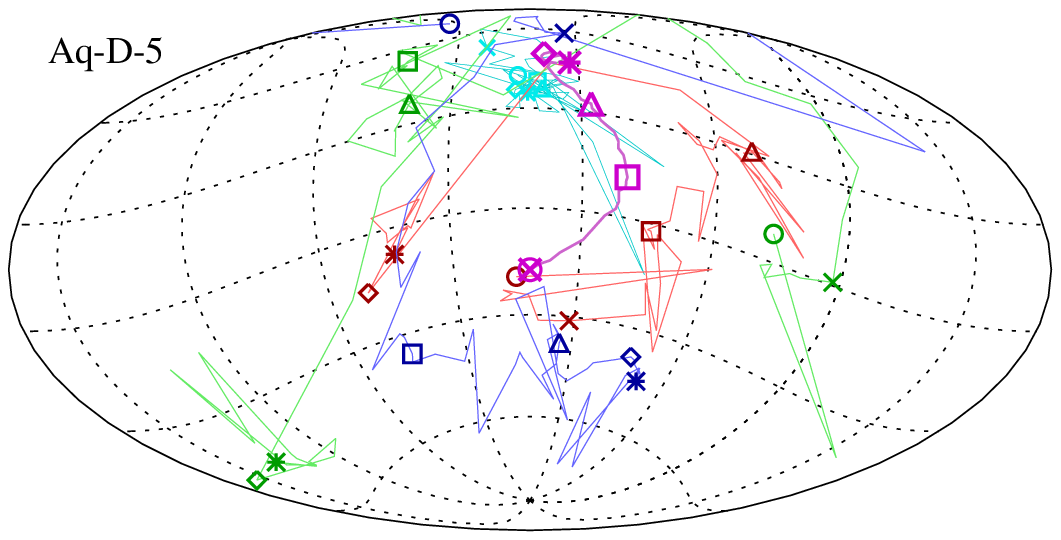}}\\%
\resizebox{8cm}{!}{\includegraphics{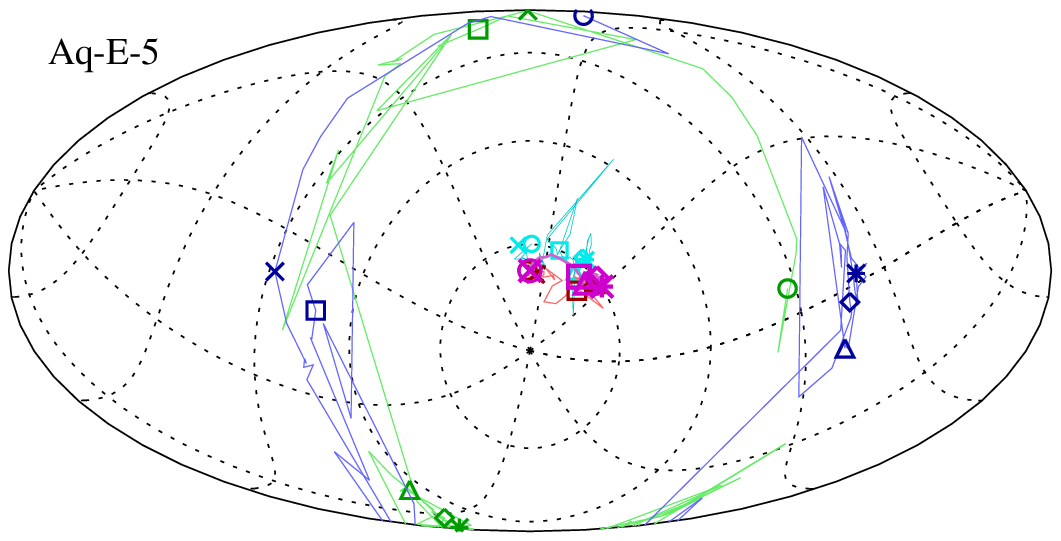}}%
\resizebox{8cm}{!}{\includegraphics{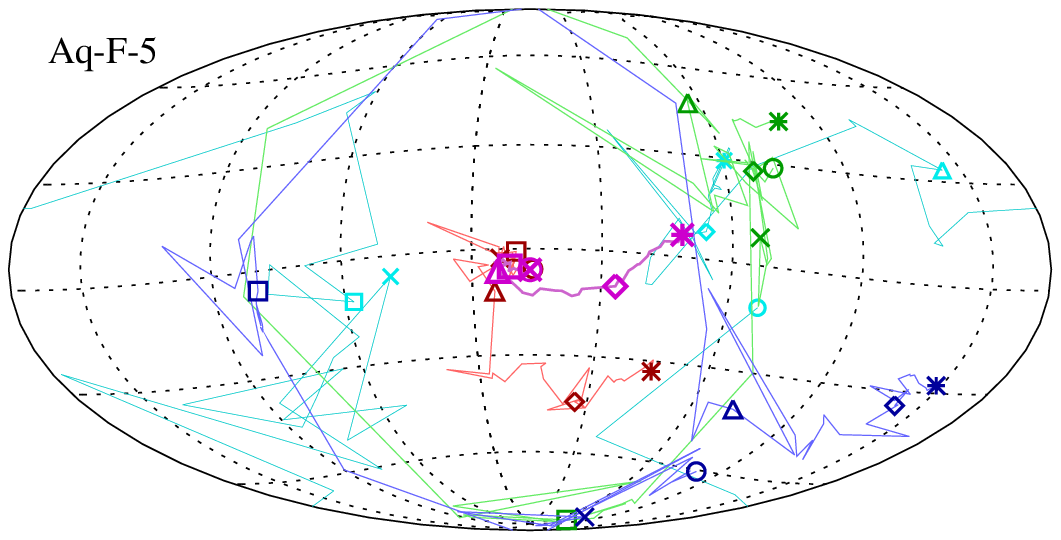}}\\%
\resizebox{8cm}{!}{\includegraphics{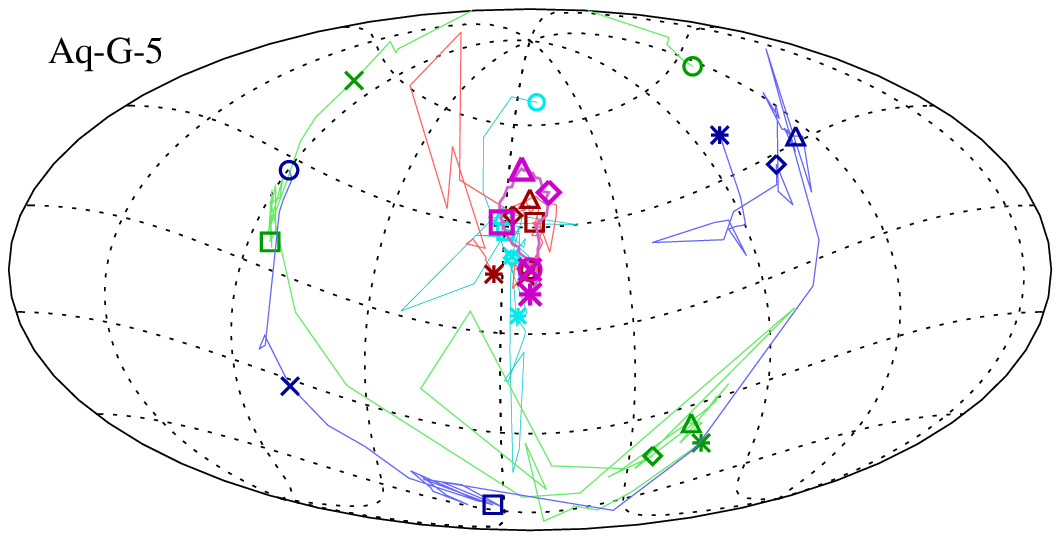}}%
\resizebox{8cm}{!}{\includegraphics{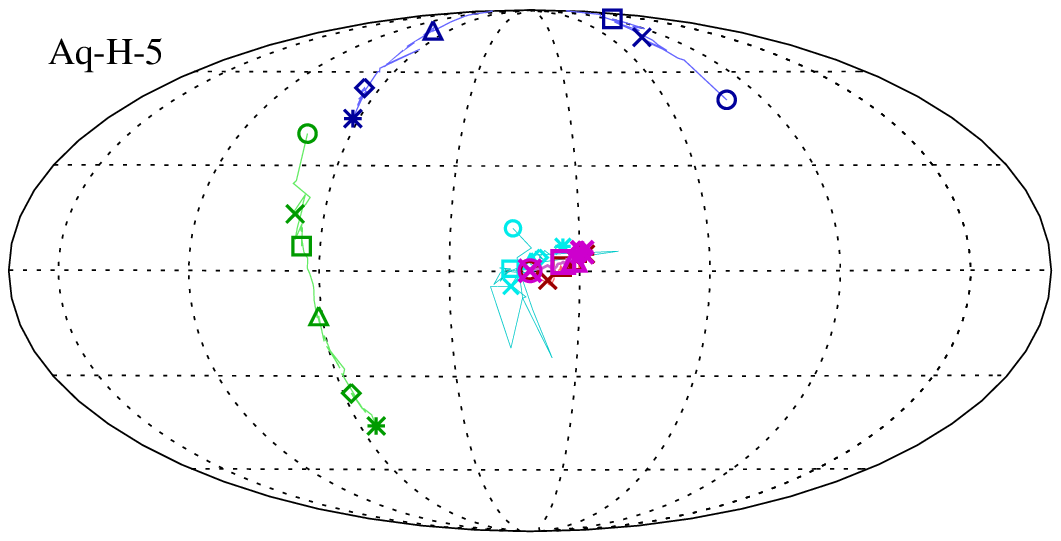}}\\%
\resizebox{16cm}{!}{\includegraphics{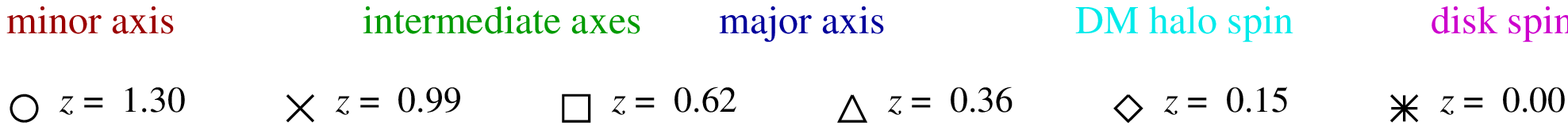}}\\%
\caption{Orientations of the principal dark matter halo axes, the
  stellar disk spin vector, and the central dark matter halo angular
  momentum as a function of time between $z=1.3$ and $z=0$ in our
  eight Aquarius halos. In each panel, a different simulation
  corresponding to our default series \#1 of runs is
  shown, as labelled. In each of the displayed runs, a pure disk is
  inserted at $z=1.3$ and grown to its final mass at $z=1.0$ keeping
  its shape and orientation fixed in time during the growth phase. 
  From $z=1.0$ to $z=0$, the disk is evolved live. 
  Different redshifts are singled out with
  symbols, as labelled. In each panel, the vector orientations are
  shown in a Mollweide projection of the unit sphere, with the initial
  orientation of the disk aligned with the centre of the map.
\label{fig:haloaxes}
}
\end{center}
\end{figure*}

\begin{figure*}
\begin{center}
\setlength{\unitlength}{1cm}
\resizebox{8cm}{!}{\includegraphics{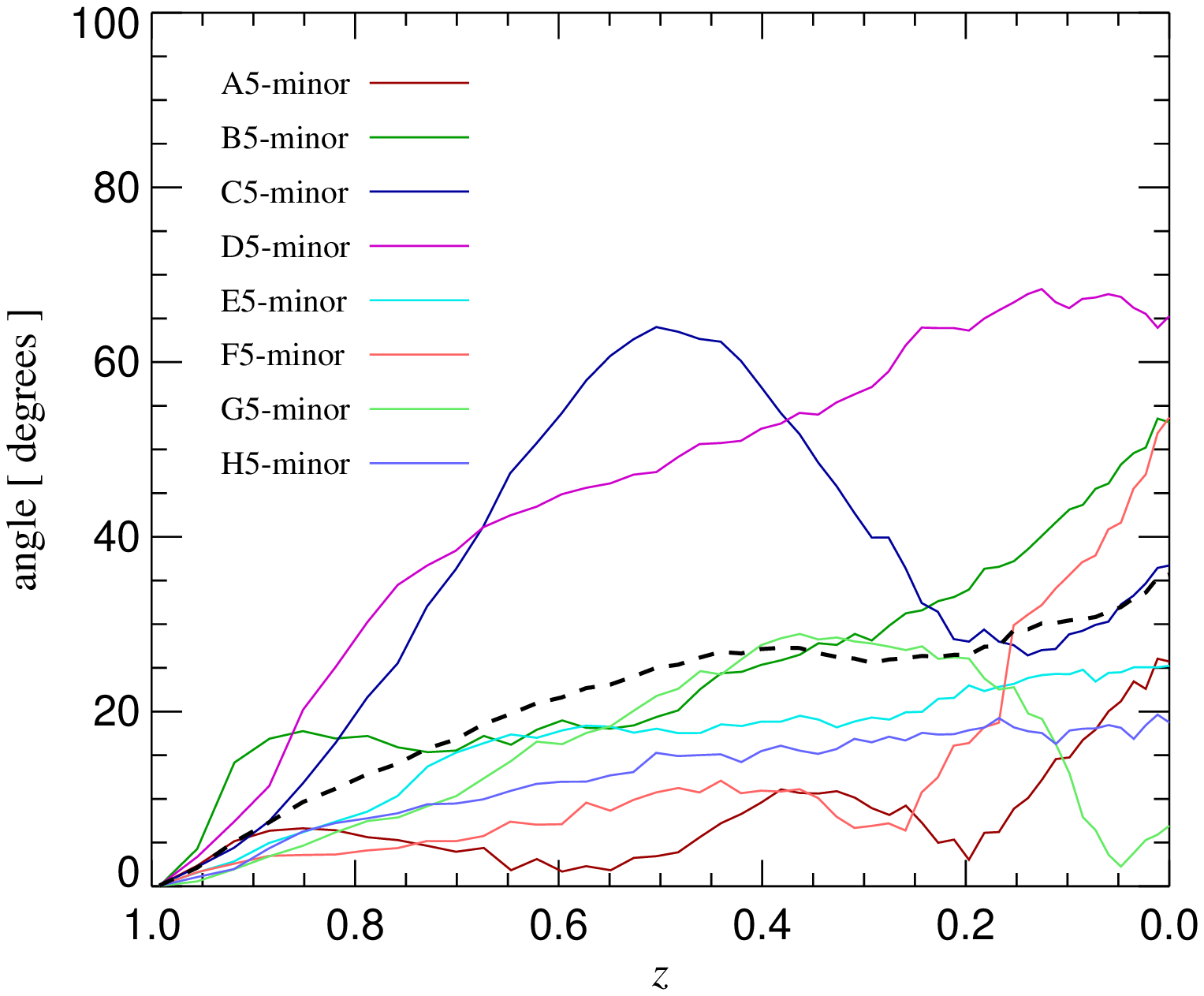}}%
\resizebox{8cm}{!}{\includegraphics{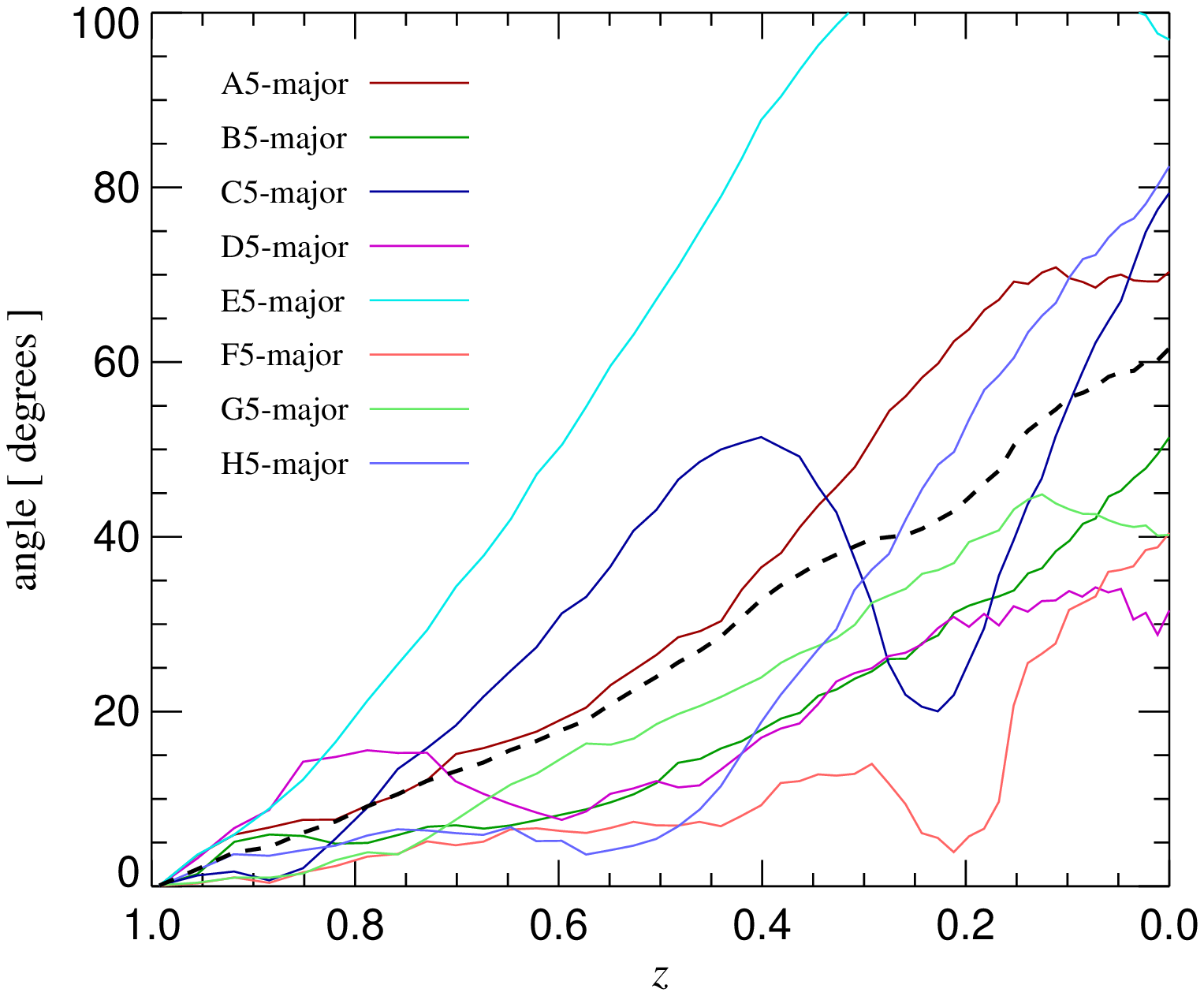}}
\caption{Angle between the current disk spin axis and the initial disk
  orientation when the disk goes live (at $z=1.0$), in our default disk
  models of simulation series \#1, as a function of time. The left
  panel shows our results if the disks are aligned along the minor axes
  of their hosting dark matter halos, the right panel is for the major axes.
  Most of the disk galaxies experience substantial tilting during the
  $7.6\,{\rm Gyr}$ of evolution from $z=1$ to $z=0$, independent of whether
  they are initially oriented along the minor or the major axis of the
  dark matter halo.
  \label{fig:disktilt}}
\end{center}
\end{figure*}

\section{Methodology}  \label{sec:methods}

In this study, we carry out resimulations of the ``Aquarius'' set of
initial conditions of Milky Way-sized dark matter halos. They have
previously been analyzed in a number of studies \citep{Springel2008,
  Springel2008b, Navarro2010, Xu2009}, where in particular the high
quality of the zoom initial conditions (created by Adrian Jenkins) was
demonstrated, allowing good convergence of all dark matter properties
of the halos. We shall mainly work with resolution `level 5' in the
nomenclature of \citet{Springel2008}, were the dark matter halos have
close to 1 million particles in the final virial radius, corresponding
to a dark matter particle mass of about $\sim 3\times 10^6\,{\rm
  M}_\odot$. For selected models, we also carry out simulations at 8
times (`level 4') and 64 times (`level 3') better resolution, reaching
up to $\sim 50$ million particles dark matter and $\sim 19$ million
star particles in the virial radius.

In order to set-up a live stellar disk in the evolving dark matter
(re-)simulations of the Aquarius halos, we proceed similarly to
\citet{DeBuhr2012}. At a certain redshift $z_{\rm insert}$ we place a
number $N_{\rm disk}$ of massless star particles into the dark matter
halo, sampling a prescribed density distribution placed at the centre
of the potential minimum and oriented along one of the principal axes
of the halo at that time. The disk mass is then grown linearly in time
to a final mass $M_d$ reached at a redshift $z_{\rm live}$. During the
growth phase, the relative distances of the disk particles with
respect to each other are kept fixed in physical coordinates, and the
whole set of disk particles is coherently moved as a solid body under
the total gravitational force experienced by all the particles of the
disk. The initial velocity of all the disk particles is set equal to
the bulk velocity of the inner dark matter halo (defined as $r <
R_{200}/4$, where $R_{200}$ is the radius enclosing a mean overdensity
200 times the critical density). While the disk mass is ramped up,
the dark matter particles start to experience the additional
gravitational force due to the disk particles, so that the dark matter
halo reacts adiabatically to the growing disk. This process of
inserting the disk ensures that the disk stays well centred in the
moving dark matter halo, and the realization of the disk potential
through particles avoids the need to make any approximation in
representing the disk potential; in fact, the density structure of the
inserted stellar system can be chosen freely as long as it is
physical.

When the disk has reached its final mass, we continue the simulation
by treating the disk particles as live, which simply means that from
this point on they are treated as ordinary collisionless particles
with independent orbits in an evolving gravitational potential. This
requires an initialization of the initial velocities of the star
particles at $z_{\rm live}$, which should be done such that the disk is
in a self-consistent dynamical equilibrium at this instant. Formally,
this corresponds to finding a stationary solution of the
Poisson-Vlasov system, which is a quite non-trivial problem for
general mass distributions \citep{Binney1987}.  In \citet{DeBuhr2012},
an approximate solution of the Jeans equation was used to initialize
the disk velocities, but the details of this procedure were not
described. Here we adopt the method of \citet{Yurin2014}, which
iteratively derives a high quality distribution function for the disk
particles. The method as implemented in the publicly available {\small
  GALIC} code integrates a large set of test particles representing
the target mass distribution in the self-consistent gravitational
potential of the system, and adjusts the velocities until the
deviation between the time-averaged density response of the particle
orbits and the target density distribution is minimized.

In order to take the actual shape of the dark matter halo into account
at the moment the disk goes live, we directly use the dark matter
particles in the cosmological simulation to compute the halo forces in
{\small GALIC}. However, as the stable version of the code is at
present restricted to axisymmetric disk models, the dark matter halo
forces are averaged in the azimuthal direction, i.e.~the dark matter
force field is ``axisymmetrized'' without actually changing the dark
matter halo. We note that this approximation could be avoided in
principle in future refinements of the method, in which case it would
then become possible to insert ellipsoidal disks. For the moment, we
stick however with inserting axisymmetric disk models with a
correspondingly axisymmetric velocity structure. Also, we note that
{\small GALIC} uses additional constraints on the velocity dispersions
in order to single out a desired target solution among the many
different degenerate solutions that are in principle possible for
reproducing the same density distribution. For example, {\small GALIC}
can realize disk models with three integrals of motion and tilted
velocity ellipsoids, similar to what is observed for the Milky Way
\citep{Sibert2008, Binney2014, Buedenbender2014}. However, for
simplicity, we have restricted ourselves to simpler models with two
integrals of motion ($E, L_z$), implying that the radial and vertical
velocity dispersions in the disk are equal.

In practice, we have used the moving-mesh code {\small AREPO}
\citep{Springel2010} for evolving our N-body systems in time. While
the hydrodynamical features of this code are not exercised in this
work, its N-body solver represents an improved and more efficient
realization of the algorithmic methods of the {\small GADGET} code
\citep{Springel2005}, which we found convenient to make use of. In
particular, the built-in parallel version of the {\small SUBFIND}
algorithm \citep{Springel2001} used to identify dark matter
substructures is more efficient in {\small AREPO} than in {\small
  GADGET}. We run {\small SUBFIND} regularly on the fly while the
N-body system is involved in order to track the masses and positions
of all halos and subhalos, as well as to measure basic dark matter
halo properties such as their shape and orientations. 

To facilitate the realization of a large number of disk insertion
simulations, we have largely automized the process of putting in live
disks into the Aquarius halos. To this end, the cosmological
simulation automatically measures the dark matter halo orientation at
$z_{\rm insert}$, inserts the initially rigid stellar system, and
continues with the disk growth phase until $z_{\rm live}$, at which
point a snapshot file is written. A special version of {\small GALIC}
is then started by a script that replaces the disk velocities in this
snapshot file with a self-consistent stationary solution, taking the
halo motion, the disk orientation, the Hubble flow at that epoch,
etc., into account. Then, the {\small AREPO} simulation of the N-body
system is continued from $z_{\rm live}$ to $z=0$.

To determine the orientation of the dark matter halo, we calculate the
principal axes of the moment-of-inertia tensor of the dark matter
particles in a spherical region of size $R_{\rm vir}/4$.  By
restricting ourselves to the inner region, we avoid strong influences
from the less well relaxed outer parts of halos. In general, analysis
of the halo shapes of CDM halos has found only mild variations of the
axis ratios with radius \citep{Allgood2006}, and quite stable
directions of the ellipsoidal axis \citep{Hayashi2007}. By measuring
the moment-of-inertia tensor in a spherical aperture the axis ratios
are biased low, but the directions of the principal axes, which is all
that matters for our purposes should be unaffected and line up well
with the principal axes of the halo potential. In our simulations we
have aligned the spin axis of the disk either with the minor or the major
axis identified in this way. The direction of the spin axis was chosen
such that the angle with the dark matter angular momentum of the inner
halo (again using $r < R_{200}/4$ for selecting this) was minimized.
To take account of the fact that we expect the formed galaxy to be
stationary in physical coordinates, we keep the gravitational
softening fixed in our runs in comoving coordinates until $z=1.5$, and
fixed in physical coordinates thereafter. The same softening length is
used for the stellar particles and the high-resolution dark matter
particles.

The above methodology can be readily generalized to stellar systems
that include a bulge component besides a disk. To this end an
additional set of particles is inserted and grown in parallel to the
disk component, and {\small GALIC} then calculates initial velocities
for the bulge component as well. 

For test simulations, we have also implemented a ``roundening''
procedure for the dark matter halo. To this end, all dark matter
particles contained in the friends-of-friends (FOF) group of the
target halo at $z_{\rm insert}$ are rotated by a random angle around
the halo centre, and their velocity vectors are randomized in
direction in the rest frame of the halo. By construction, this
procedure makes the halo spherically symmetric and smooth with
isotropic distribution function while keeping the spherically averaged
density profile and kinetic energy in random motions unchanged. All
substructure in the FOF halo are eliminated as well.  However, as a
result of this procedure the halo will be slightly out of equilibrium
initially. It has however enough time to relax again during the disk
growth phase, so that when the disk goes live it does so in a
stationary halo that is spherical apart from asymmetries induced by
the disk growth itself.  In a variant of this procedure, we restrict
the rounding operation to just those particles bound in substructures
(which amount to a few percent of the total mass of the halo). This
allows the creation of dark matter host halos that retain the
cosmological triaxiality but are largely pruned of dark matter
substructures, at least at times close to $z_{\rm insert}$. Later, new
substructures will fall in due to halo growth.

\section{Simulation set} \label{sec:simset}

For definiteness and ease of comparison with \citet{DeBuhr2012}, we
adopt for our default disk insertion runs their choice of disk mass,
$M_d = 5\times 10^{10}\,{\rm M}_\odot$, and disk scale length, $R_d =
3\,{\rm kpc}$. Also, we use in these default models their choices of
$z_{\rm insert} = 1.3$, $z_{\rm live}=1.0$, and $N_{\rm
  disk}=200,000$, combined with a canonical thickness of $0.2$ times
the scale length. For our adopted cosmology\footnote{The cosmology
  adopted in the Aquarius project is the same one used in the
  Millennium simulation \citep{SpringelMS2005}, and is characterized
  by $\Omega_0=0.25$, $\Omega_\Lambda=0.75$, $\sigma_8 = 0.9$,
  $n_s=1.0$, and a Hubble constant of $H_0 = 73\,{\rm
    km\,s^{-1}Mpc^{-1}}$, consistent with the WMAP-1 and WMAP-5
  cosmological constraints. The small offset of the cosmological
  parameters with the most recent determinations by Planck does not
  matter for the purposes of this study.}, the growth period of the
disk then lasts $T_{\rm growth} \simeq 1\,{\rm Gyr}$, and the live
evolution of the disk system proceeds for $T_{\rm live} \simeq 7.6\,{\rm
  Gyr}$.

We have carried out simulations for eight different Aquarius dark
matter halos, labelled A to H, following the notation of
\citet{Springel2008} and \citet{Scannapieco2009}. In all of these, the
pure disk models were run both with minor and major axes orientations,
forming a set of 16 default models. The simulations are so-called
`zoom-simulations' where the mass resolution of the initial particle
load has a strong spatial resolution; a high-resolution sampling of
the Lagrangian region of the target halo is surrounded by shells of
progressively more massive particles, so that the target halo feels
the same gravitational tidal fields as if it was forming in a
simulation where the whole periodic box of side-length $137\,{\rm
  Mpc}$ was uniformly followed at the high resolution.

In additional simulation sets, either the structural properties of the
inserted galaxy models were modified, the time of the disk insertion
was varied, or additional experiments like a rounding of the dark
matter halo or a (partial) elimination of substructure were carried
out. Most of these additional runs were only done for the minor
orientation, as we generally find only small systematic differences
between the minor and major orientations, with a small preference for
a higher stability of the minor orientation. Finally, we have done
runs at higher resolution for selected models of the A-halo, the pure
disk model with the minor rotation, and also the default disk plus
bulge with minor axis orientation.

Besides the default disk series, our other main series of runs
consists of disk plus bulge models where the total stellar mass was
kept fixed at $M_\star = 5 \times 10^{10}\,{\rm M}_\odot$, but one
third of the stellar mass was moved to a spherical stellar bulge,
modelled with a \citet{Hernquist1990} profile with scale length $a=2\,
{\rm kpc}$, with the rest staying in a disk with exponential surface
density profile. Note that these systems still have roughly the right
stellar mass expected based on abundance matching arguments for halos
of this size \citep[e.g.][]{Guo2010, Moster2010}, and the
disk-to-bulge mass ratio of 2:1 is still reasonably large. In another
series we have made this ratio more extreme, by exchanging the masses
of disk and bulge, yielding a disk-to-bulge mass ratio of 1:2.  In
addition, we have considered a series of runs where only the disk mass
was reduced by one third relative to our default disk model (``light
disk'' models) and the bulge was omitted.

Table~\ref{tab:sims} gives an overview of these different simulation
sets and lists some of their most important numerical parameters. In
all the runs, we have used conservative integration settings for the
tree force accuracy and timestep size in order to ensure
that all simulations are unaffected by orbit integration errors.

\section{Results for pure disk models} \label{sec:default}

\subsection{Disk orientation and visual morphology}

In Figure~\ref{fig:defaultruns_faceon} we show projected images of the
time evolution of the stellar disk material in our eight Aquarius
halos, where a pure disk of mass $5\times 10^{10}\,{\rm M}_\odot$ and
scale length $R_d=3 \,{\rm kpc}$ is inserted along the minor axis of
the halos. In each individual panel of the figure, the disk stars have
been turned to a face-on orientation, taking the spin angular momentum
of the stars in the central region of the disk (within $5\,{\rm kpc}$)
to define the disk normal. All the models almost immediately form very
strong bars, consistent with the findings of \citet{DeBuhr2012} for
halos A-D. This bar instability can be understood as
 an early response of the disk to the asymmetry of the dark matter distribution
  \citep{Romano-Diaz2008}.

\begin{figure}
\begin{center}
\resizebox{8cm}{!}{\includegraphics{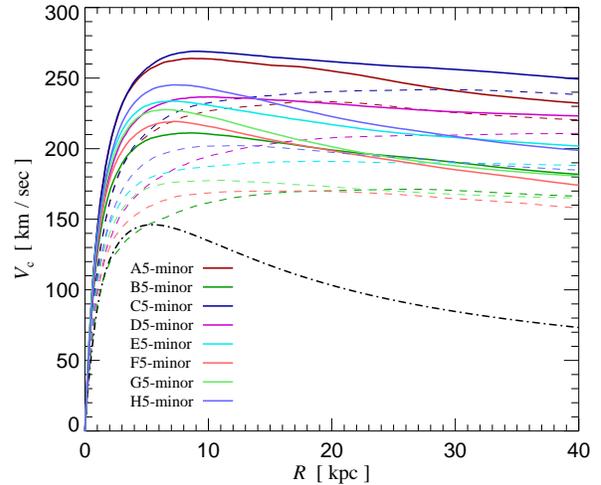}}%
\caption{Rotation curves of our default disk models (series \#1) at
  redshift $z=1$, when the disk goes live. The rotation curve velocity
  is here defined based on the enclosed mass at a given radius. Note
  that the disk is kept the same in all these eight simulations, hence
  its contribution to the rotation curve (dot-dashed line) is always
  the same. The dashed lines show the contributions of the different
  dark matter halos, while the solid curves give the total rotation
  curves of the eight systems, with different colours as specified in
  the legend.
  \label{fig:default_rotcurve}}
\end{center}
\end{figure}

\begin{figure}
\begin{center}
\resizebox{8cm}{!}{\includegraphics{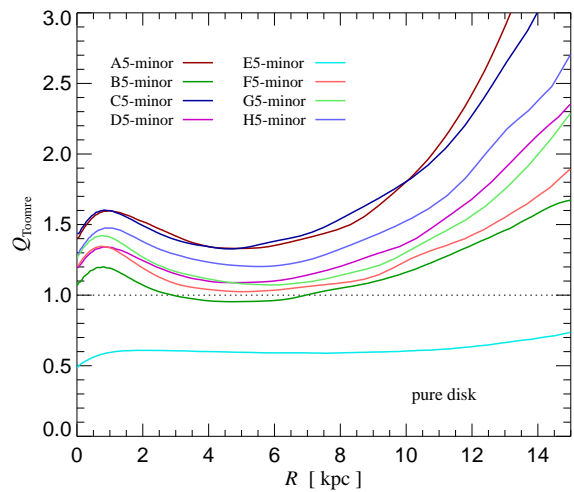}}%
\caption{Radial run of Toomre's Q-parameter \citep{Toomre1964} 
for axisymmetric stability
  for our disk simulations of series \#1, at $z_{\rm live} =1$. 
Model B5 is expected to be marginally unstable against axisymmetric
  instabilities, whereas E5 should be strongly unstable. Evidence for
  this is in fact seen in Figure~\ref{fig:defaultruns_faceon}, where
  E5 shows residual axisymmetric ring-like features in its disk at
  $z=0.85$ that are not present in this form in the other runs.
  \label{fig:default_toomre}}
\end{center}
\end{figure}

When viewed in an edge-orientation, as shown in
Figure~\ref{fig:defaultruns_edgeon}, a planar disk-like distribution
of the majority of stars is maintained in all the cases,
but the presence of the strong bars is clearly revealed by pronounced
X-shaped features in the centre of the galaxies. It is also evident
that the disks are thickened to different degrees. Halo E sports a
particularly thick disk at the end, and some systems, notably halos B
and F, show substantial bending in the periphery of the
disks. Nevertheless, the amount of stellar material significantly
outside the disk plane seems to be rather limited, somewhat different
from the findings of \citet{DeBuhr2012} for halos A-D.

\begin{figure*}
\begin{center}
\setlength{\unitlength}{1cm}
\resizebox{4.5cm}{!}{\includegraphics{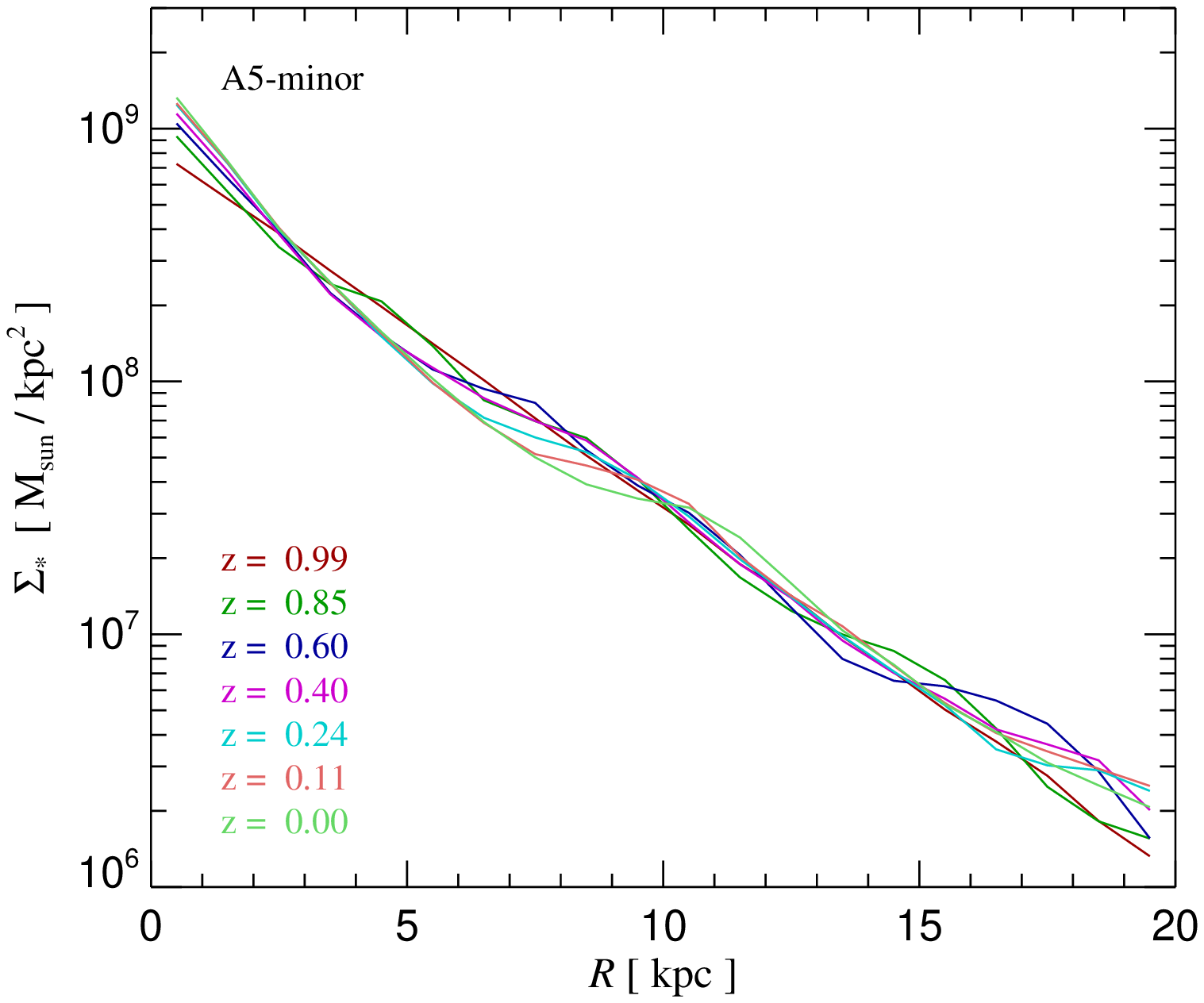}}%
\resizebox{4.5cm}{!}{\includegraphics{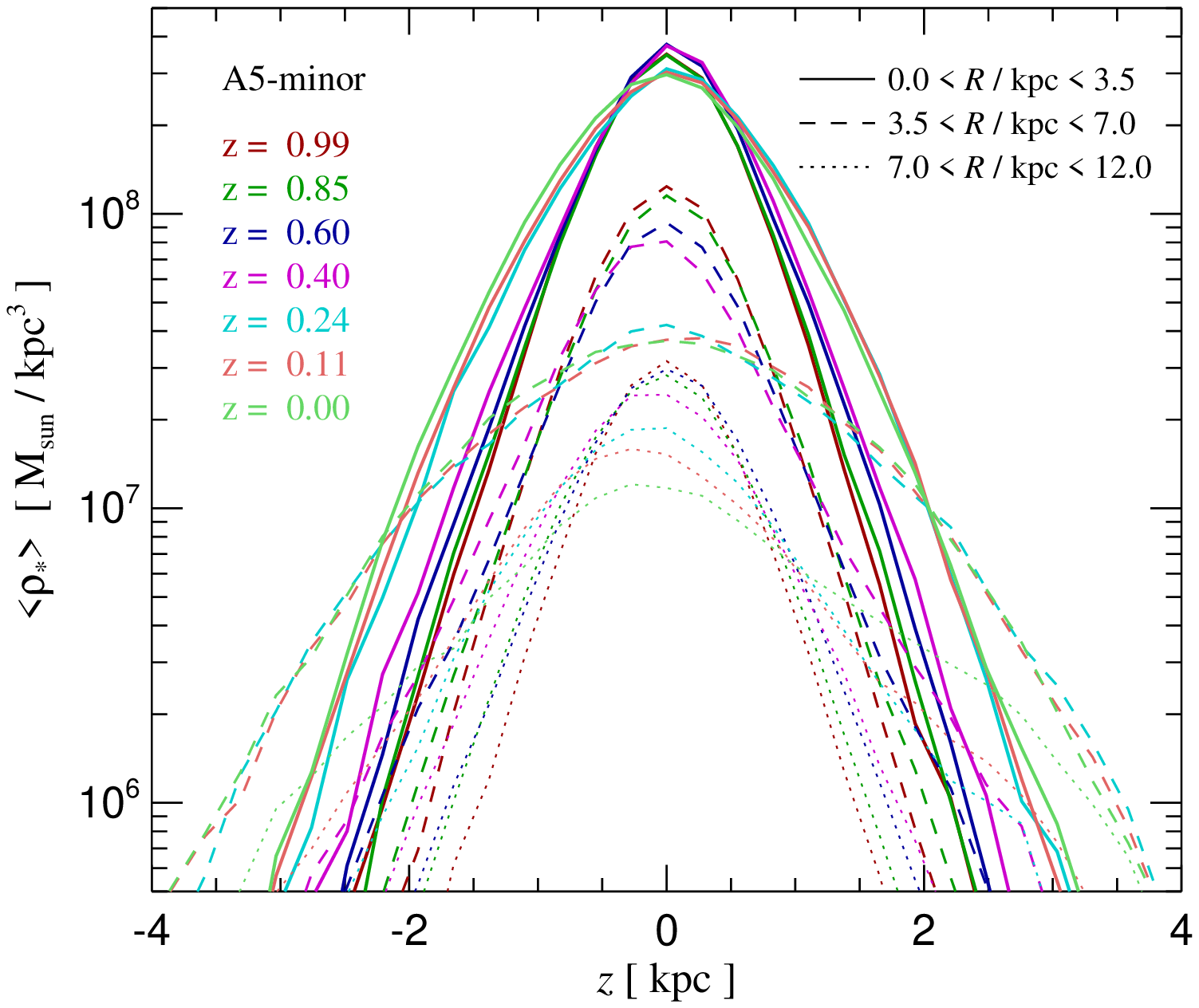}}%
\resizebox{4.5cm}{!}{\includegraphics{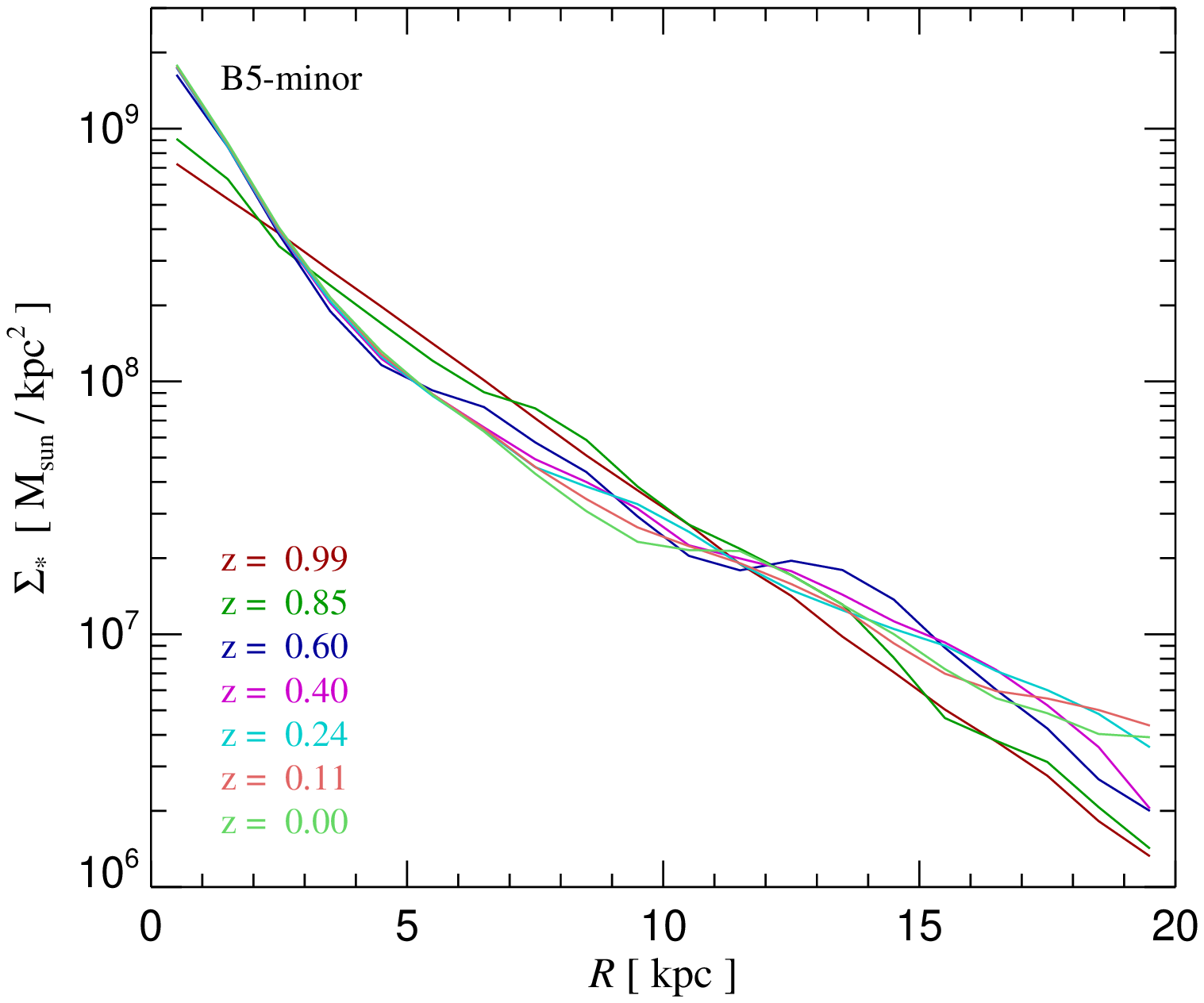}}%
\resizebox{4.5cm}{!}{\includegraphics{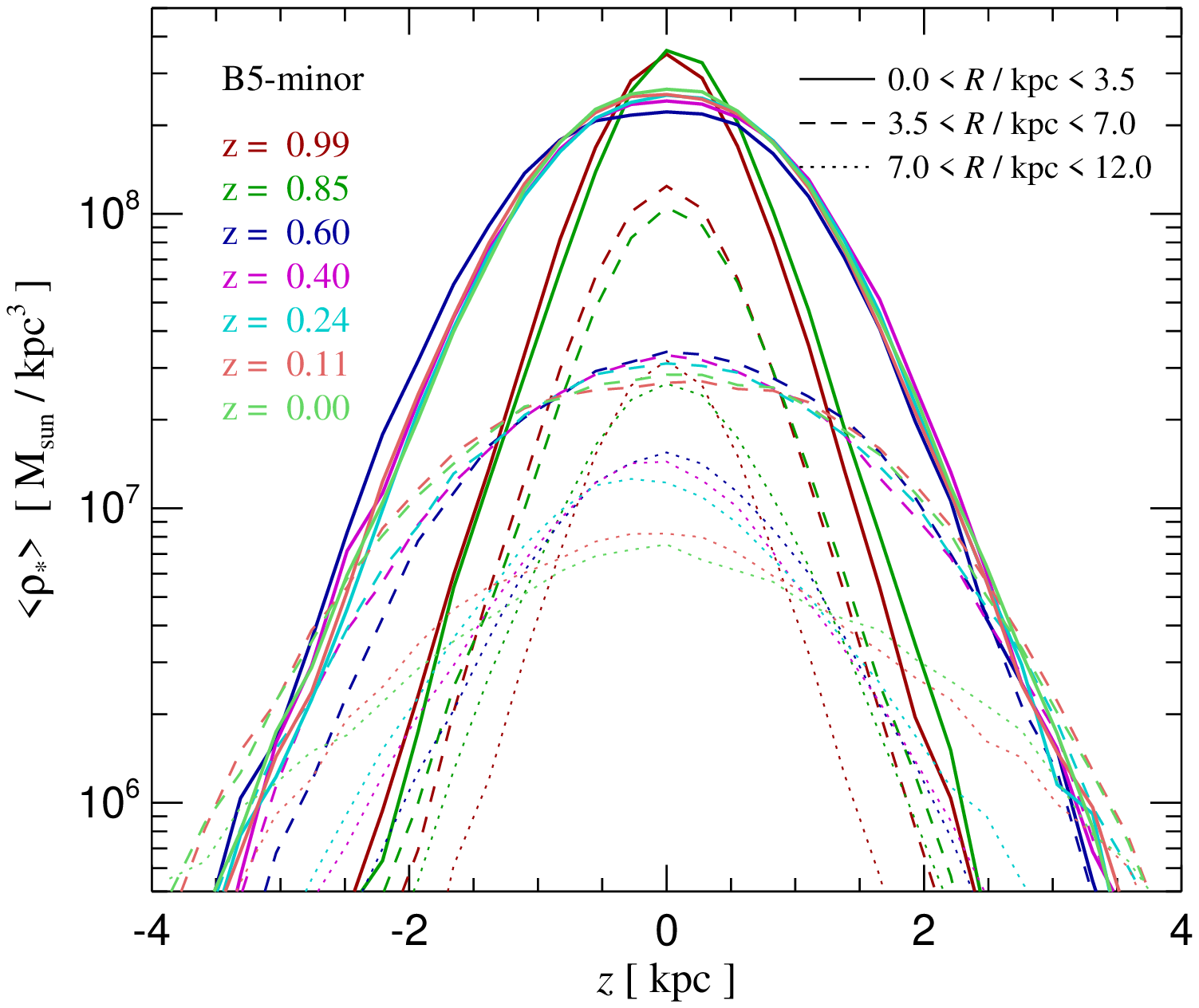}}\\%
\resizebox{4.5cm}{!}{\includegraphics{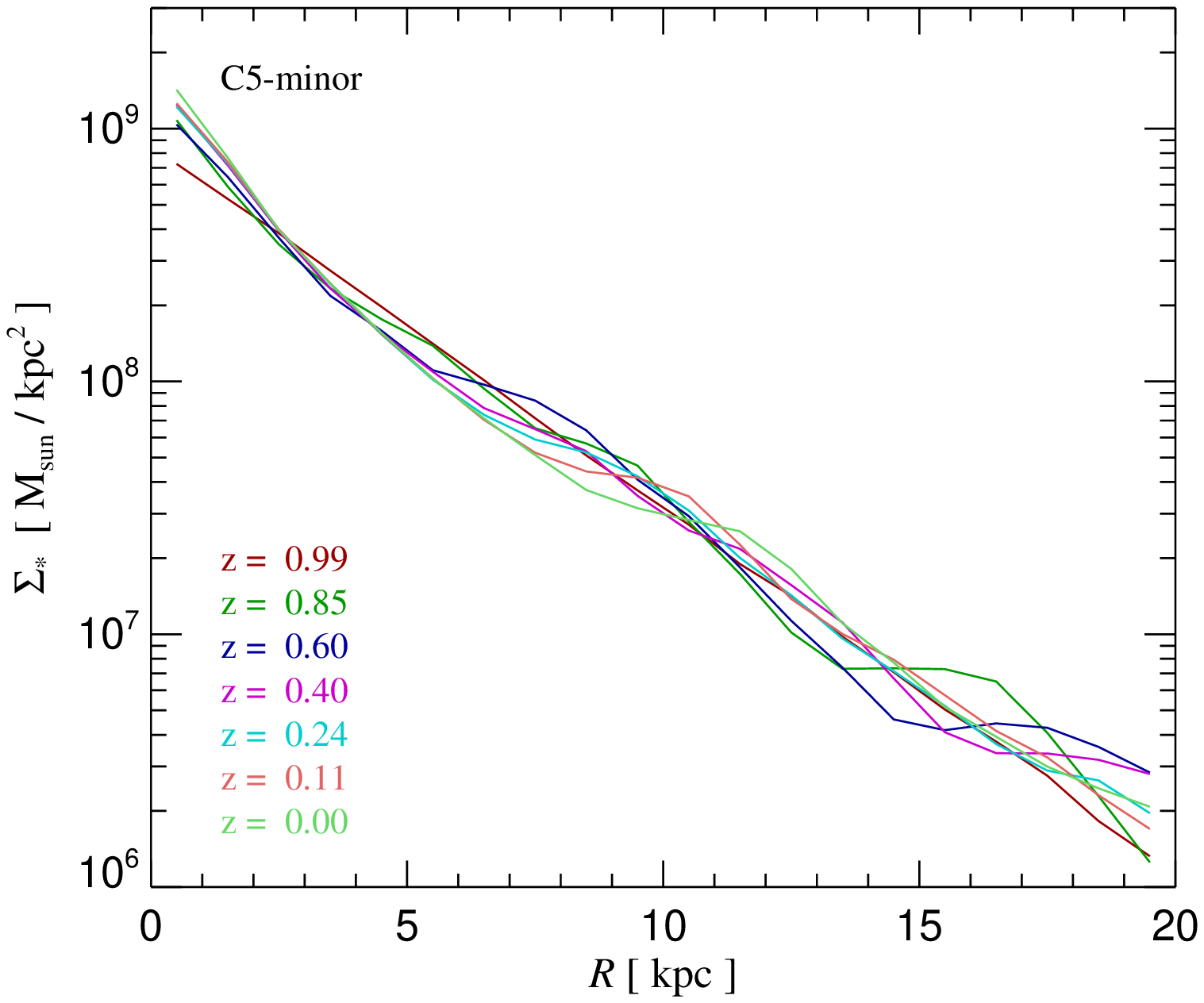}}%
\resizebox{4.5cm}{!}{\includegraphics{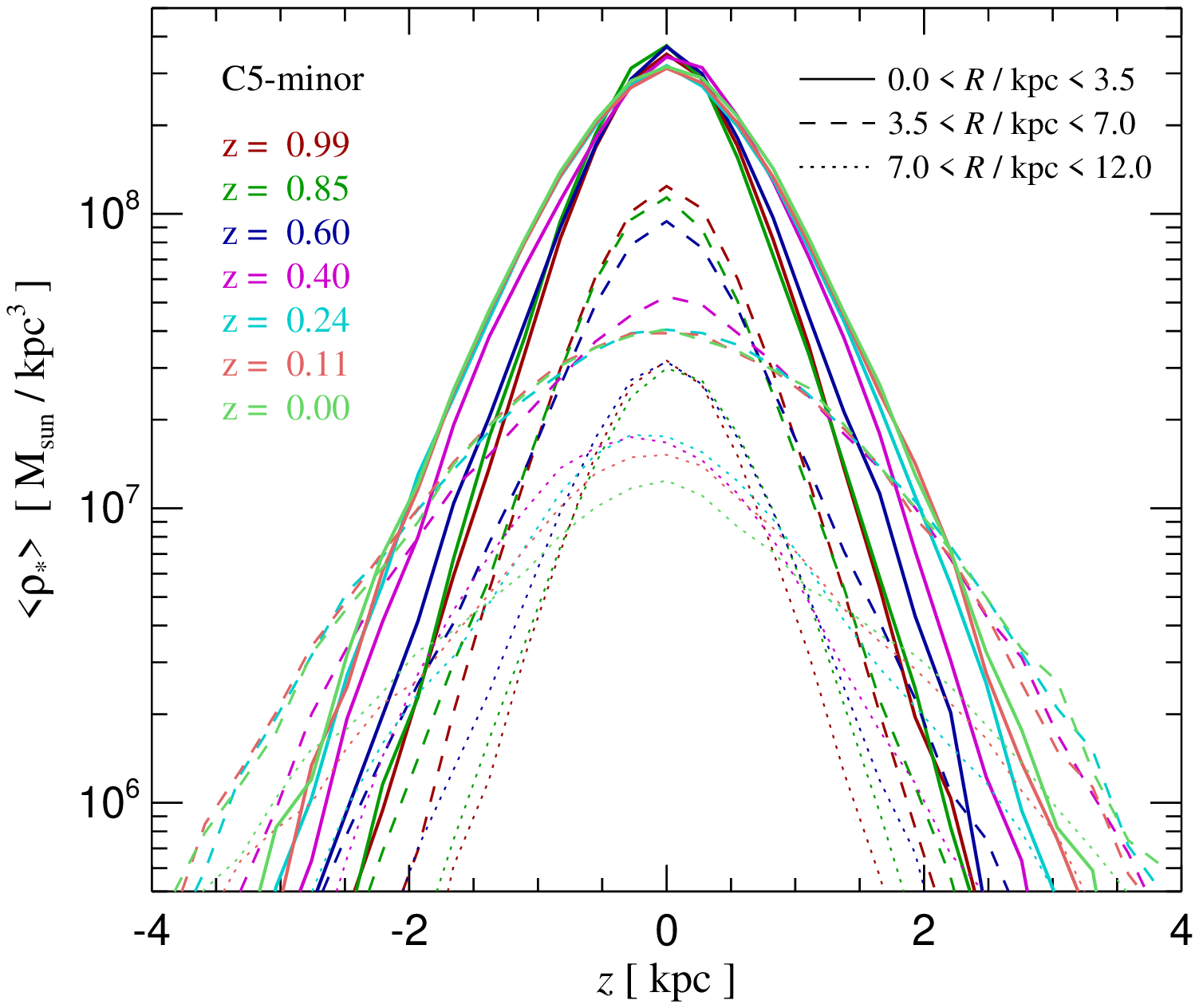}}%
\resizebox{4.5cm}{!}{\includegraphics{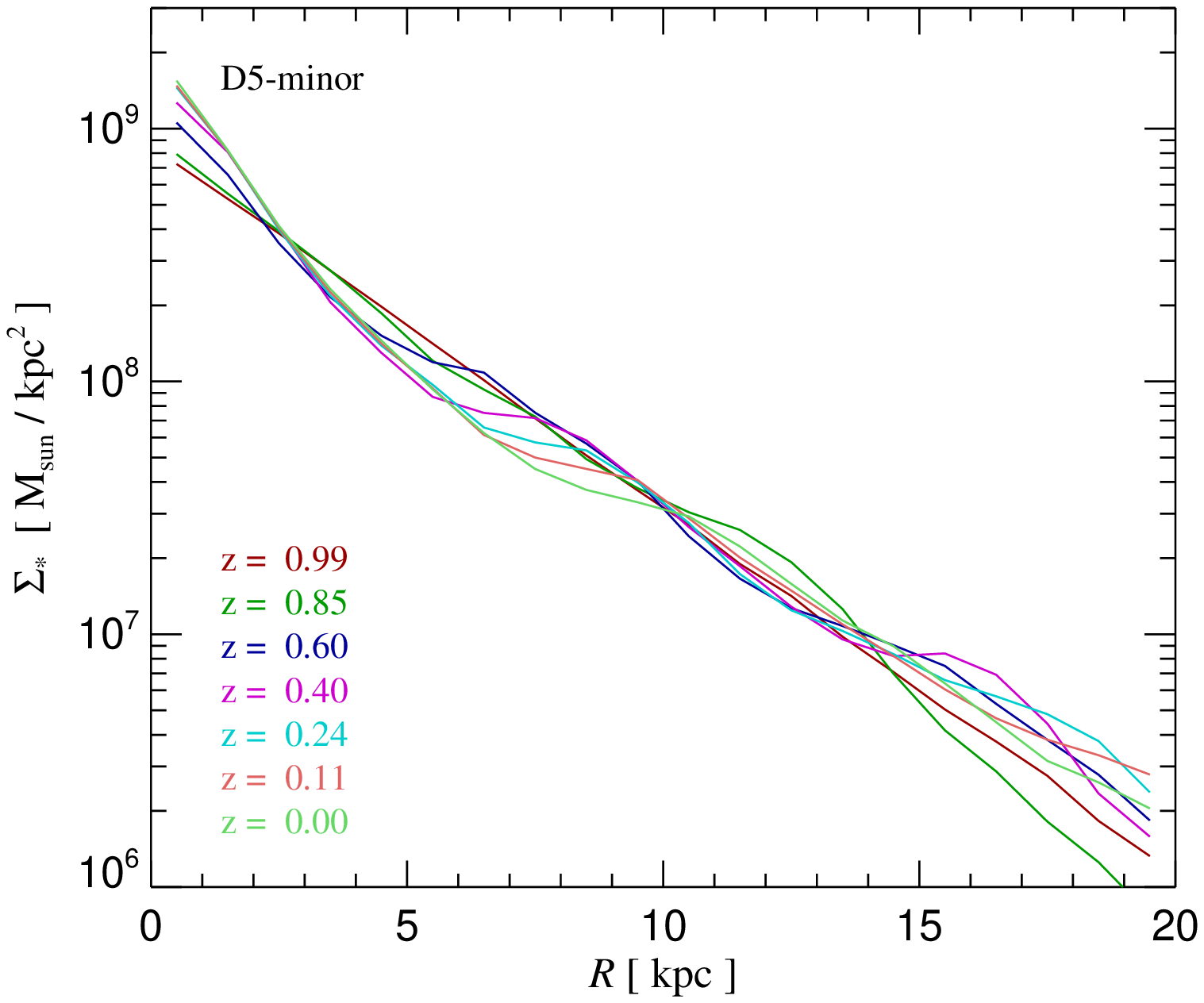}}%
\resizebox{4.5cm}{!}{\includegraphics{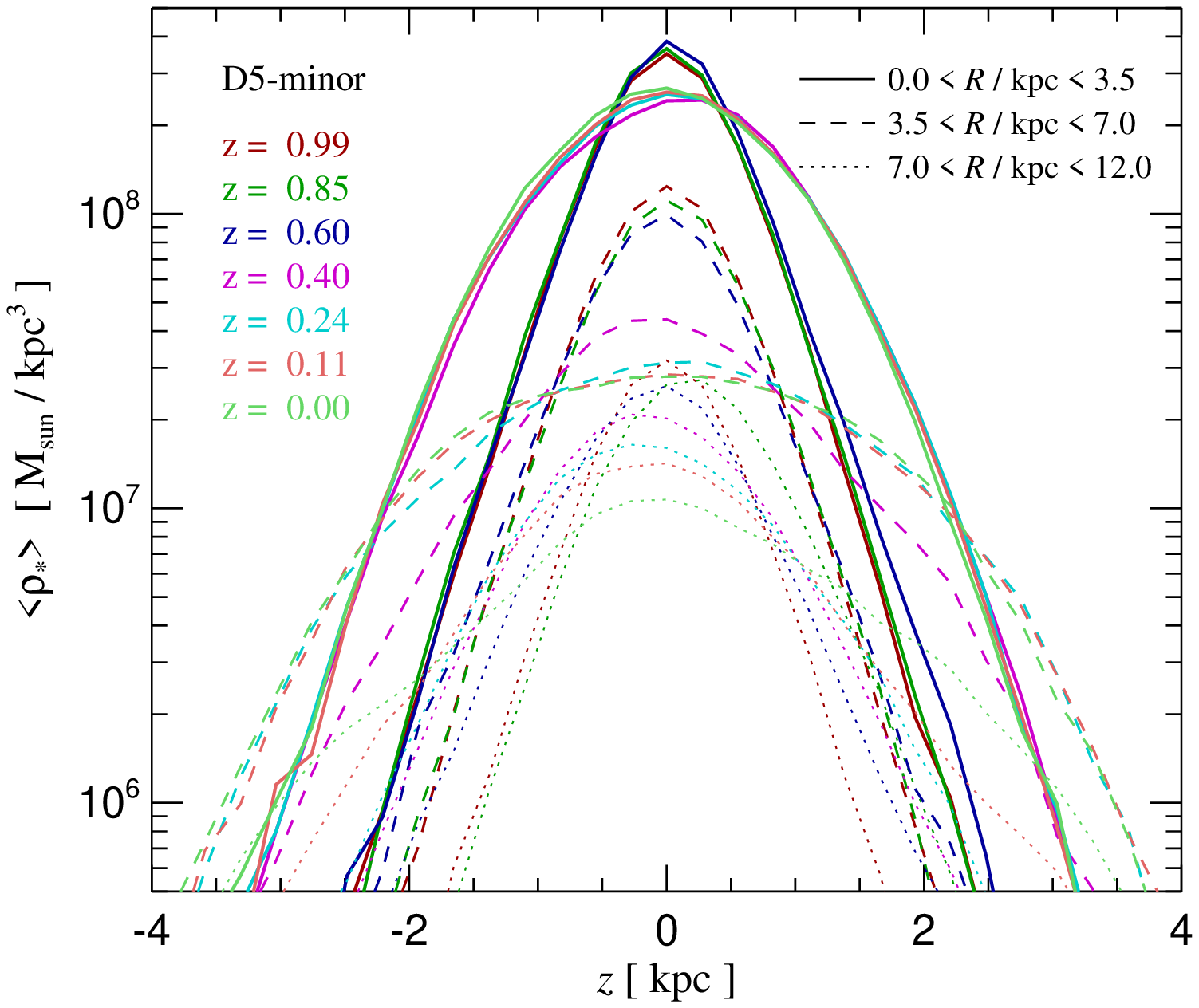}}\\%
\resizebox{4.5cm}{!}{\includegraphics{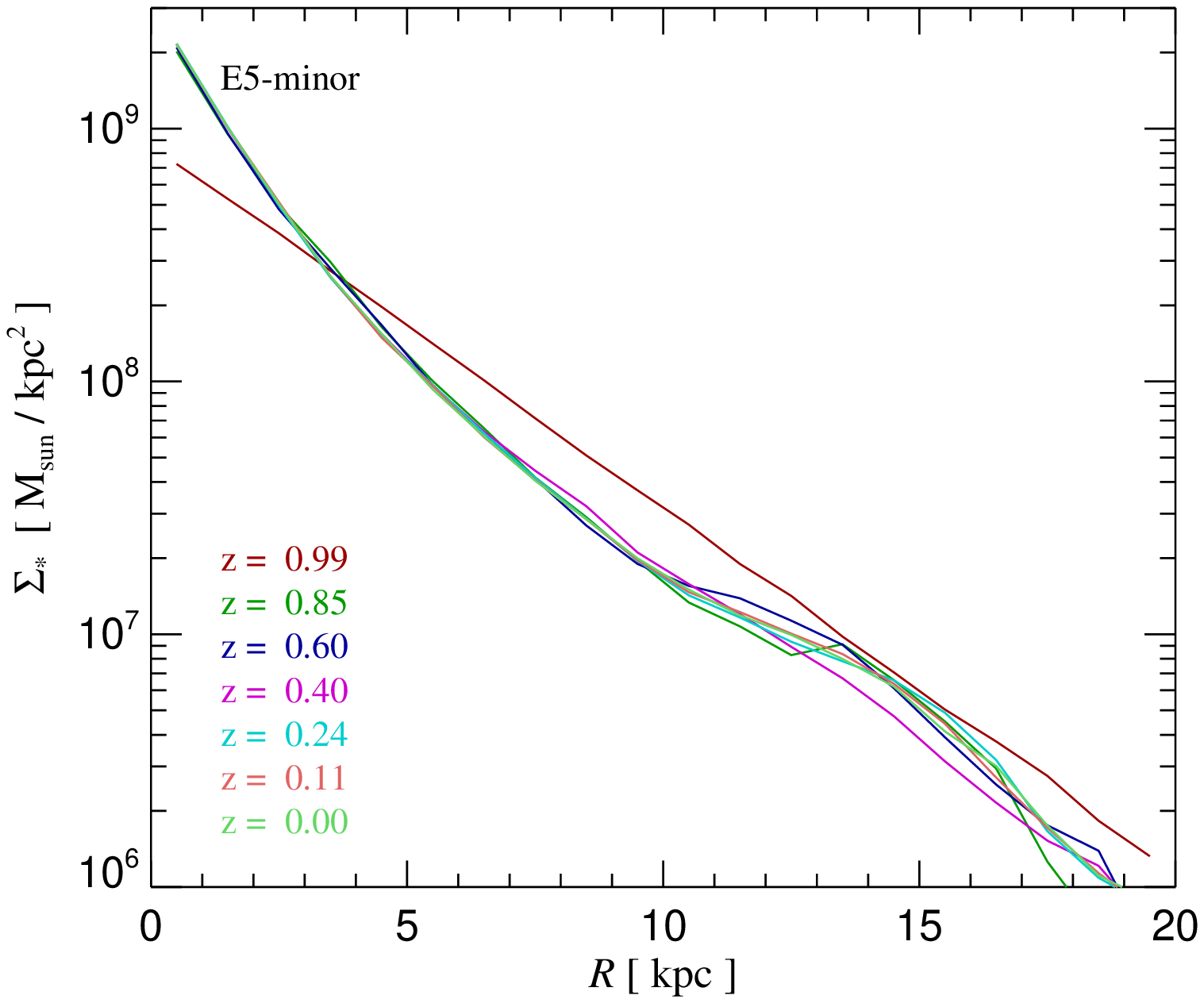}}%
\resizebox{4.5cm}{!}{\includegraphics{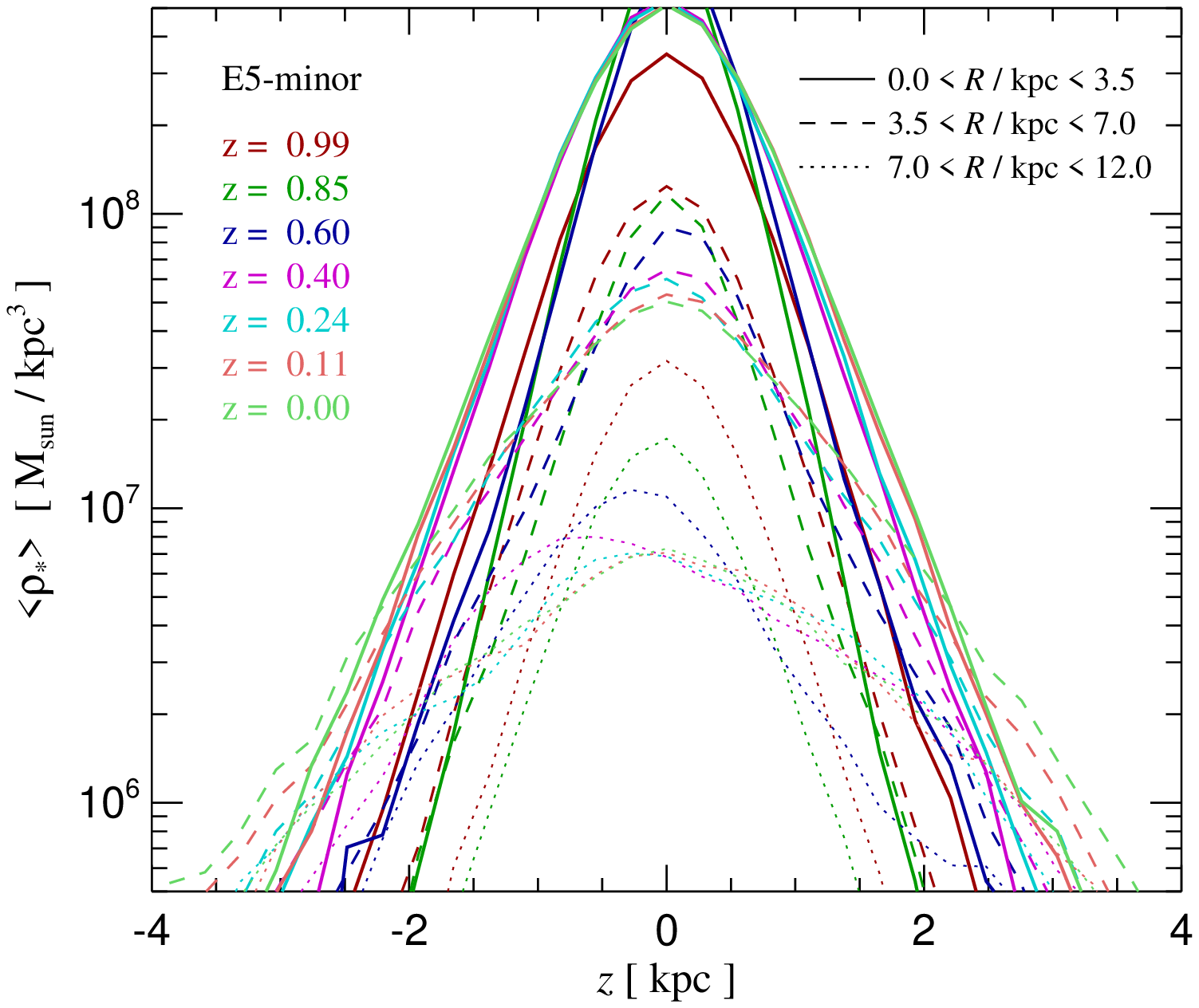}}%
\resizebox{4.5cm}{!}{\includegraphics{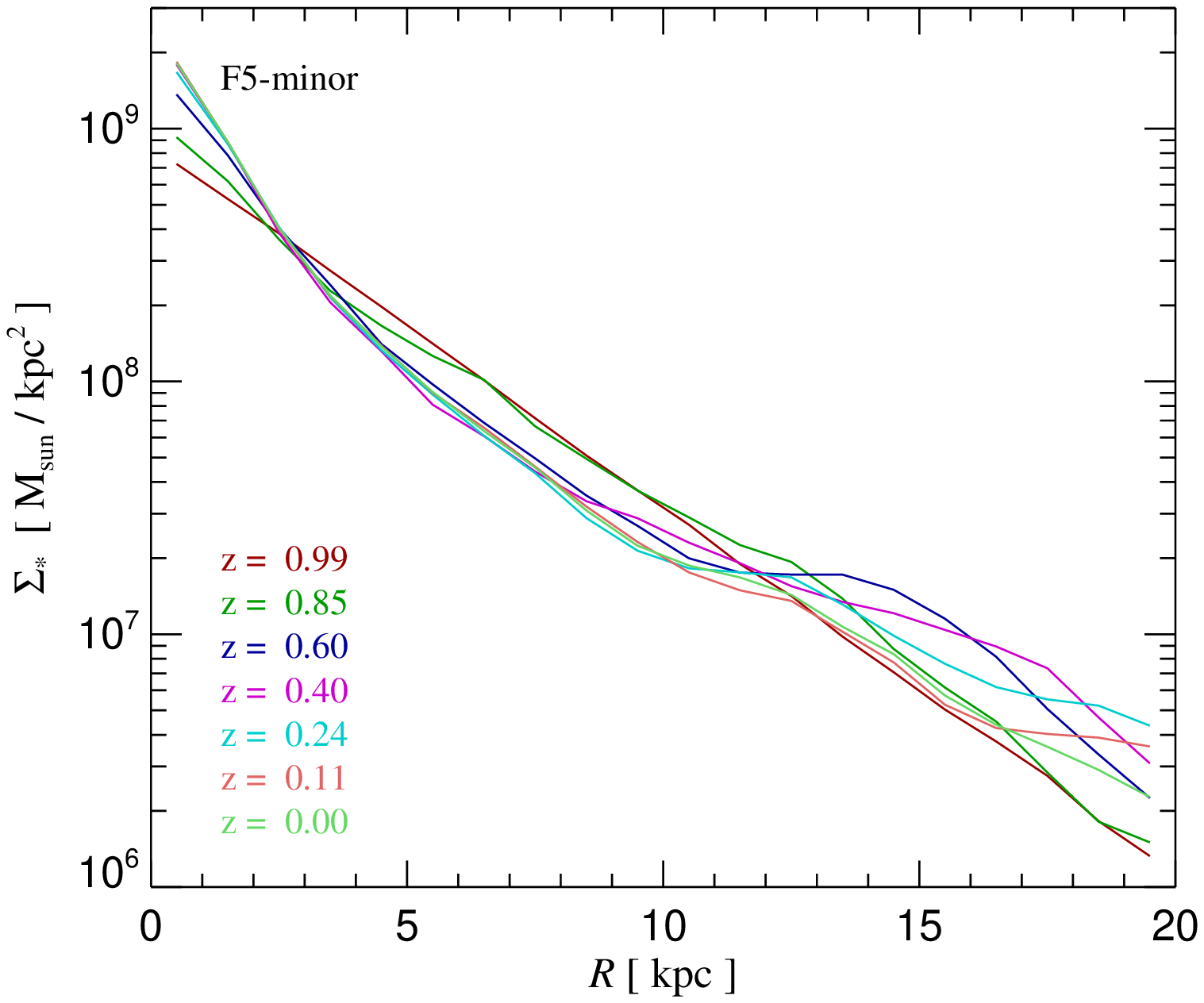}}%
\resizebox{4.5cm}{!}{\includegraphics{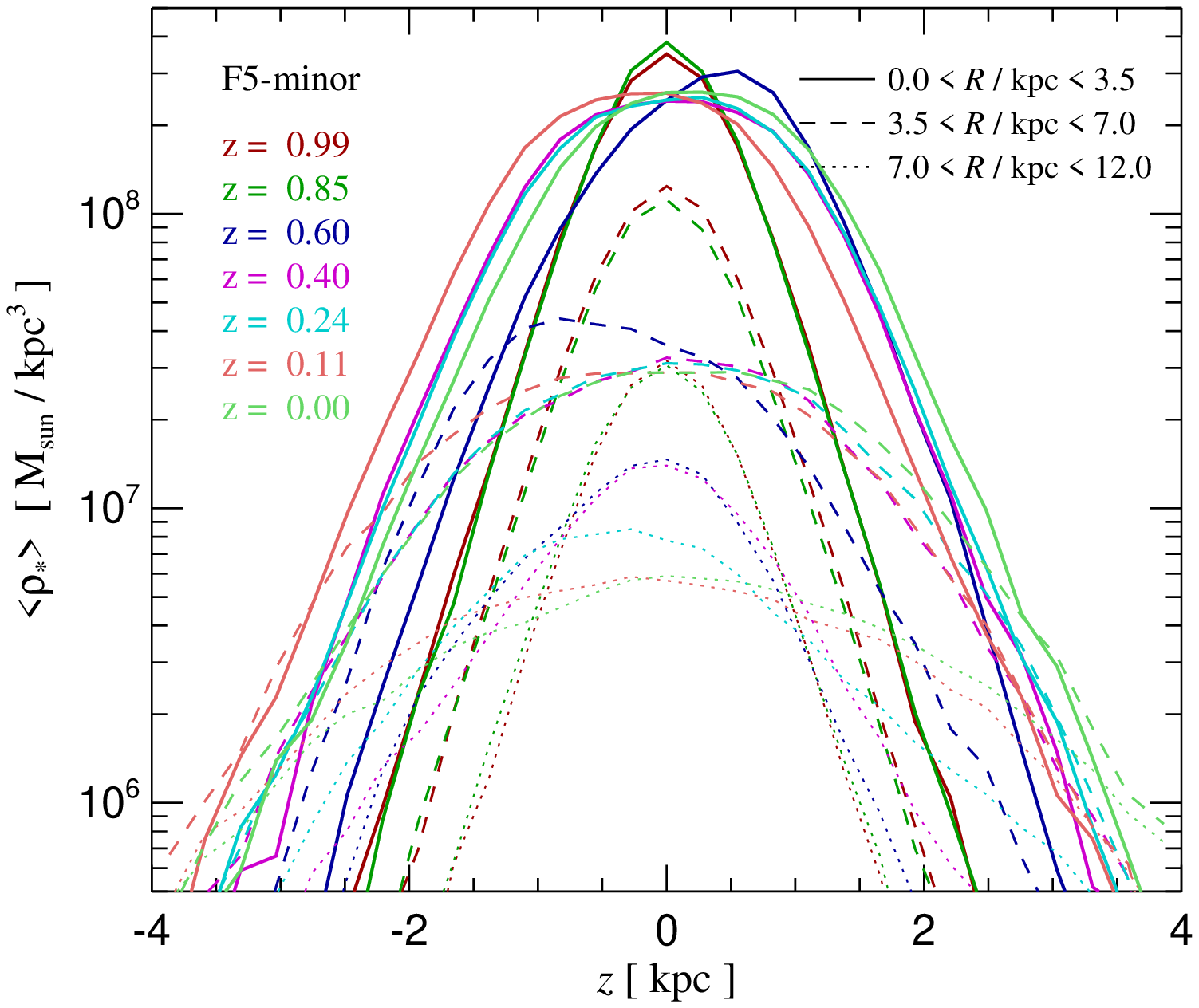}}\\%
\resizebox{4.5cm}{!}{\includegraphics{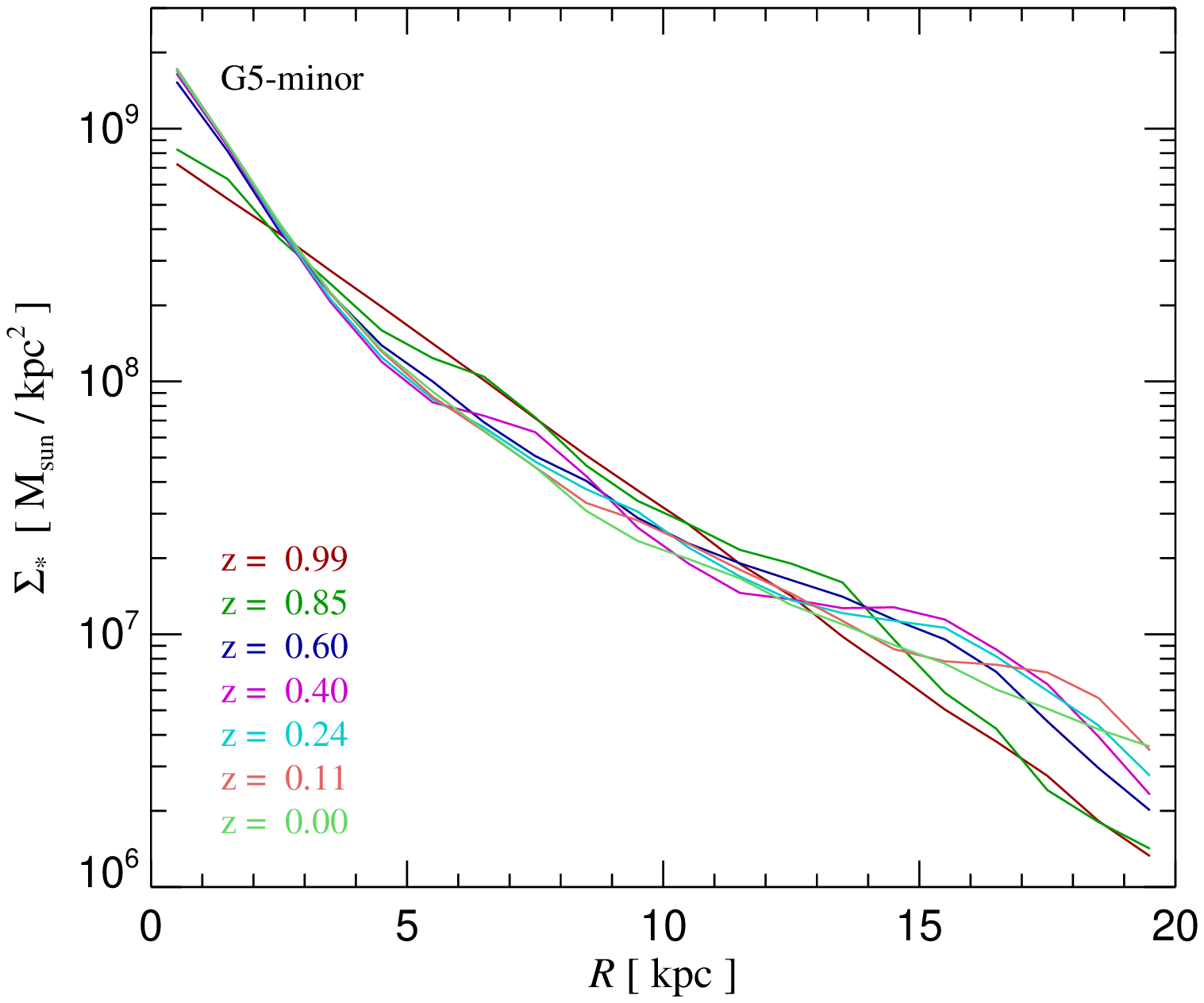}}%
\resizebox{4.5cm}{!}{\includegraphics{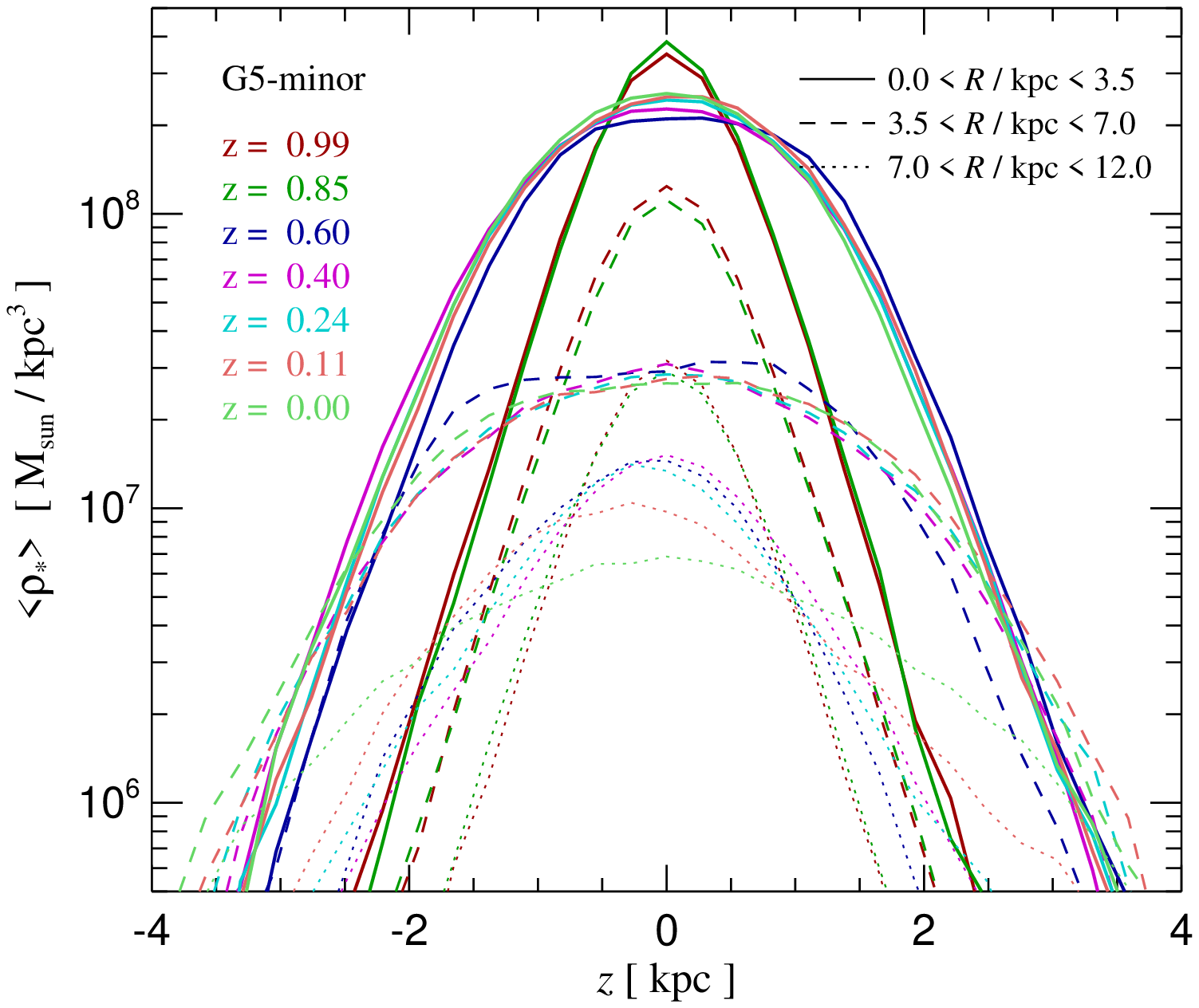}}%
\resizebox{4.5cm}{!}{\includegraphics{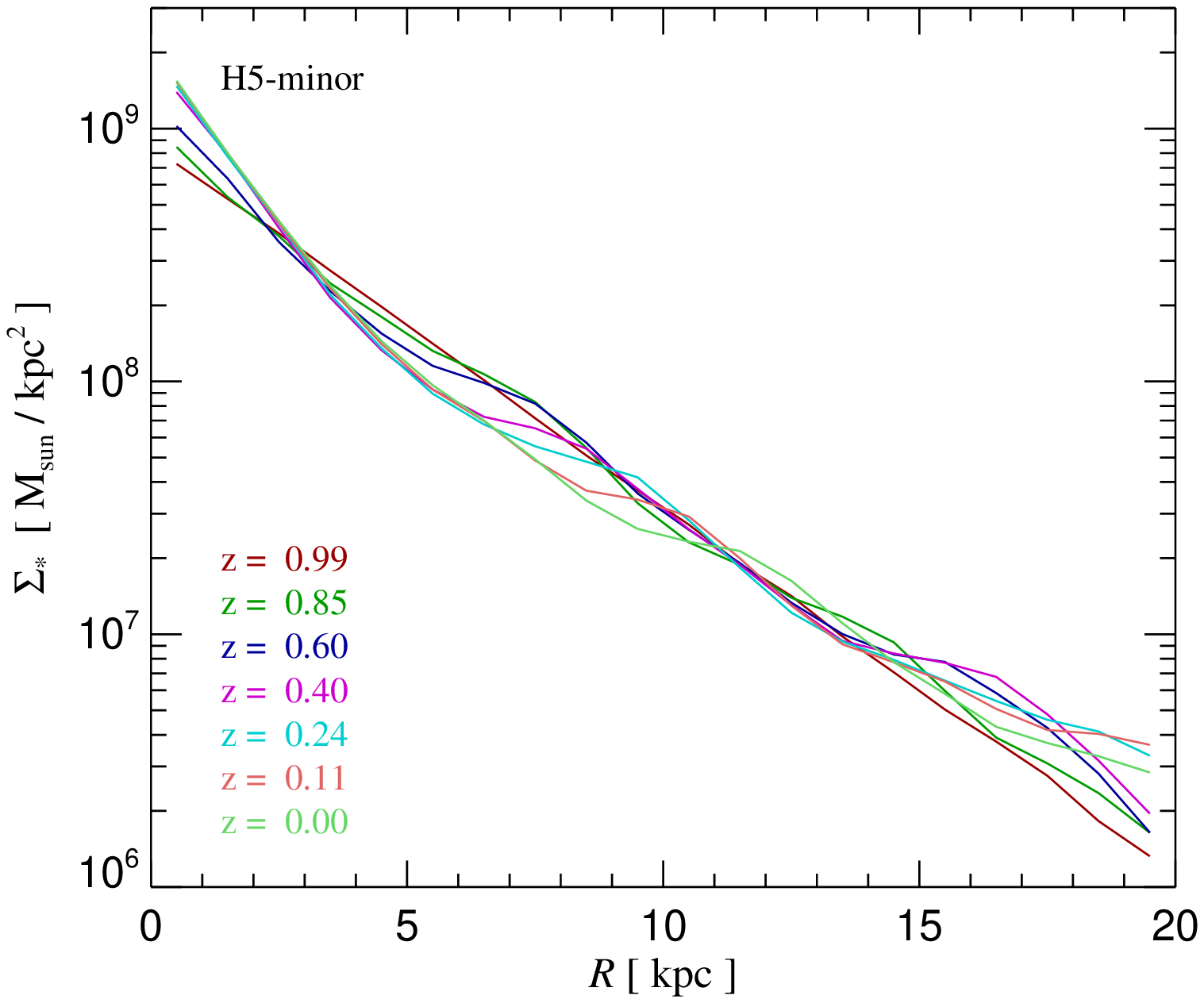}}%
\resizebox{4.5cm}{!}{\includegraphics{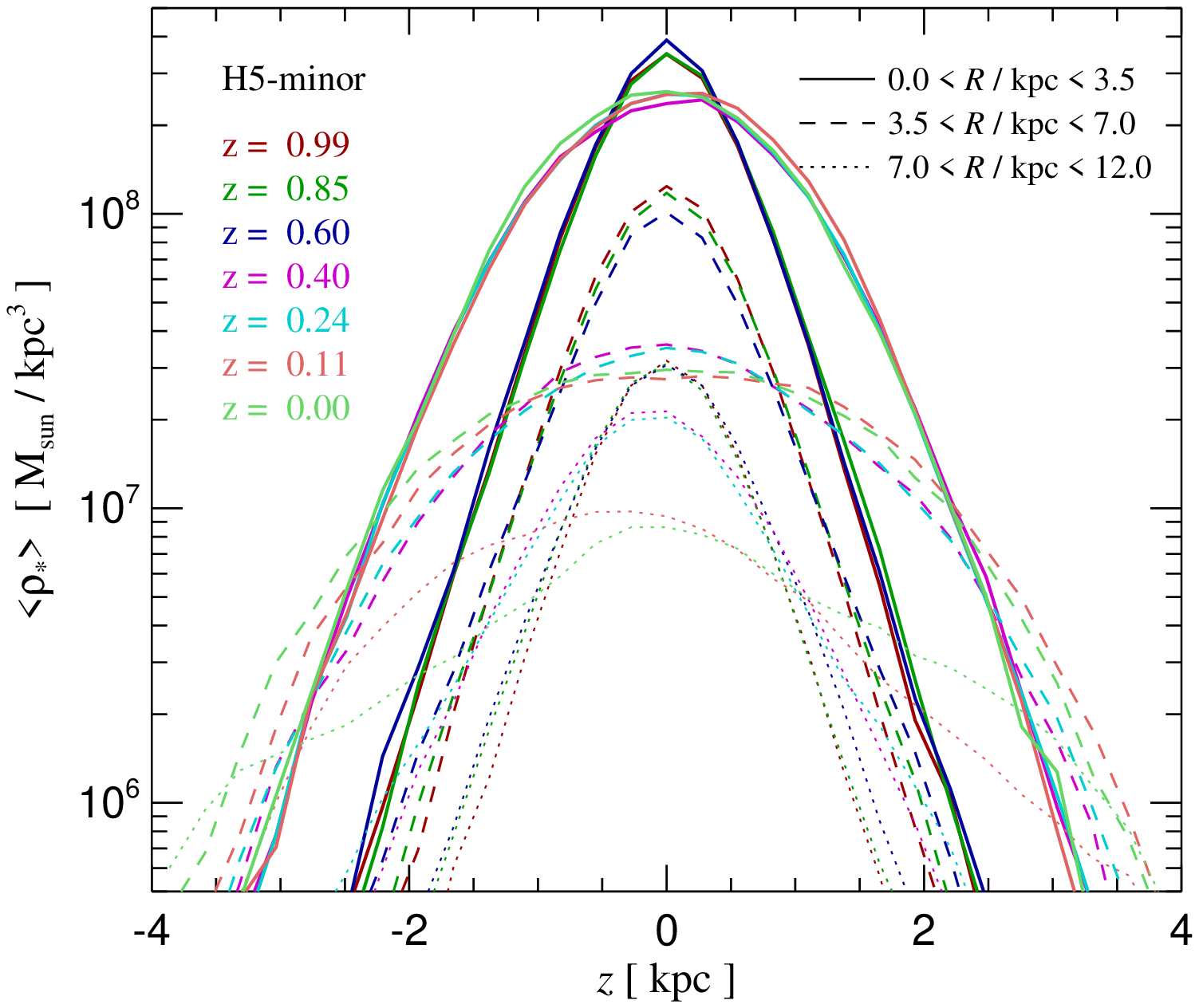}}%
\caption{Radial and vertical disk density profiles for our default
  disk simulations (series \#1) as a function of time. We show results
  for all of our eight Aquarius halos, in each case with a pair of
  panels where the surface density profile is shown on the left, and the
  vertical density profiles on the right. For the latter, three
  families of curves are shown, corresponding to averages over
  different radial ranges ($R < 3.5\,{\rm kpc}$, $3.5\,{\rm kpc} < R <
  7\,{\rm kpc}$, and $7\,{\rm kpc} < R <
  12\,{\rm kpc}$, as labelled). In all panels, the initial density
  distribution when the disk goes live is shown together with 6
  subsequent times down to $z=0$, and at each time, the disk plane has
  been defined based on the angular momentum of the stars in the inner
  $5\,{\rm kpc}$.
\label{fig:default_radialprofiles}
}
\end{center}
\end{figure*}

\begin{figure*}
\begin{center}
\setlength{\unitlength}{1cm}
\resizebox{4.5cm}{!}{\includegraphics{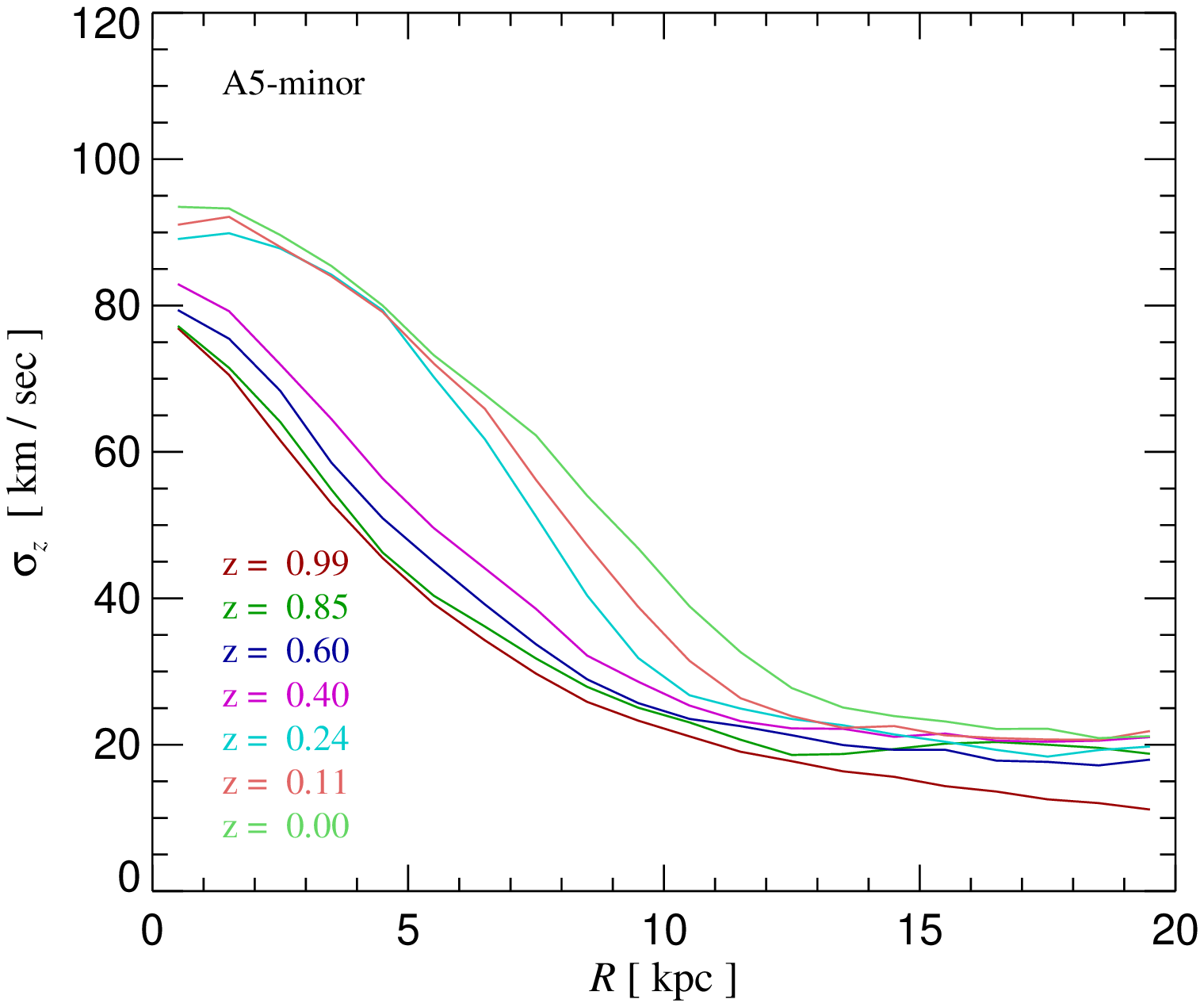}}%
\resizebox{4.5cm}{!}{\includegraphics{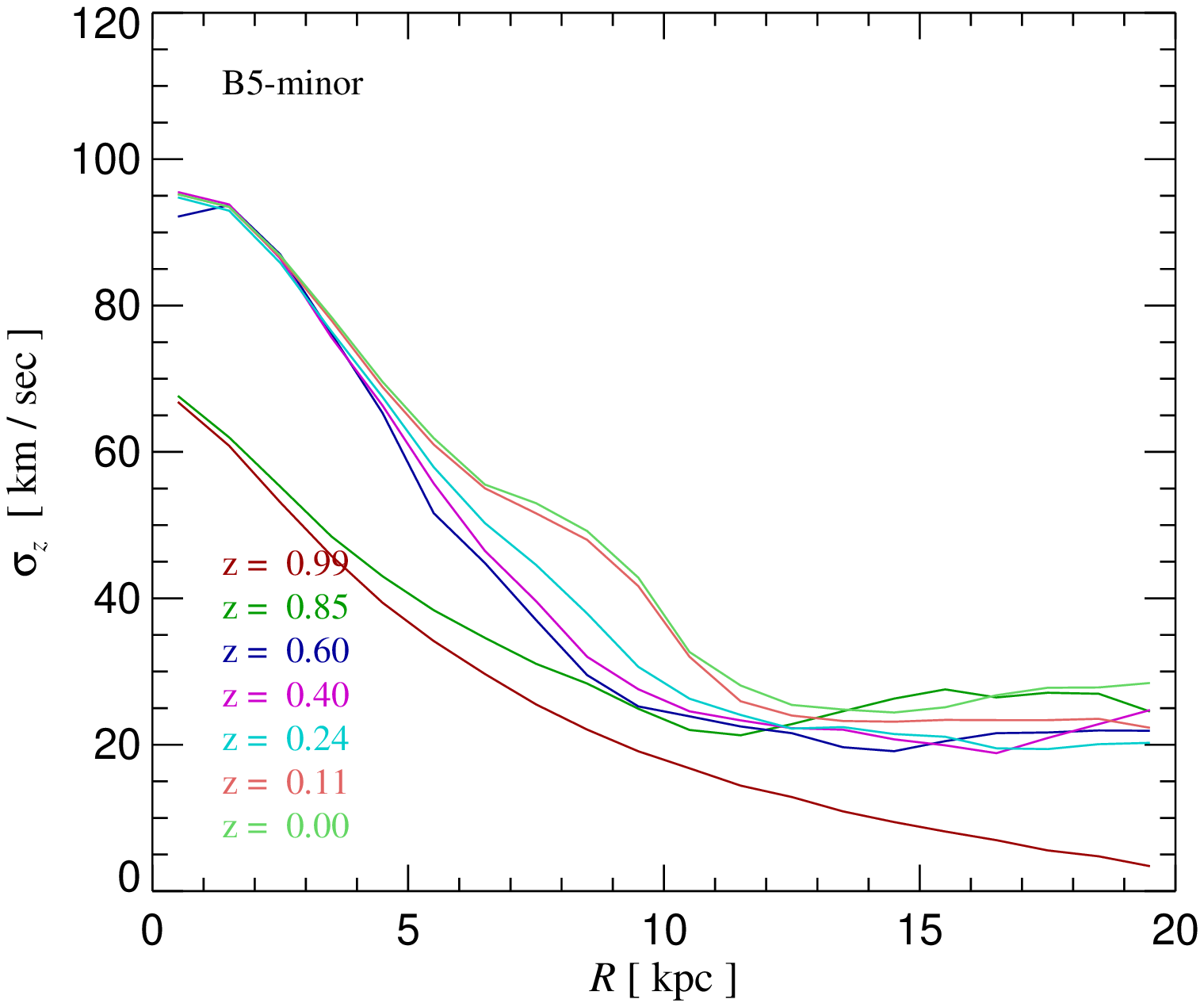}}%
\resizebox{4.5cm}{!}{\includegraphics{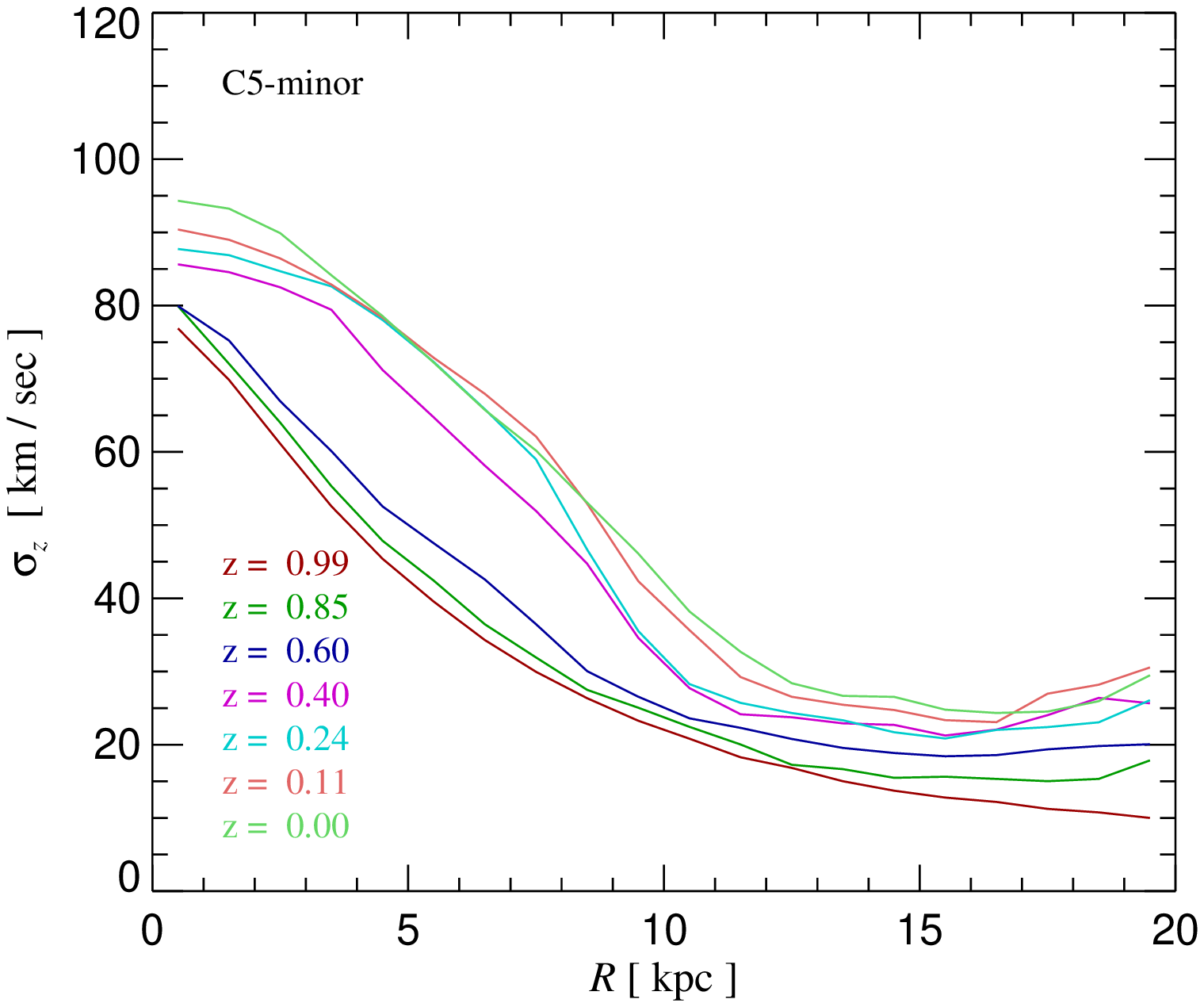}}%
\resizebox{4.5cm}{!}{\includegraphics{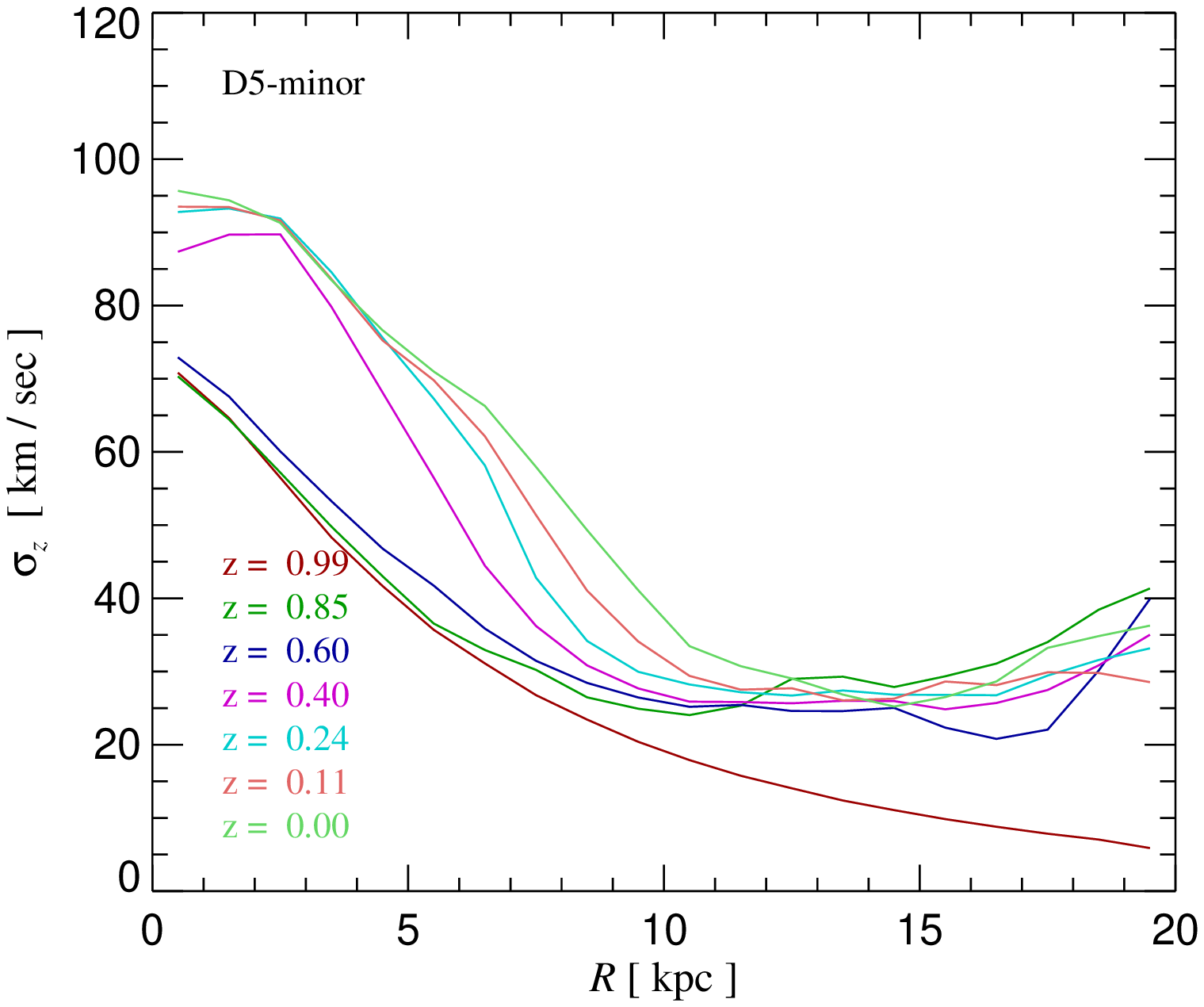}}\\%
\resizebox{4.5cm}{!}{\includegraphics{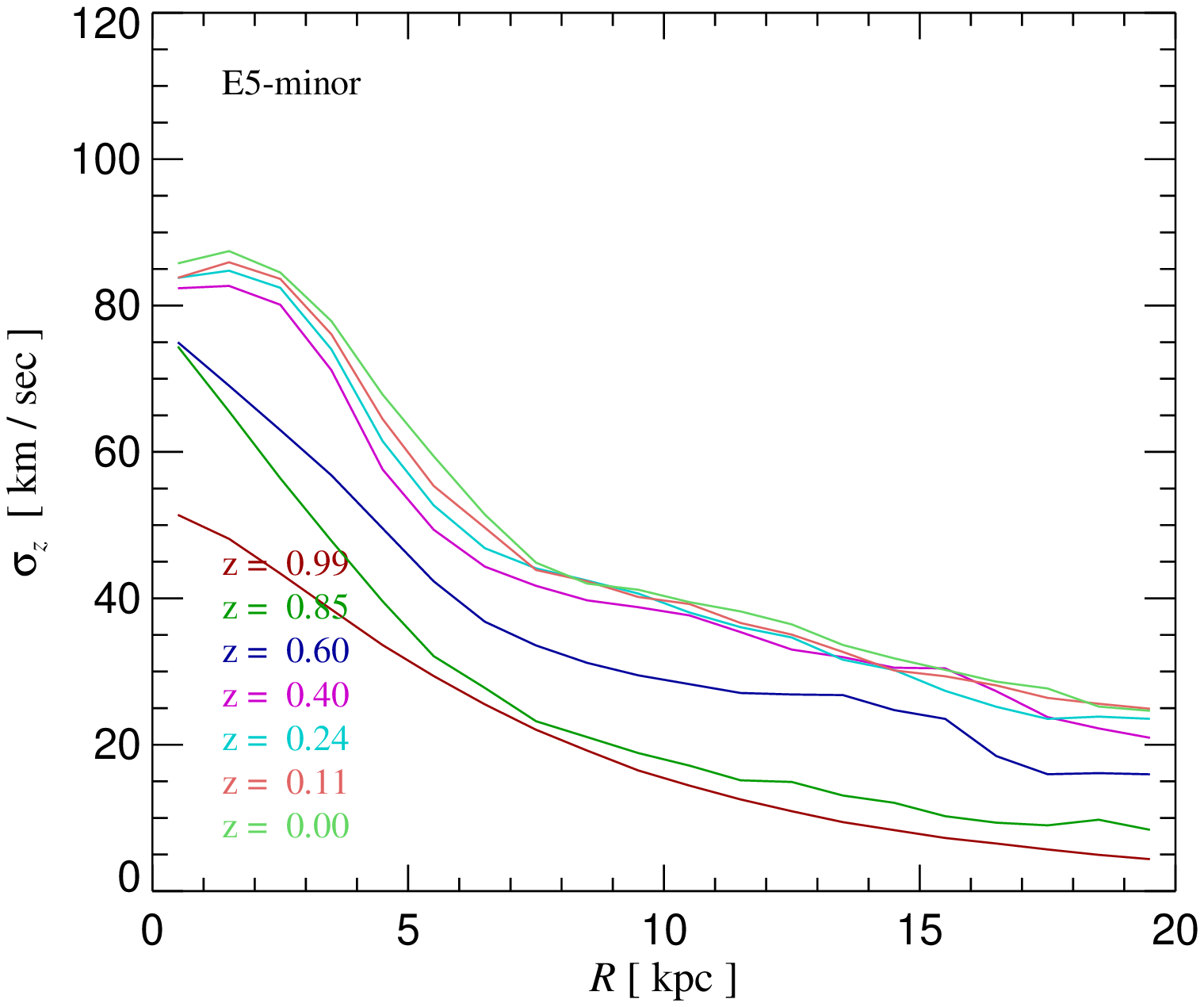}}%
\resizebox{4.5cm}{!}{\includegraphics{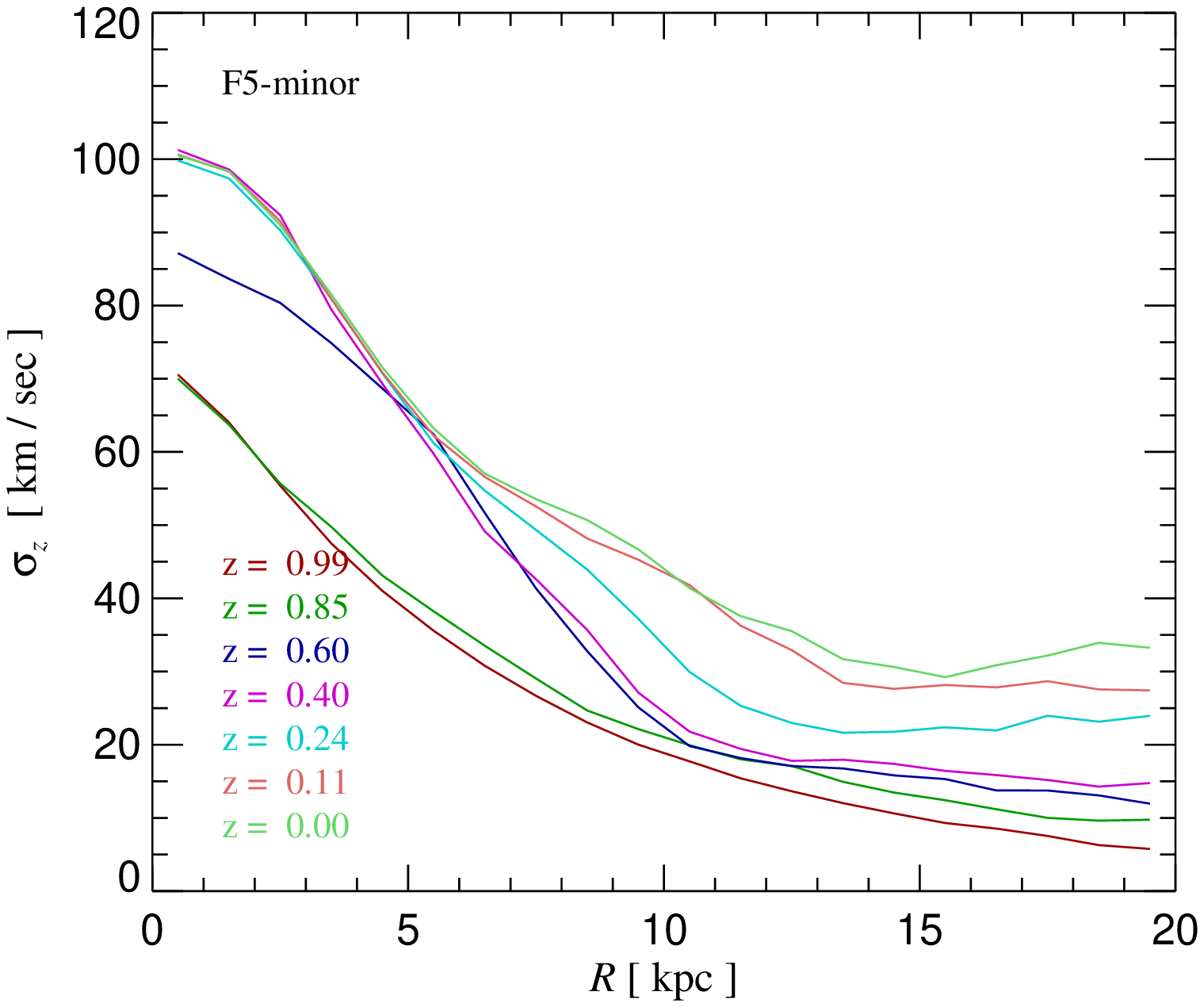}}%
\resizebox{4.5cm}{!}{\includegraphics{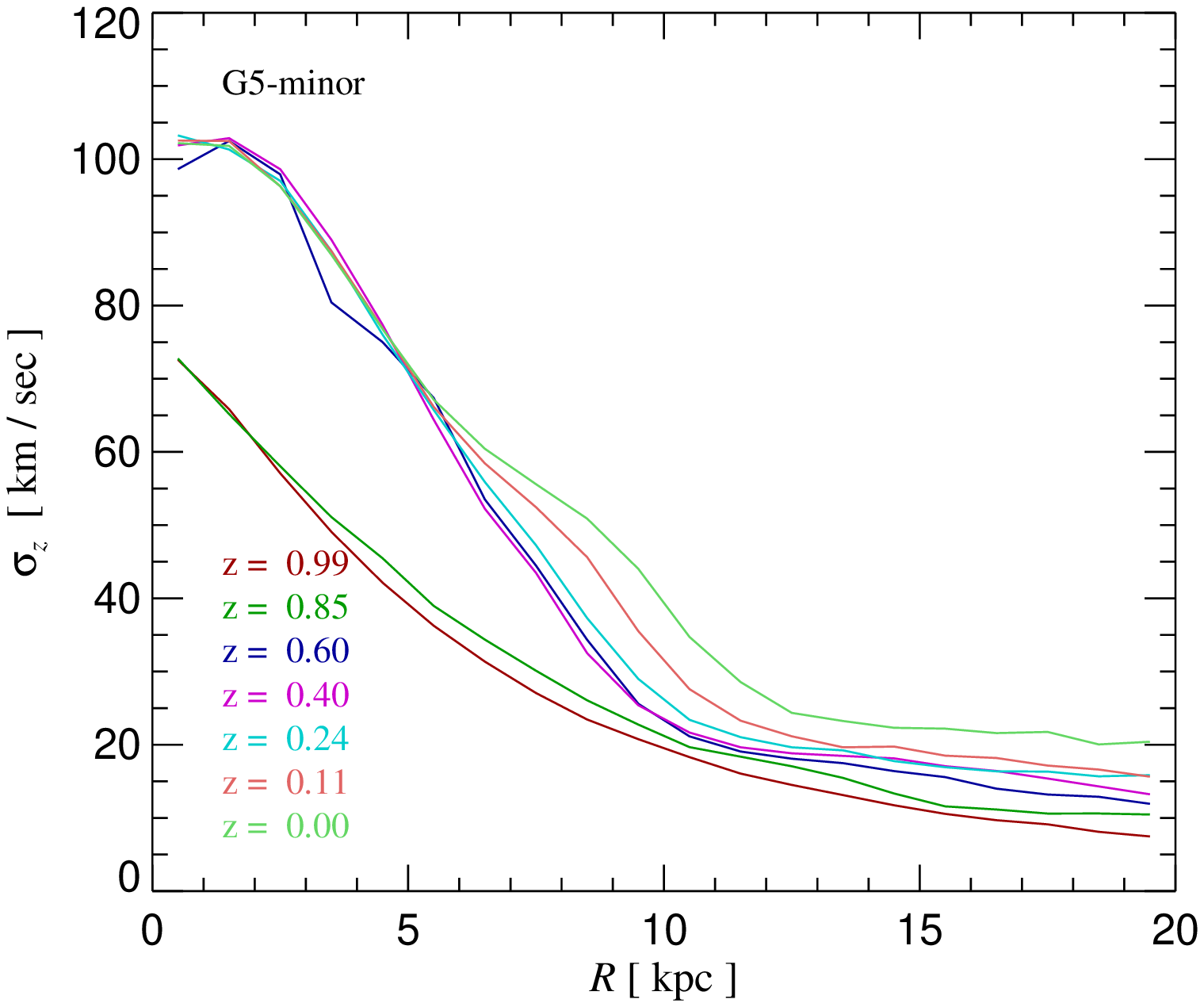}}%
\resizebox{4.5cm}{!}{\includegraphics{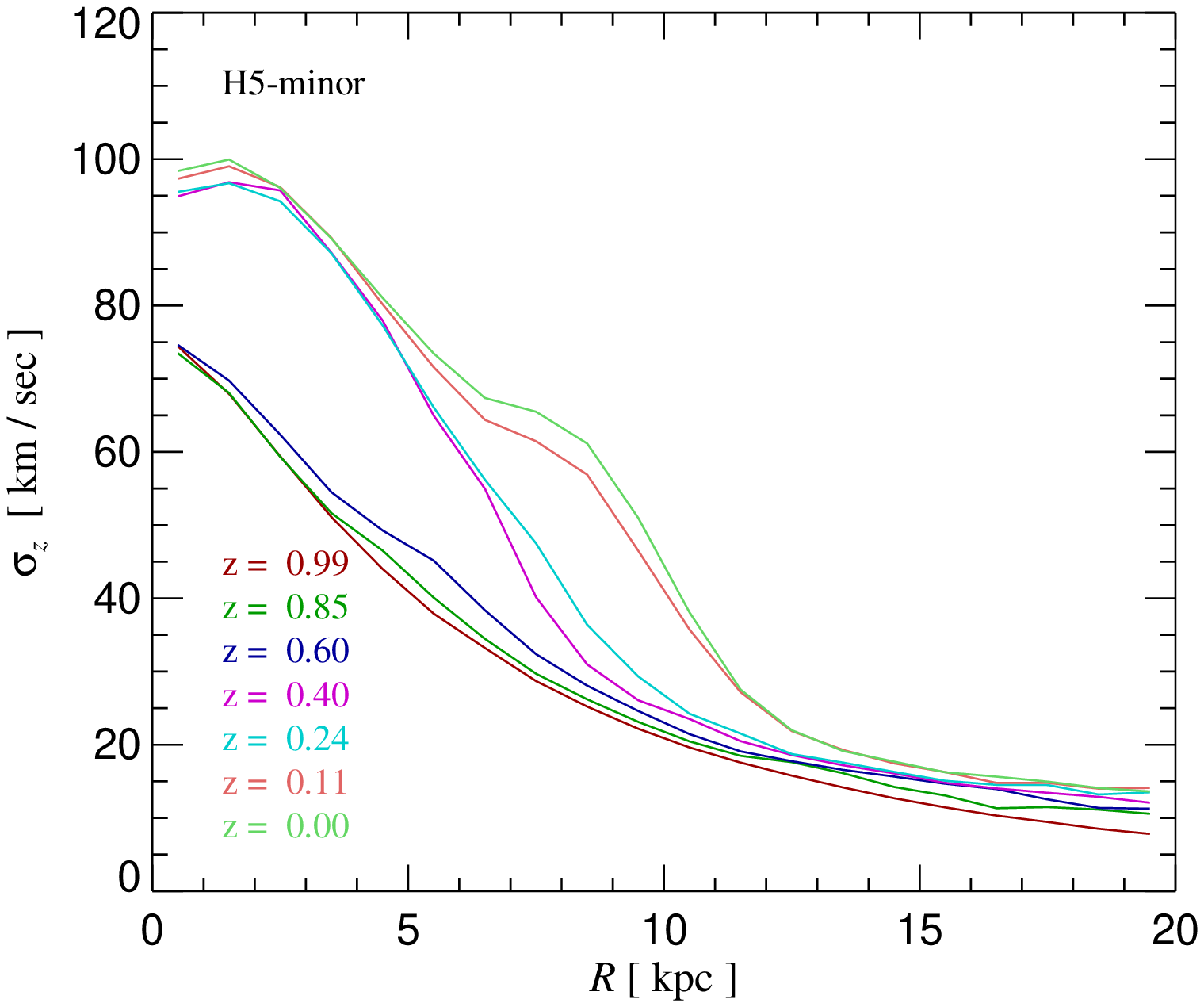}}\\%
\resizebox{4.5cm}{!}{\includegraphics{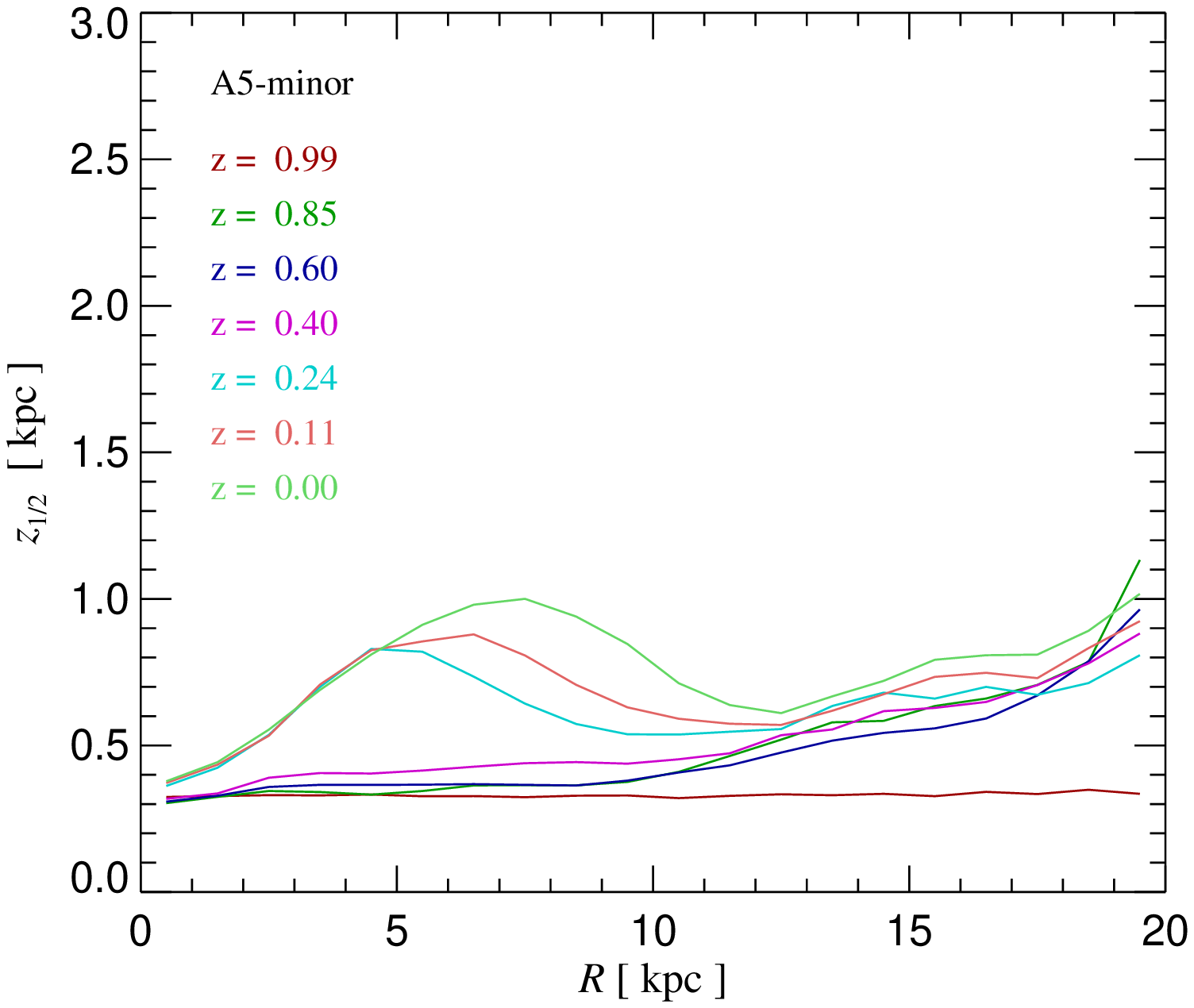}}%
\resizebox{4.5cm}{!}{\includegraphics{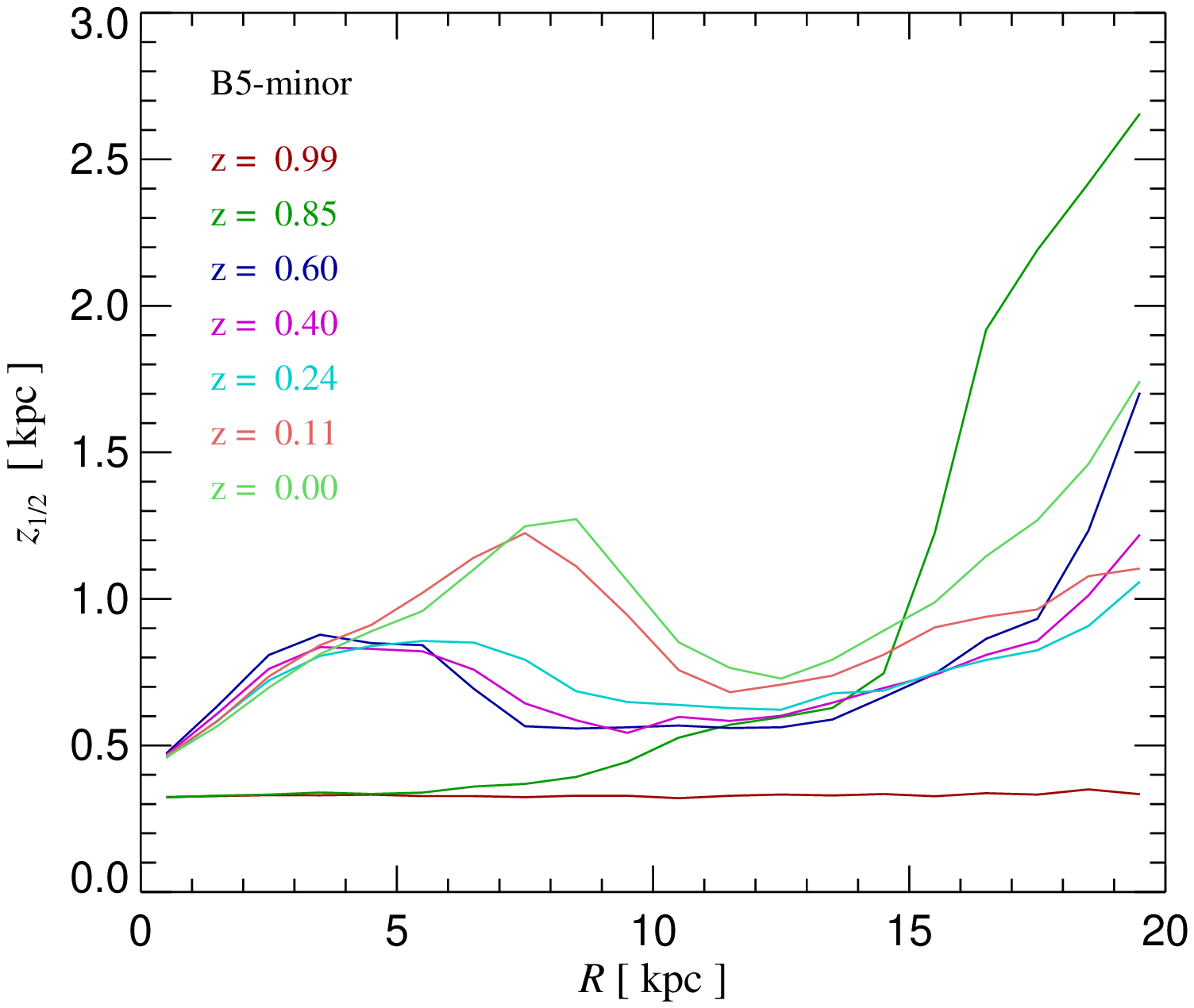}}%
\resizebox{4.5cm}{!}{\includegraphics{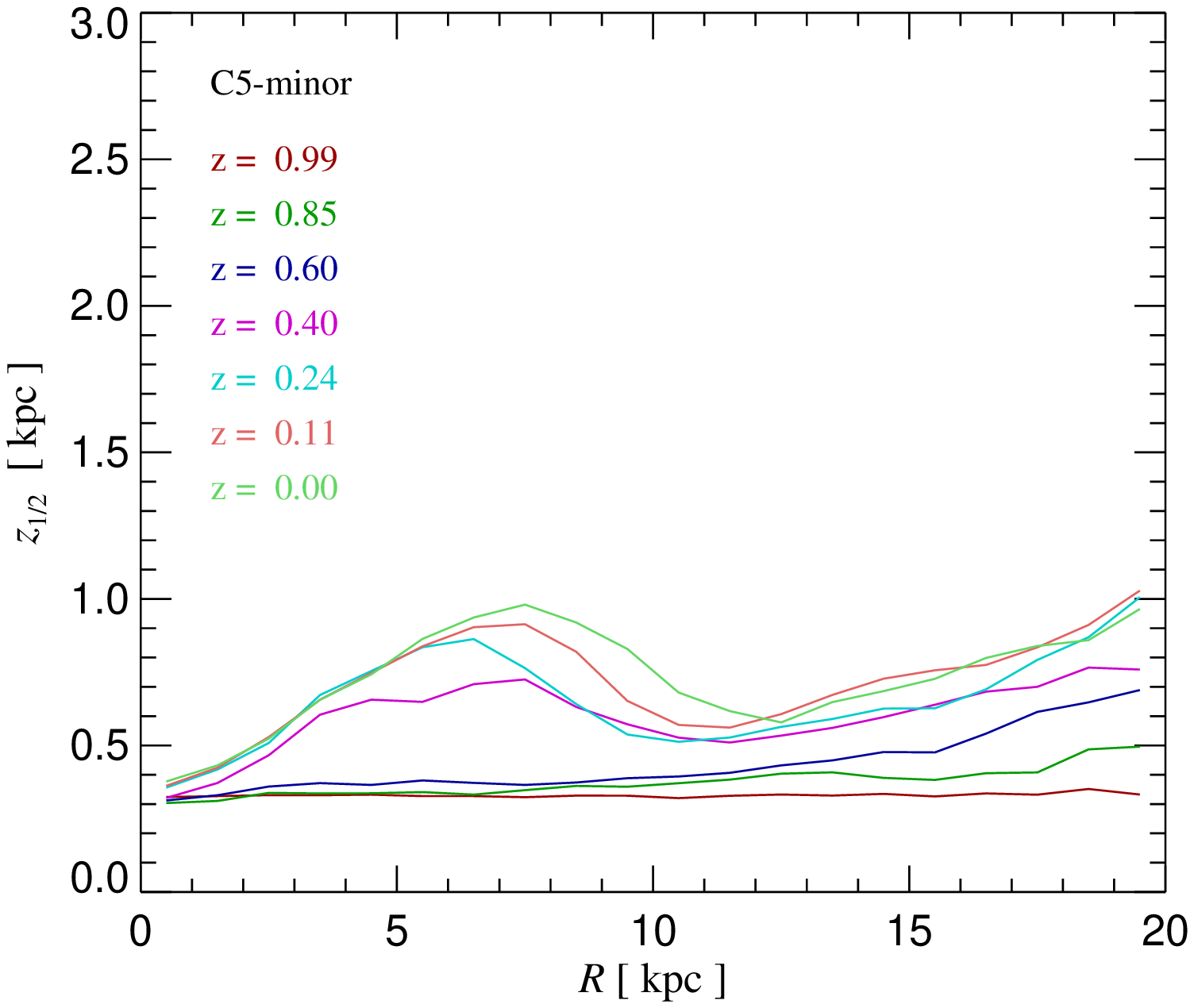}}%
\resizebox{4.5cm}{!}{\includegraphics{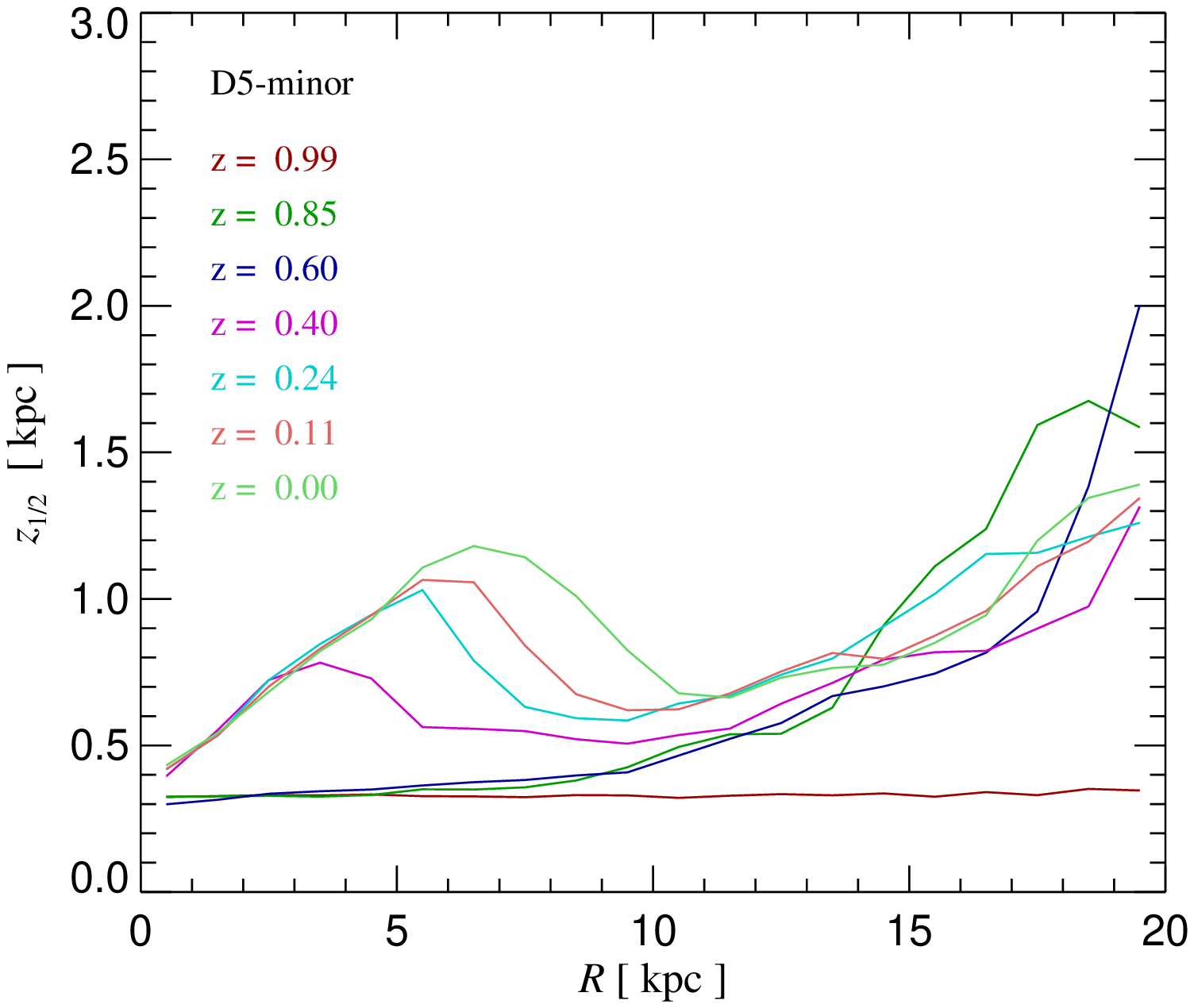}}\\%
\resizebox{4.5cm}{!}{\includegraphics{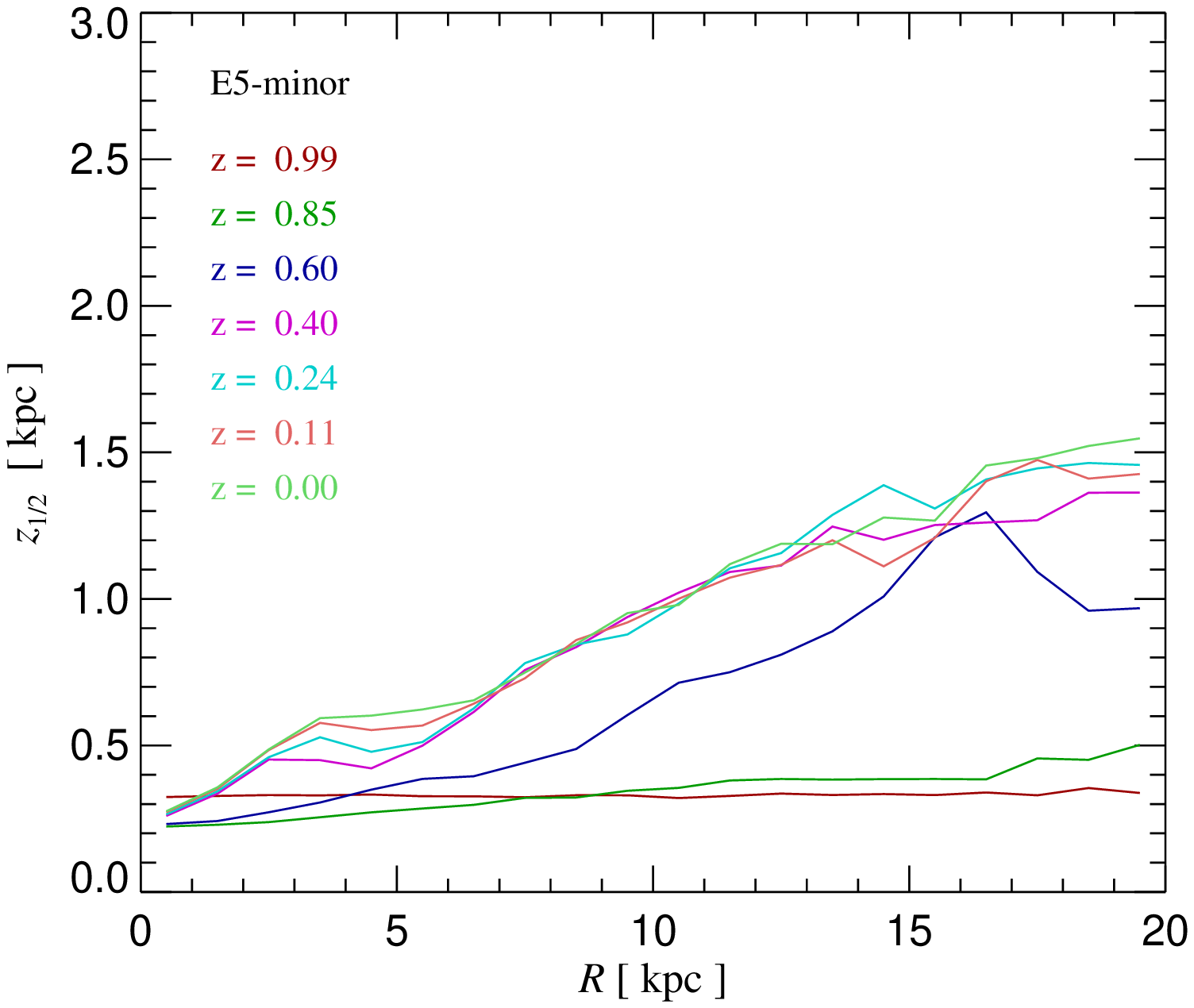}}%
\resizebox{4.5cm}{!}{\includegraphics{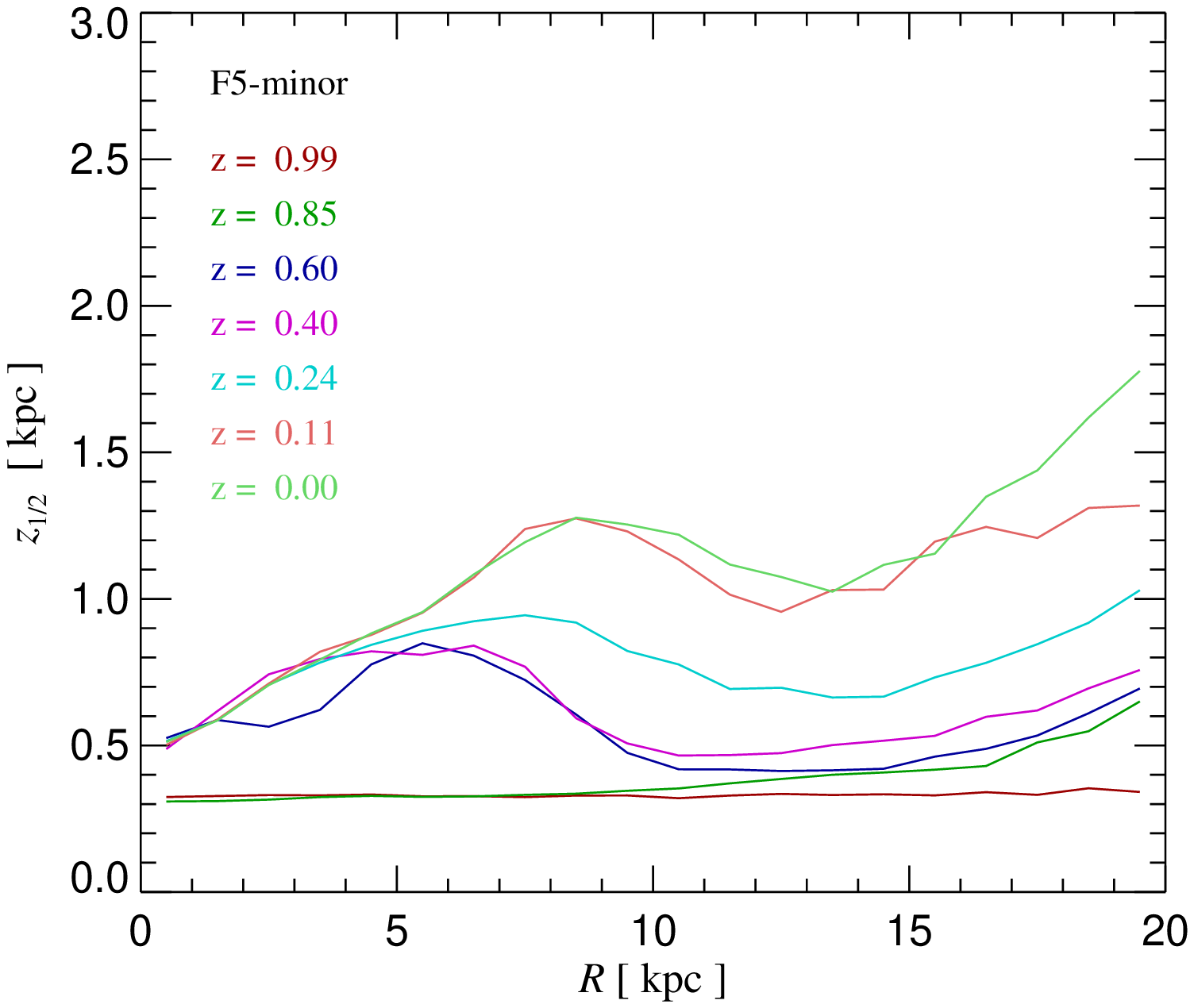}}%
\resizebox{4.5cm}{!}{\includegraphics{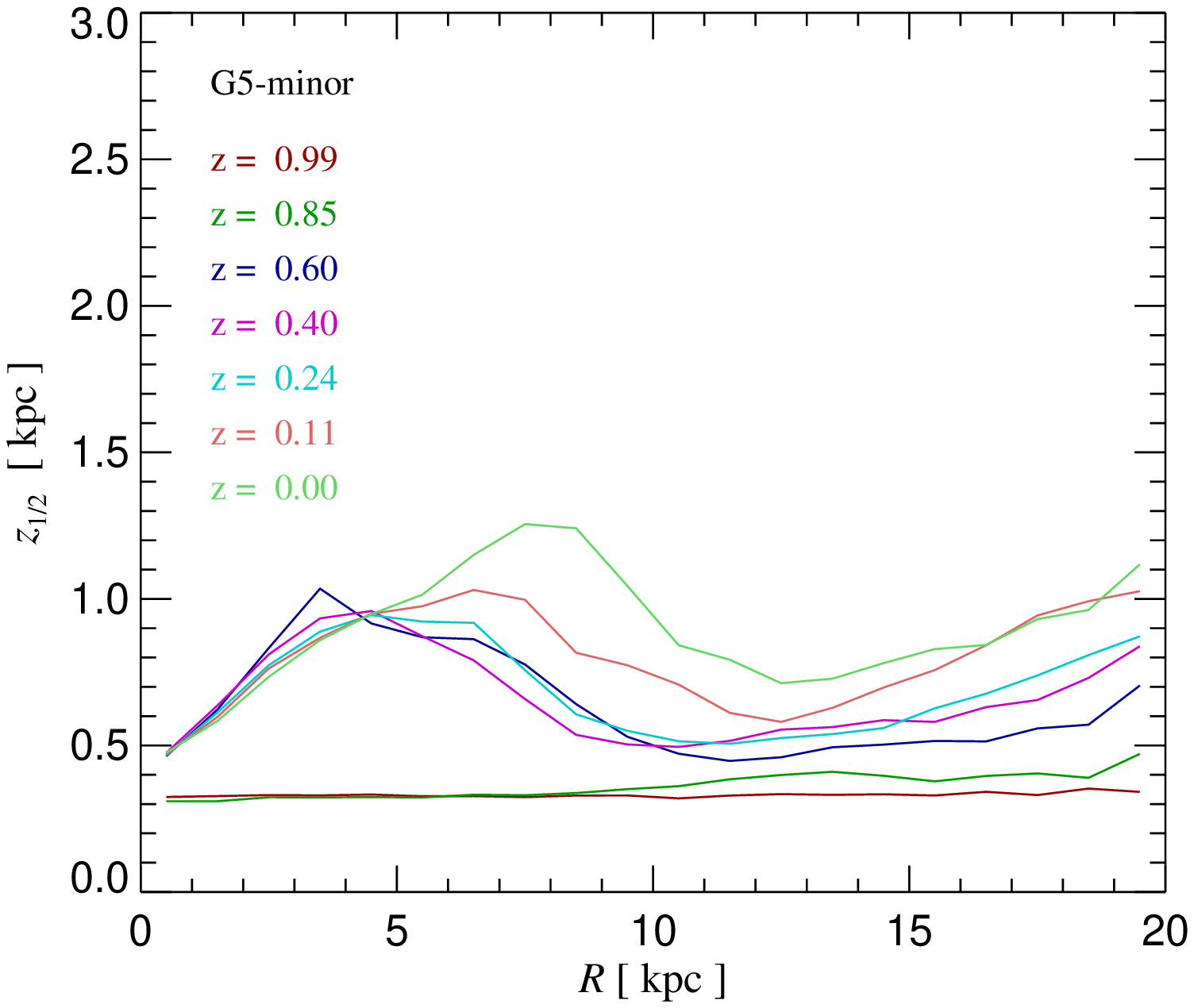}}%
\resizebox{4.5cm}{!}{\includegraphics{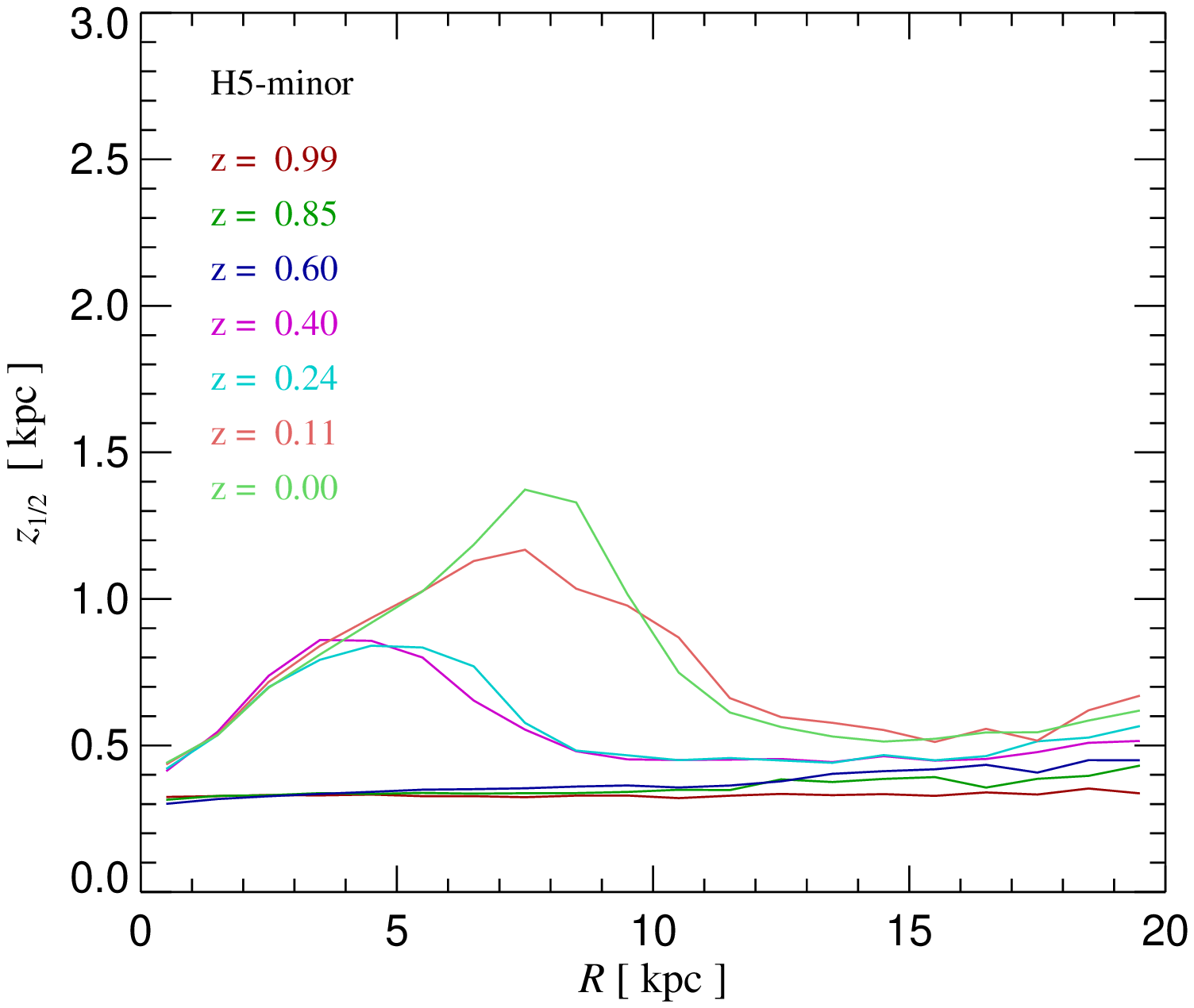}}\\%
\caption{Evolution of the vertical velocity dispersion profile and of
  the half-mass height profile of the disk particles in our default
  disk insertion simulations (series \#1).
 The top panels give the expectation value of the vertical stellar
 velocity $\sigma_z = \left< v_z^2\right>^{1/2}$ measured in different cylindrical 
 shells aligned with the stellar spin axis of the stars at the
 corresponding times. The height profiles give the median of $|z|$
 relative to the disk plane, i.e.~half the stars have distances
 below/above $z_{1/2}$ from the disk plane.  
\label{fig:default_heightprofiles}
}
\end{center}
\end{figure*}

\begin{figure*}
\begin{center}
\setlength{\unitlength}{1cm}
\resizebox{8.5cm}{!}{\includegraphics{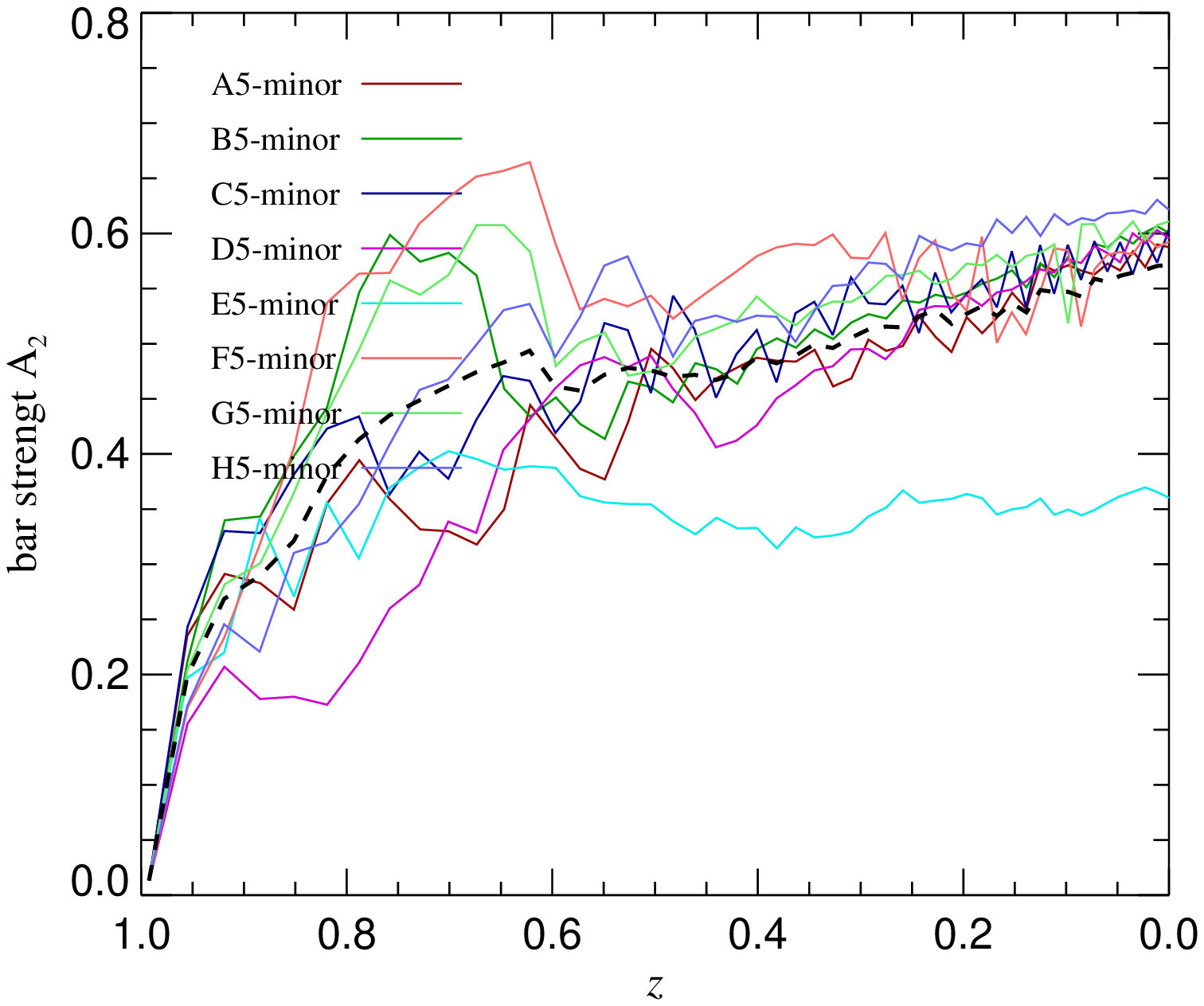}}%
\resizebox{8.5cm}{!}{\includegraphics{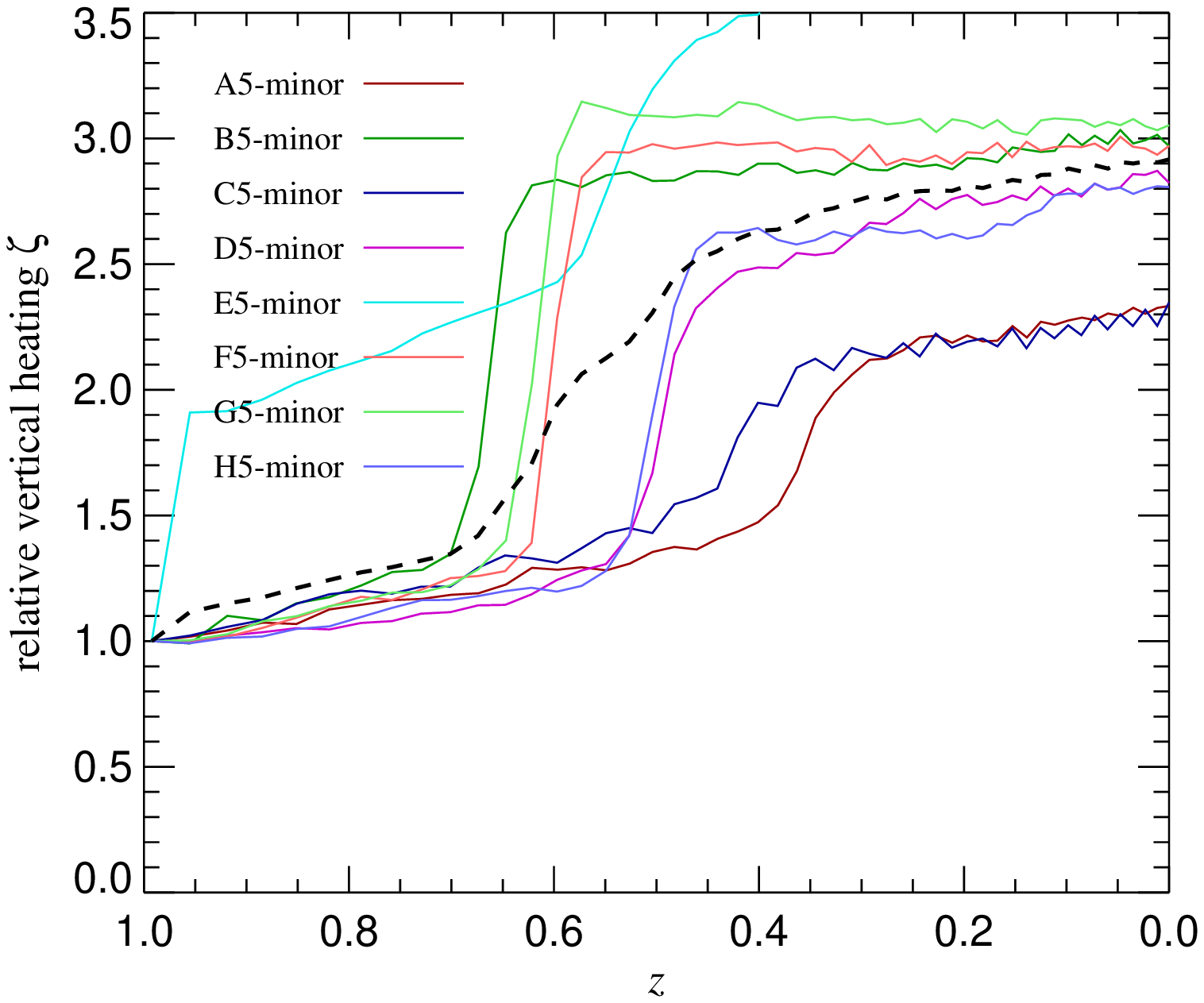}}%
\caption{Bar strength parameter $A_2$ (left panel) and relative
  vertical heating (right panel) in the simulations of our default
  pure disk insertion simulations (series \#1), as a function of
  time. It is clearly seen that all the models develop a strong bar
  characterized by $A_2\simeq 0.6$, except for model E, which yields
  $A_2\sim 0.3$ at the end. The latter model is special as it shows
  substantial vertical heating right after the disk becomes live. This
  is because this system is instable against axisymmetric
  instabilities (see Fig.~\ref{fig:default_toomre}). The dashed lines
  in the two panels illustrate the average simulation behaviour.
  \label{fig:default_barstrength}}
\end{center}
\end{figure*}

\begin{figure*}
\begin{center}
\resizebox{14cm}{!}{\includegraphics{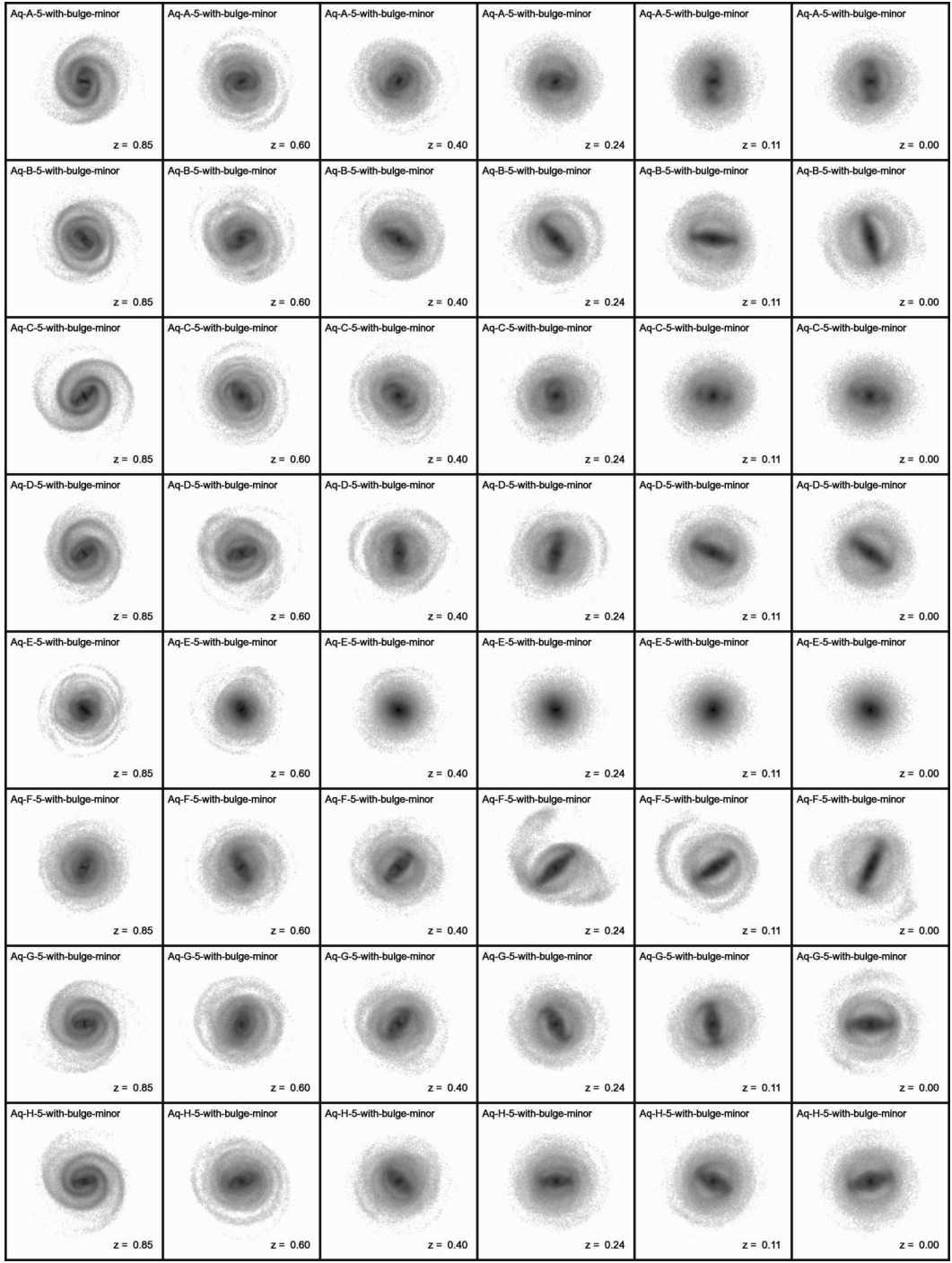}}
\caption{Face-on projections of the stellar mass of the disks in our
runs with bulges (series \#3), where one third of the stellar mass is
moved to a central spheroidal bulge and two thirds are kept in the disk.
Otherwise, the outline of the figures and the image generation method
corresponds exactly to that of Fig.~\ref{fig:defaultruns_faceon}.
\label{fig:bulgeruns_faceon}
}
\end{center}
\end{figure*}

\begin{figure*}
\begin{center}
\resizebox{14cm}{!}{\includegraphics{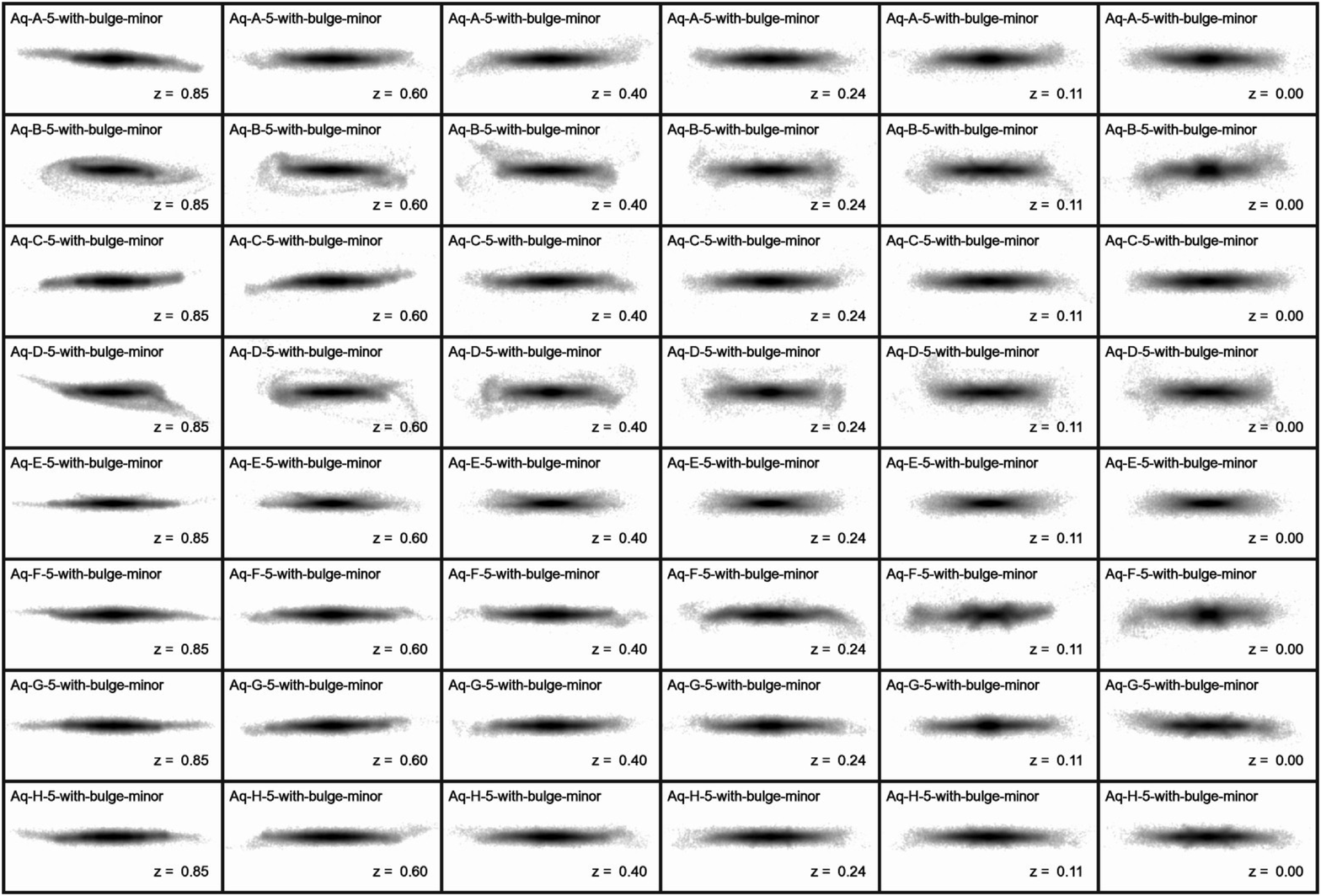}}\\%
\caption{Edge-on projections of the stellar disks in our runs with
  bulges, giving the side-views of the
  corresponding images shown in
  Figure~\ref{fig:bulgeruns_faceon}. As in the corresponding images
  of Fig.~\ref{fig:defaultruns_edgeon}, each panel has
   an unchanged physical size in the horizontal
  direction ($54\,{\rm kpc}$) and uses an identical colour-scale as in the
  face-on images.
\label{fig:bulgeruns_edgeon}
}
\end{center}
\end{figure*}

\begin{figure*}
\begin{center}
\resizebox{8cm}{!}{\includegraphics{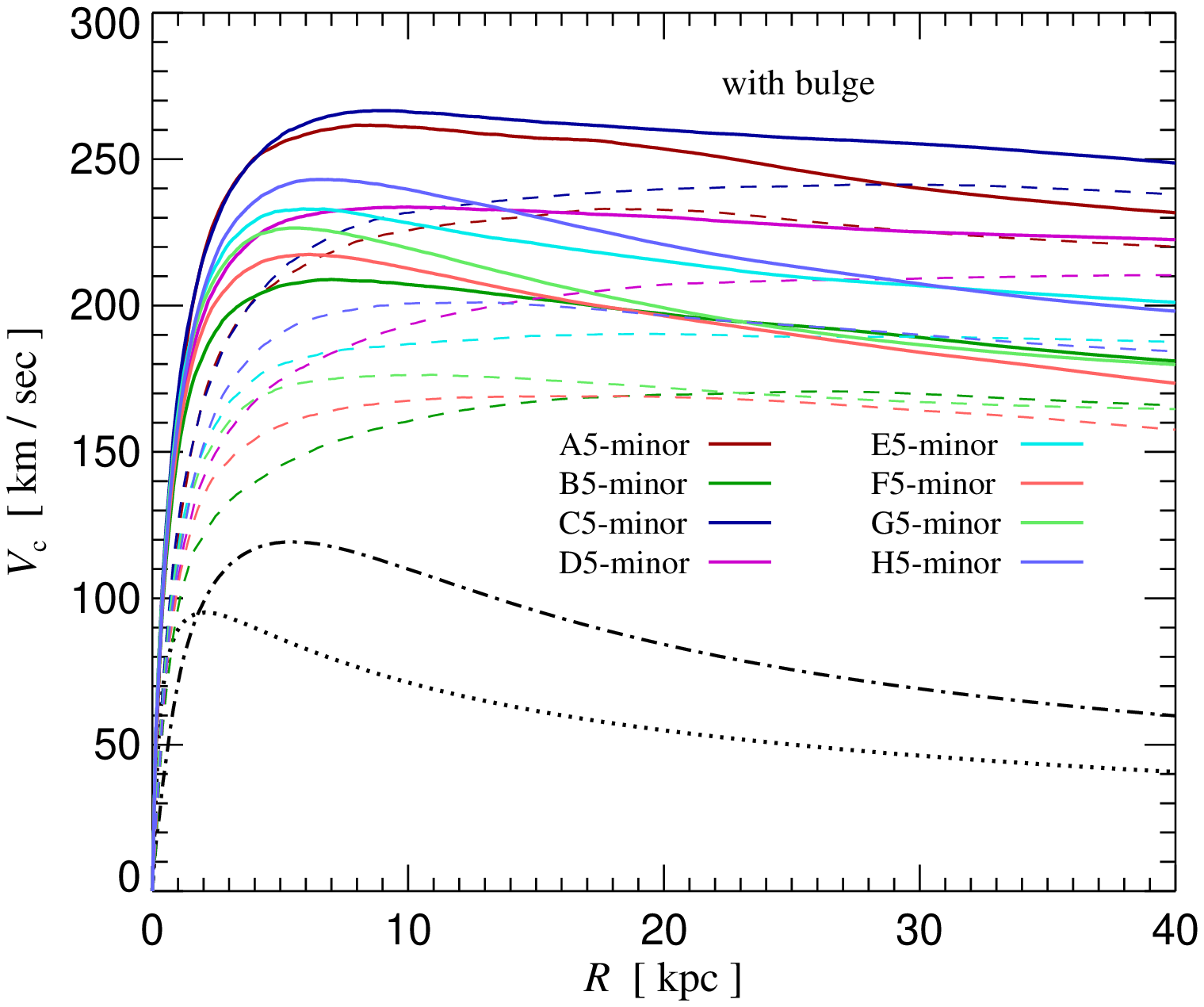}}%
\resizebox{8cm}{!}{\includegraphics{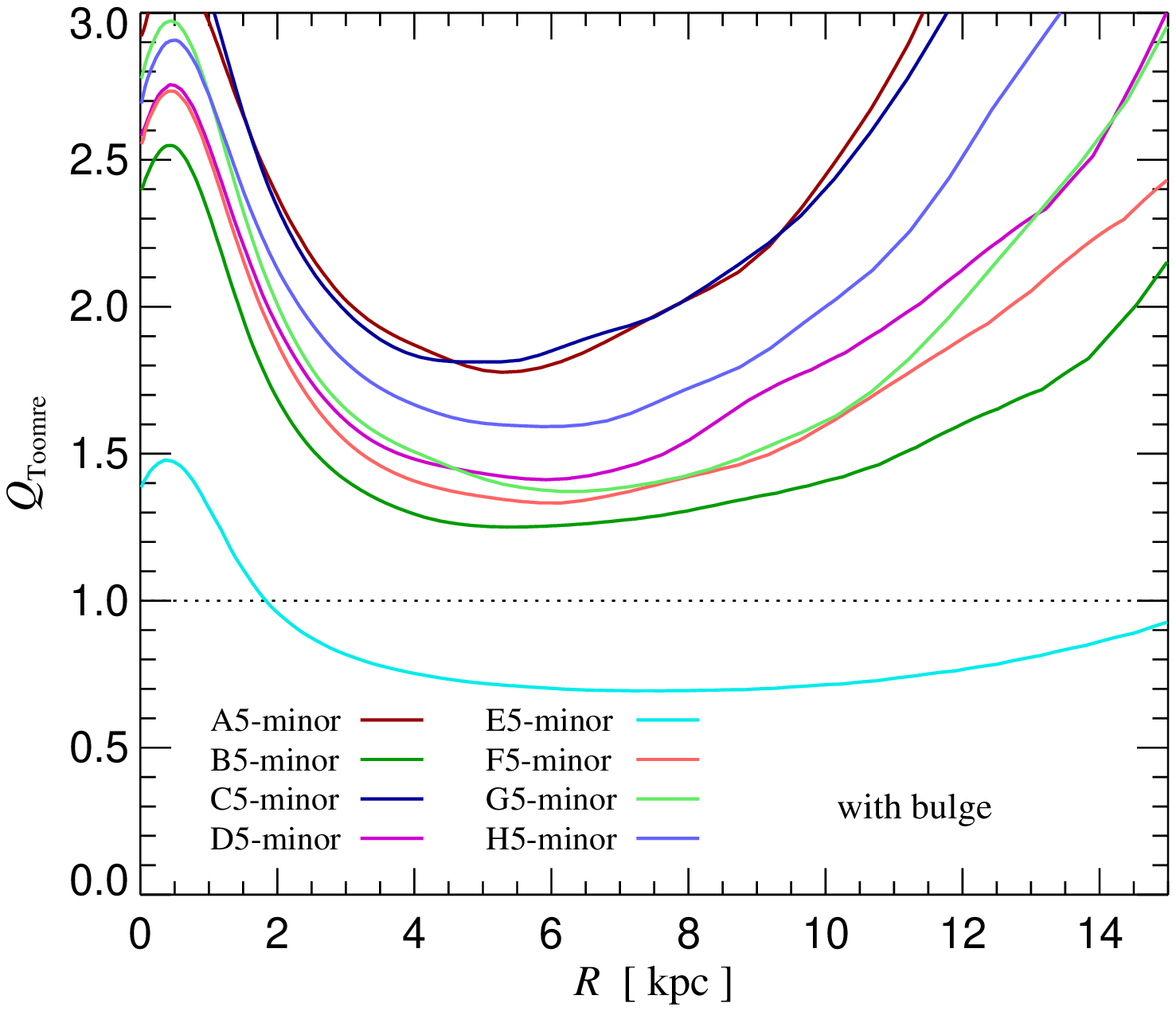}}%
\caption{Rotation curves and Toomre-Q stability parameter of our
  default simulations with bulges (series \#3), as a function of
  radius.  The left panel shows the rotation curves, where the disk
  (dot-dashed) and bulge (dotted) always make the same contributions
  due to their constant parameters. The different dark matter halo
  contributions are shown by dashed lines, and the total rotation
  curves by solid lines.  The panel on the right illustrates the
  expected stability against axisymmetric instabilities. While the
  bulge has a substantial stabilizing influence (compare to
  Fig.~\ref{fig:default_toomre}), model E5 is still found to be
  unstable for the adopted stellar parameters of disk and bulge.
\label{fig:rotcurve_toomre_withbulges}}
\end{center}
\end{figure*}

\begin{figure*}
\begin{center}
\setlength{\unitlength}{1cm}
\resizebox{8.5cm}{!}{\includegraphics{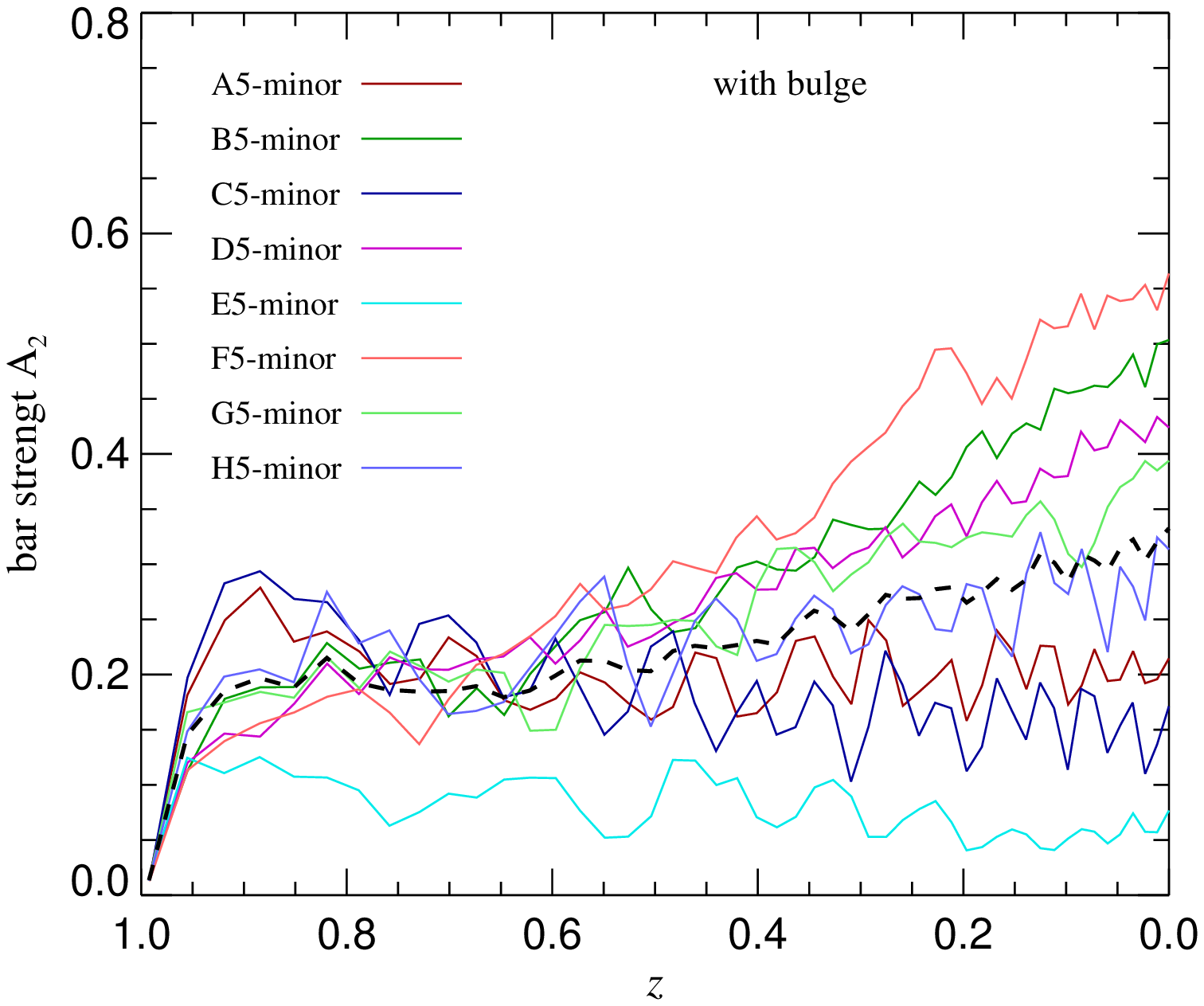}}%
\resizebox{8.5cm}{!}{\includegraphics{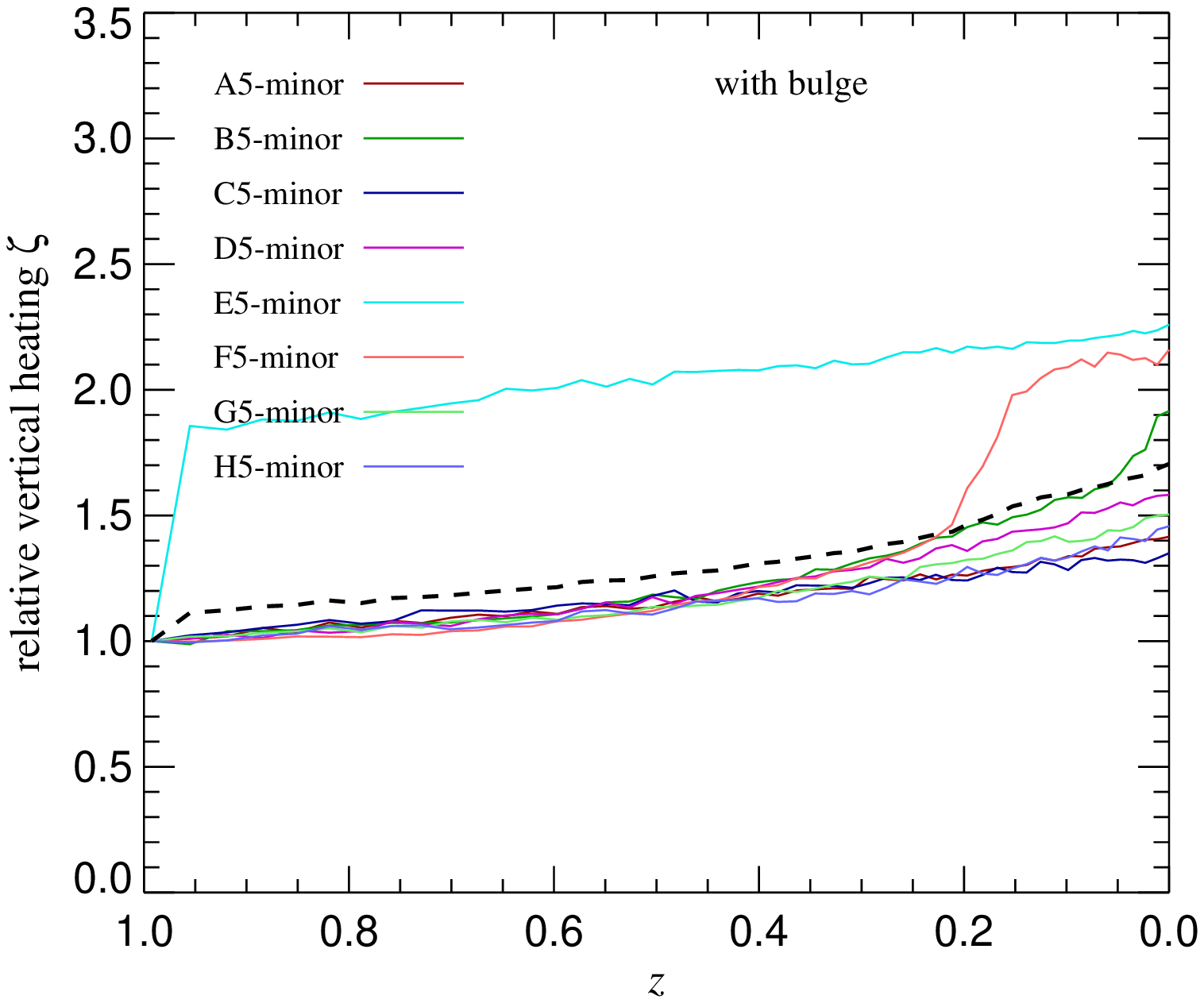}}%
\caption{
Bar strength parameter $A_2$ (left panel) and relative
  vertical heating (right panel) in our default disk+bulge 
simulations (series \#3), as a function of
  time. These evolutions can be directly compared to the corresponding 
results for the pure disk case shown in Fig.~\ref{fig:default_barstrength}. 
Now only a subset of the systems develops bars, and even if this happens,
the bars are weaker and form later. Model E still appears as an
outlier, caused by its instability against axisymmetric
perturbations. We note that the early growth of the $A_2$ indicator to
values of around $\sim 0.2$ does not really measure a bar; it is
presumably caused by a quick distortion of the spherical disk into an
ellipsoidal disk due to the residual asymmetry of the dark matter
halo potential.
\label{fig:barstrength_withbulge}}
\end{center}
\end{figure*}

\begin{figure*}
\begin{center}
\setlength{\unitlength}{1cm}
\resizebox{4.5cm}{!}{\includegraphics{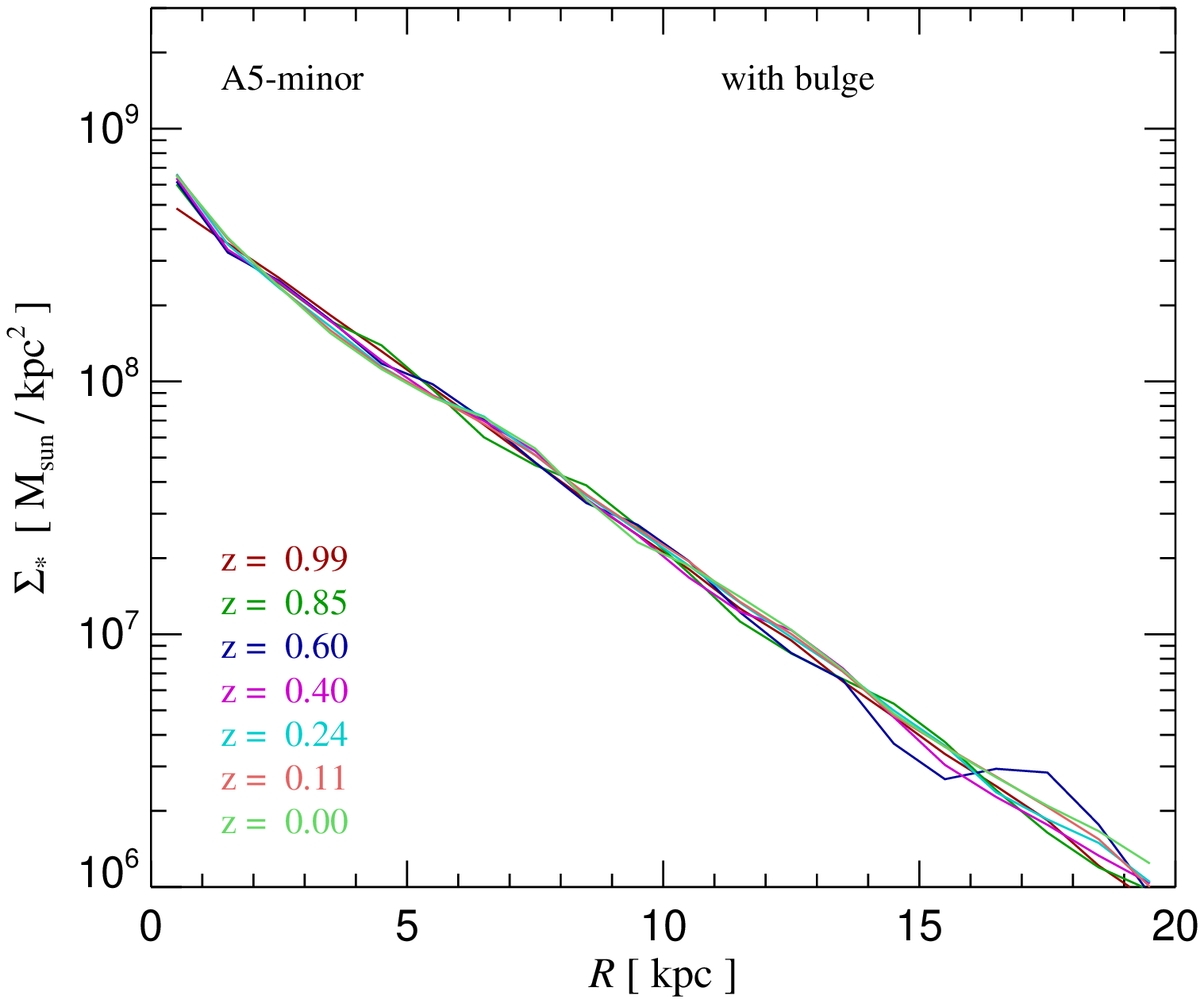}}%
\resizebox{4.5cm}{!}{\includegraphics{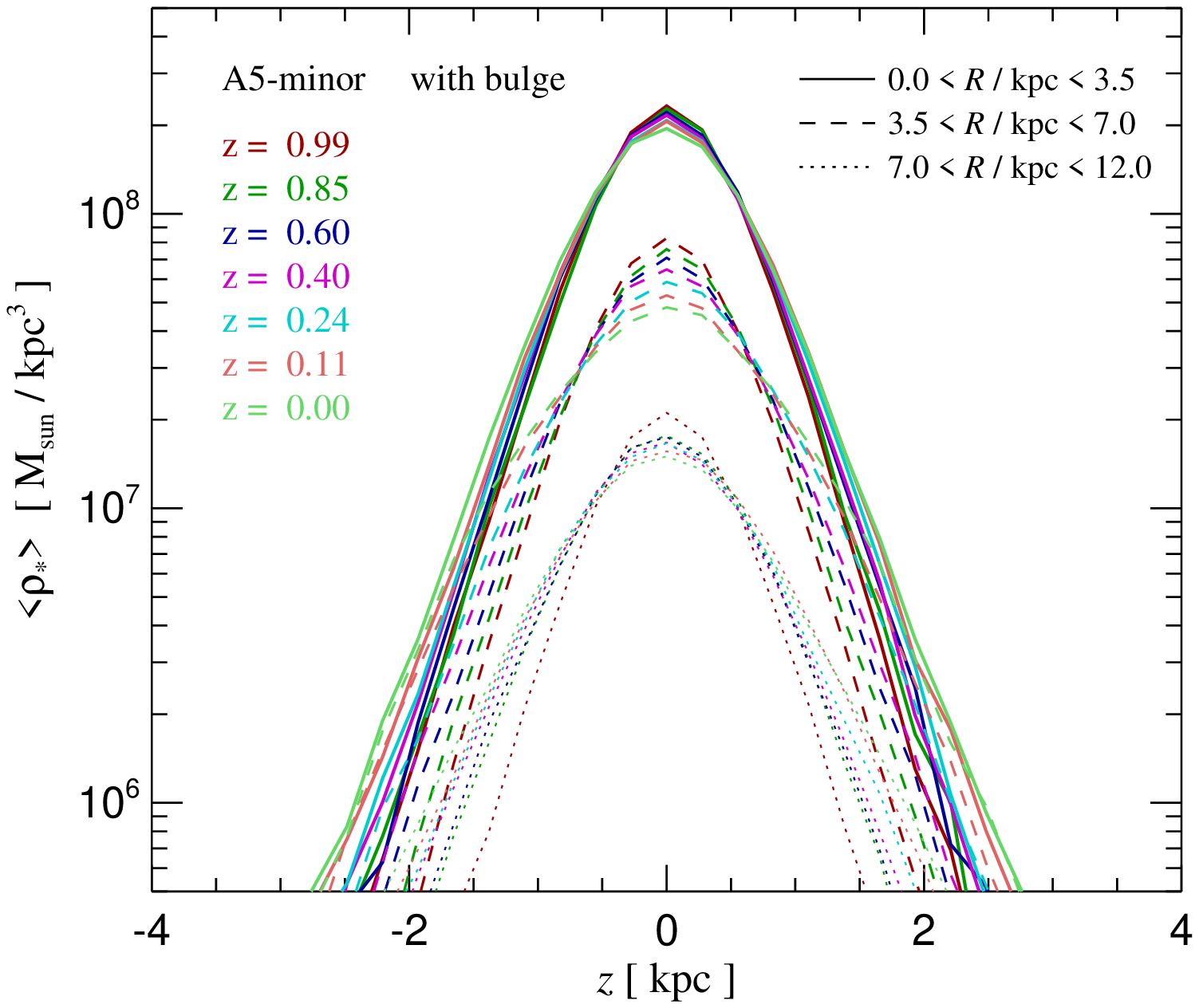}}%
\resizebox{4.5cm}{!}{\includegraphics{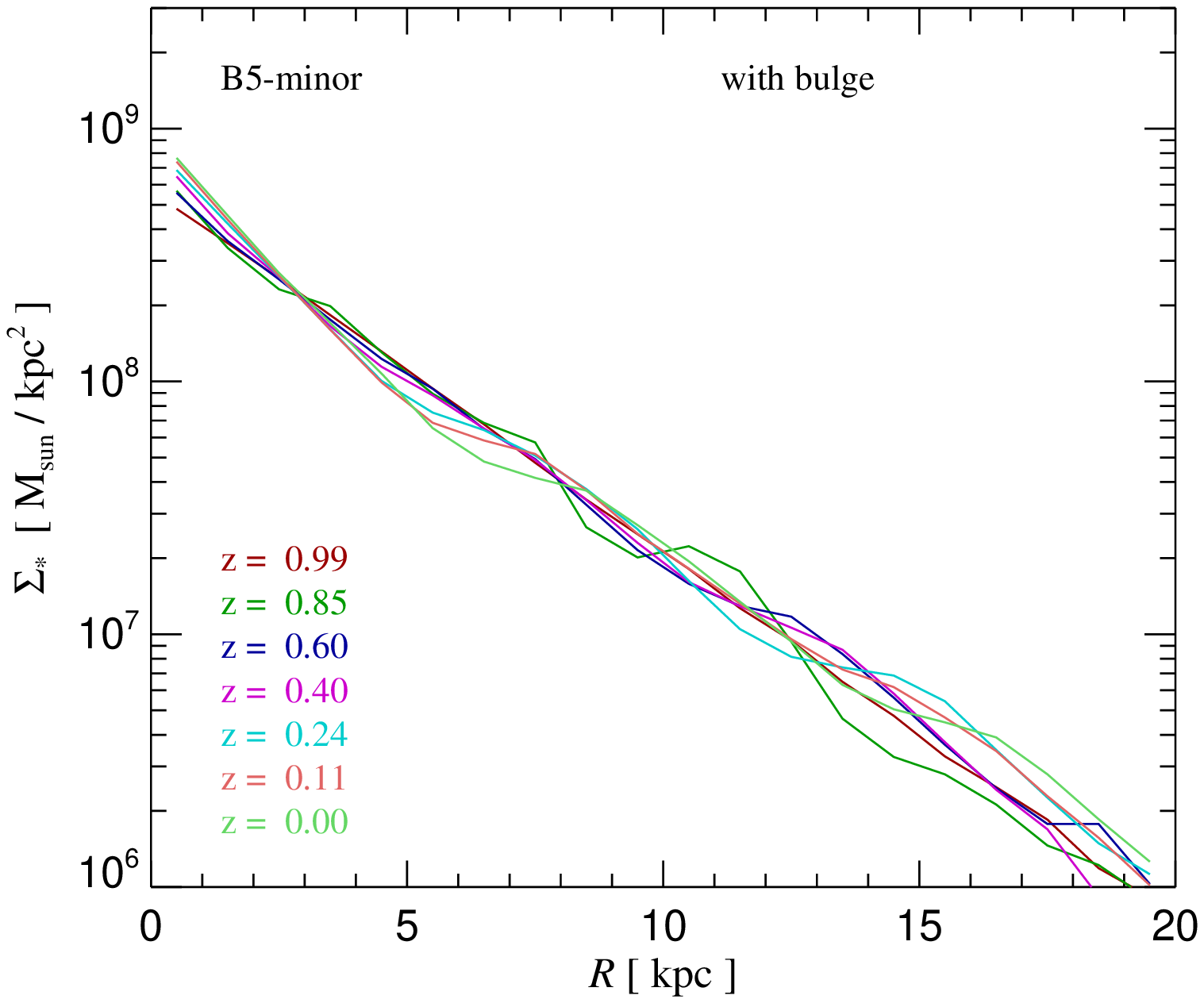}}%
\resizebox{4.5cm}{!}{\includegraphics{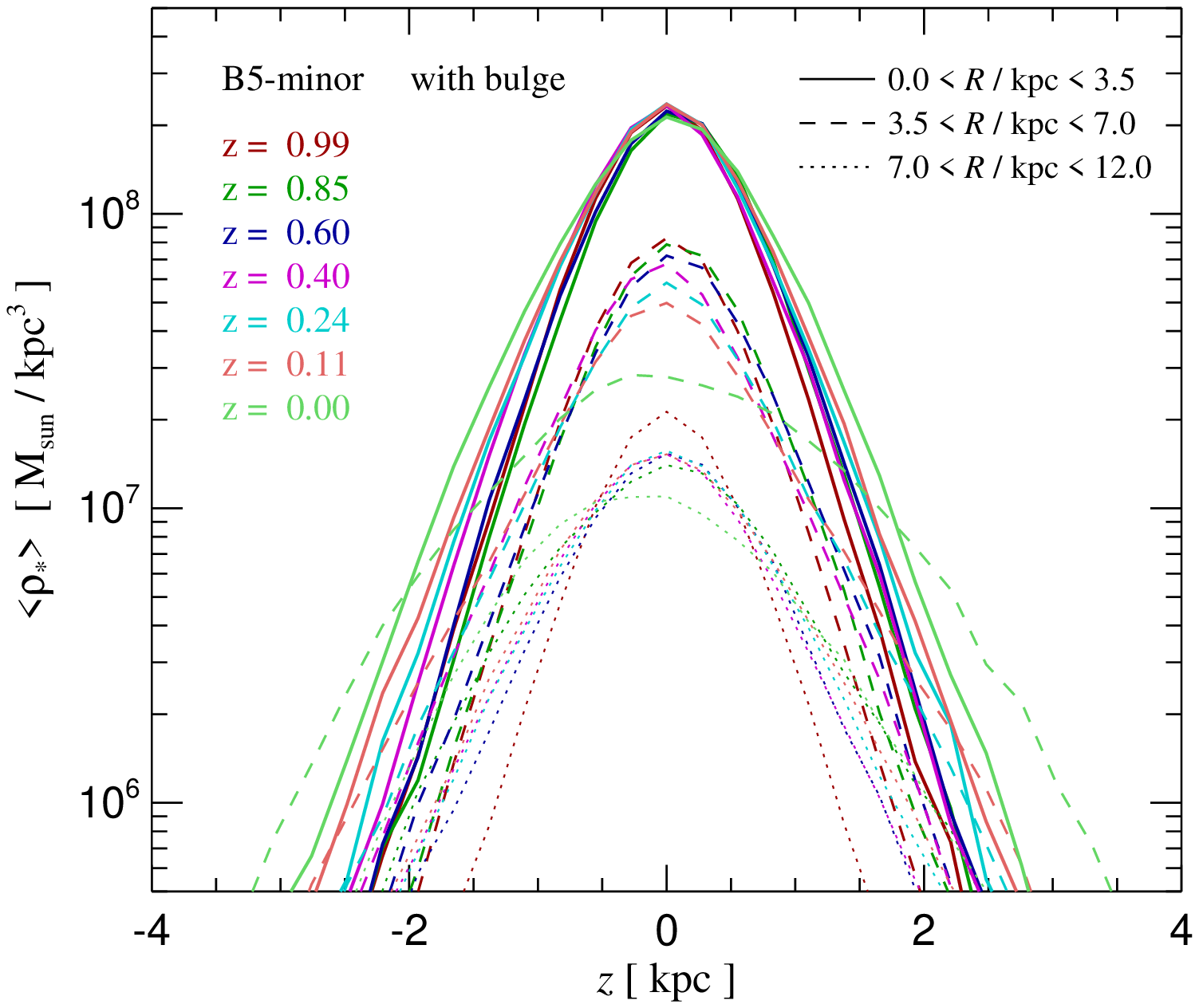}}\\%
\resizebox{4.5cm}{!}{\includegraphics{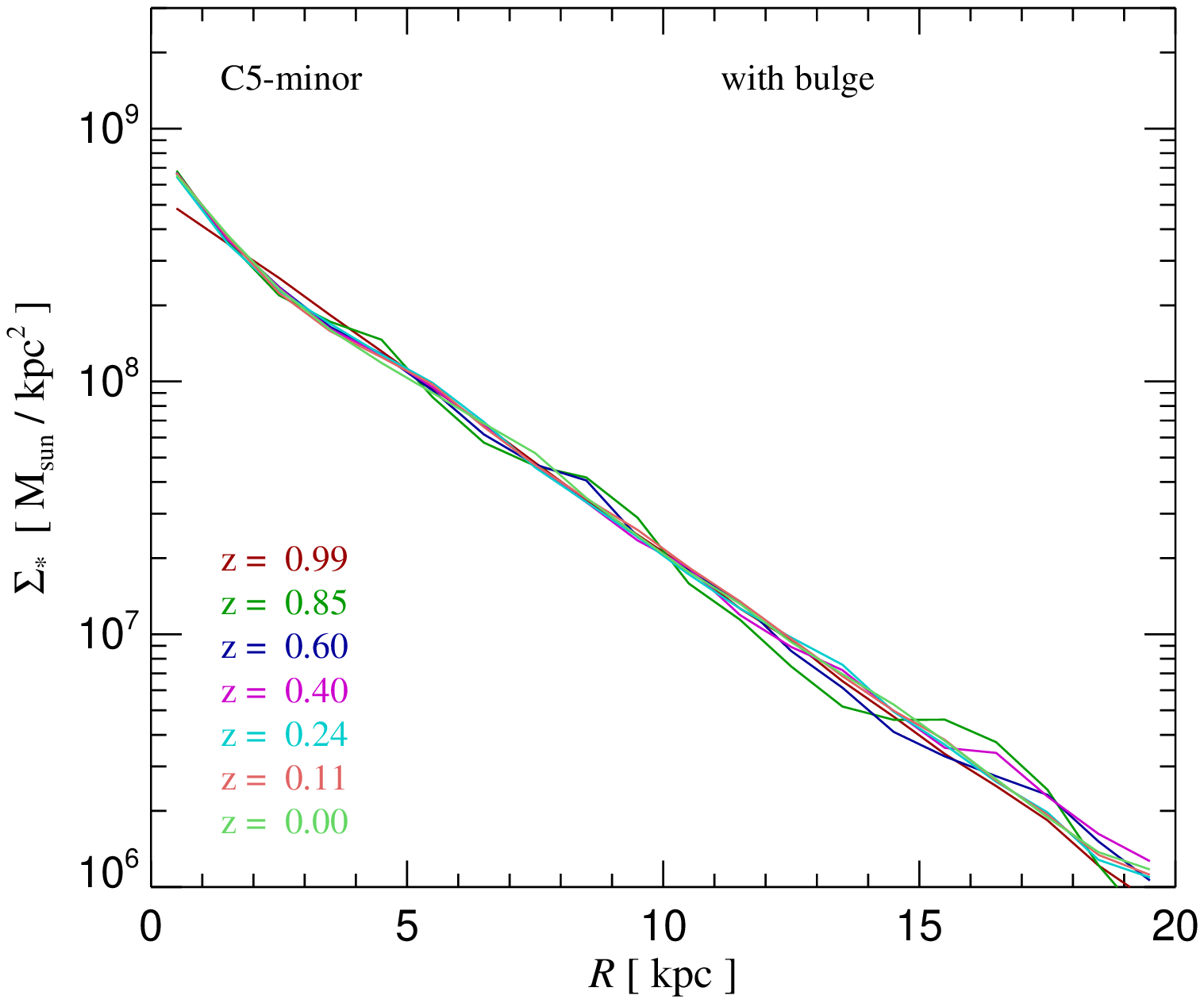}}%
\resizebox{4.5cm}{!}{\includegraphics{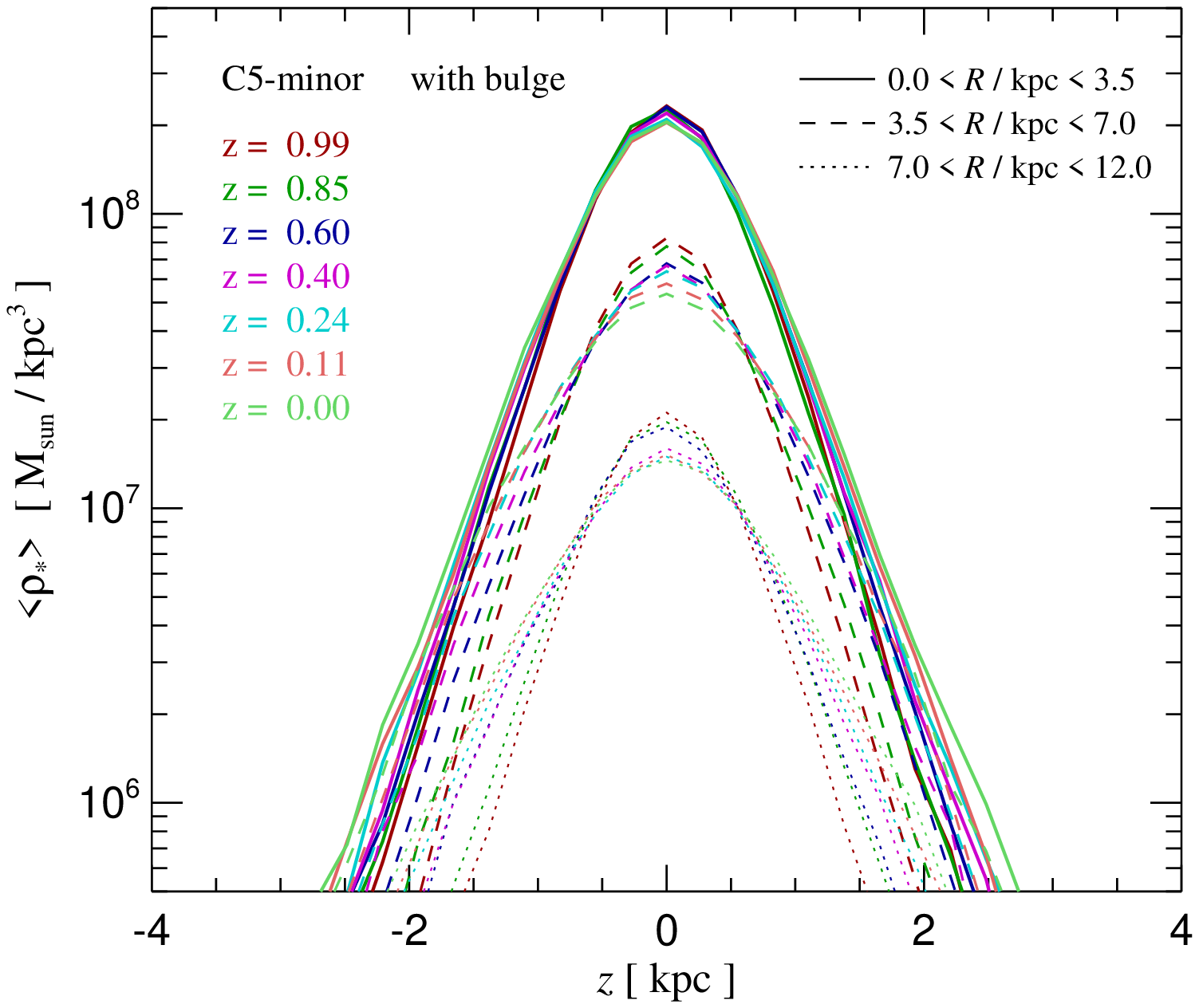}}%
\resizebox{4.5cm}{!}{\includegraphics{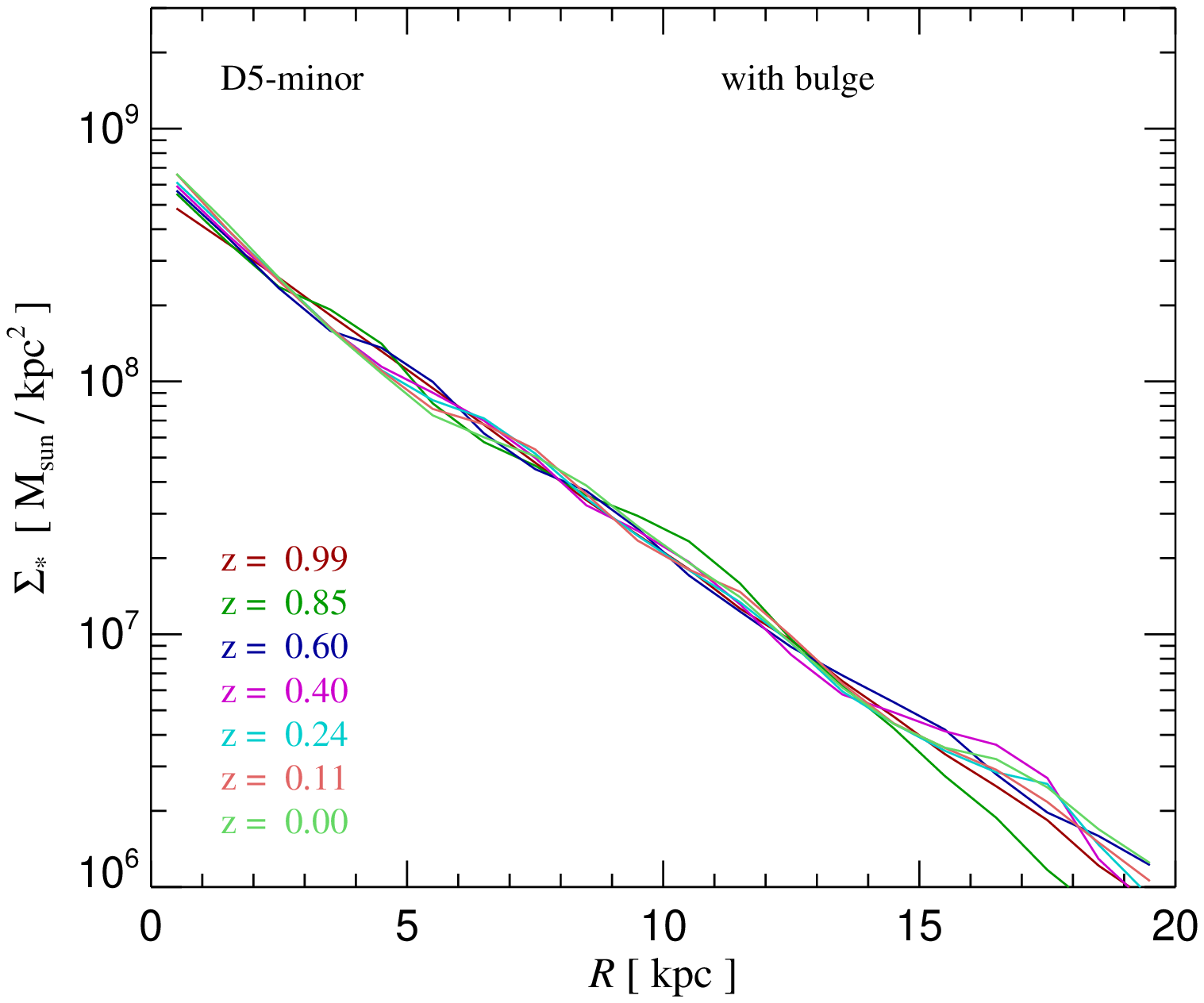}}%
\resizebox{4.5cm}{!}{\includegraphics{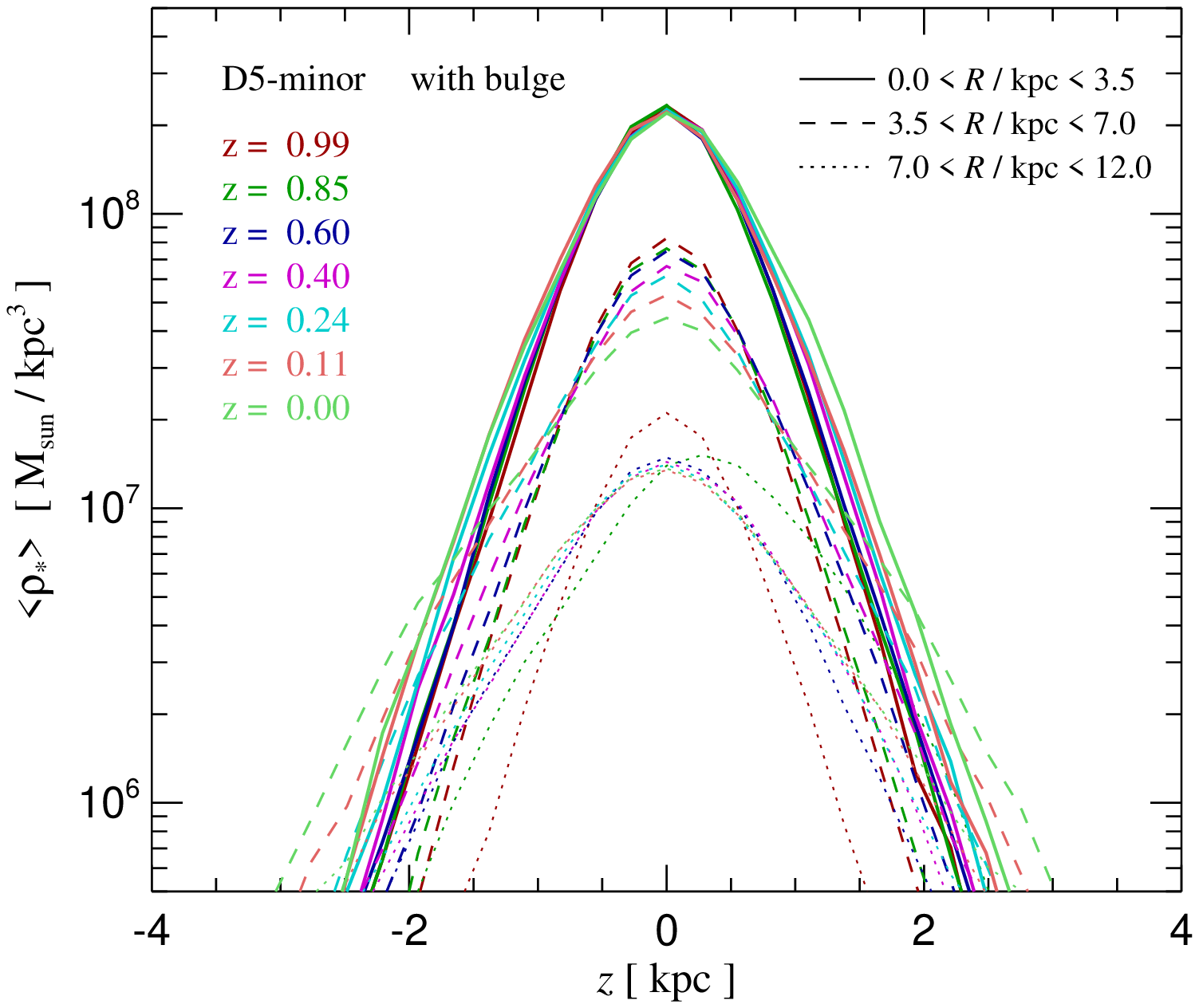}}\\%
\resizebox{4.5cm}{!}{\includegraphics{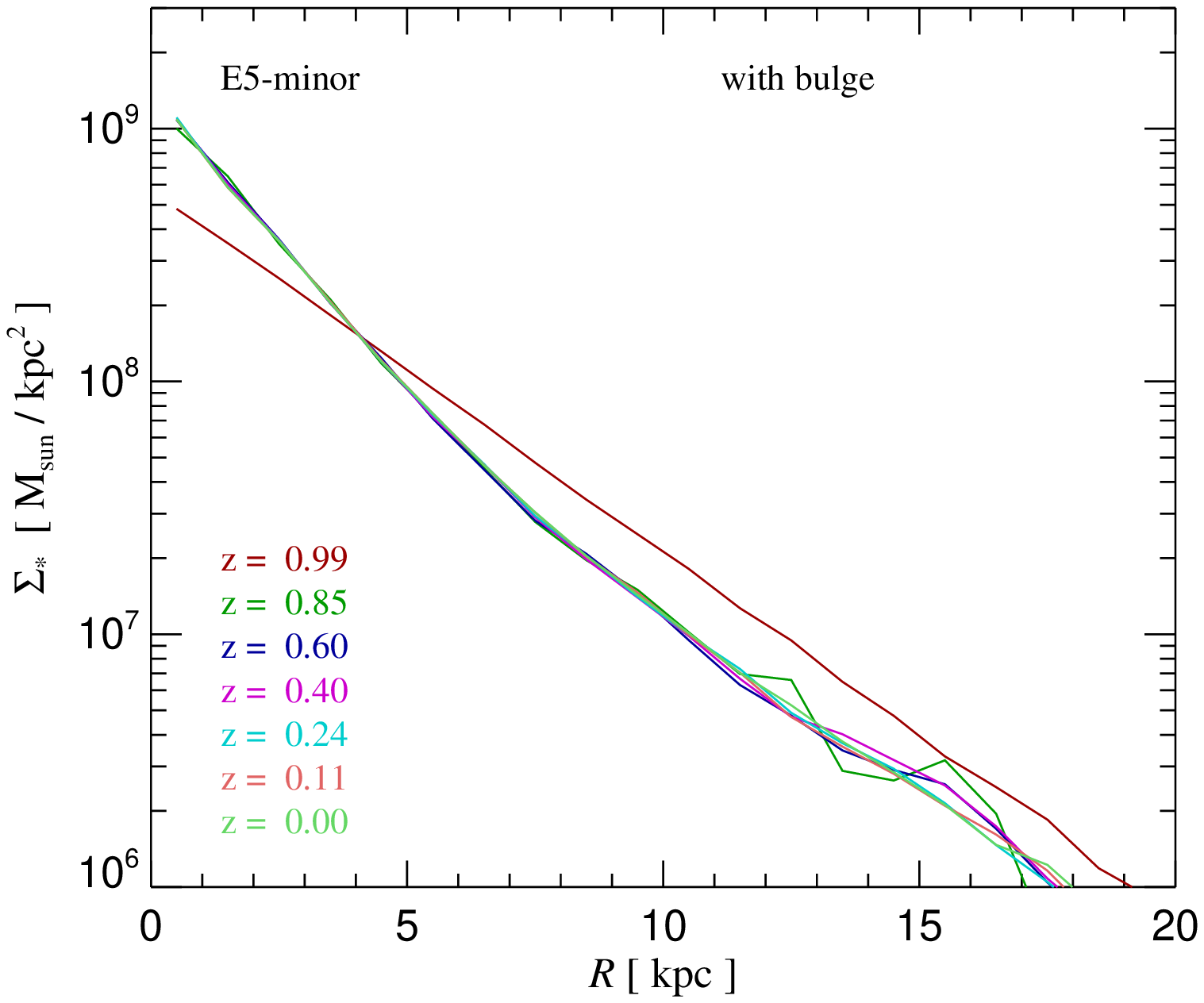}}%
\resizebox{4.5cm}{!}{\includegraphics{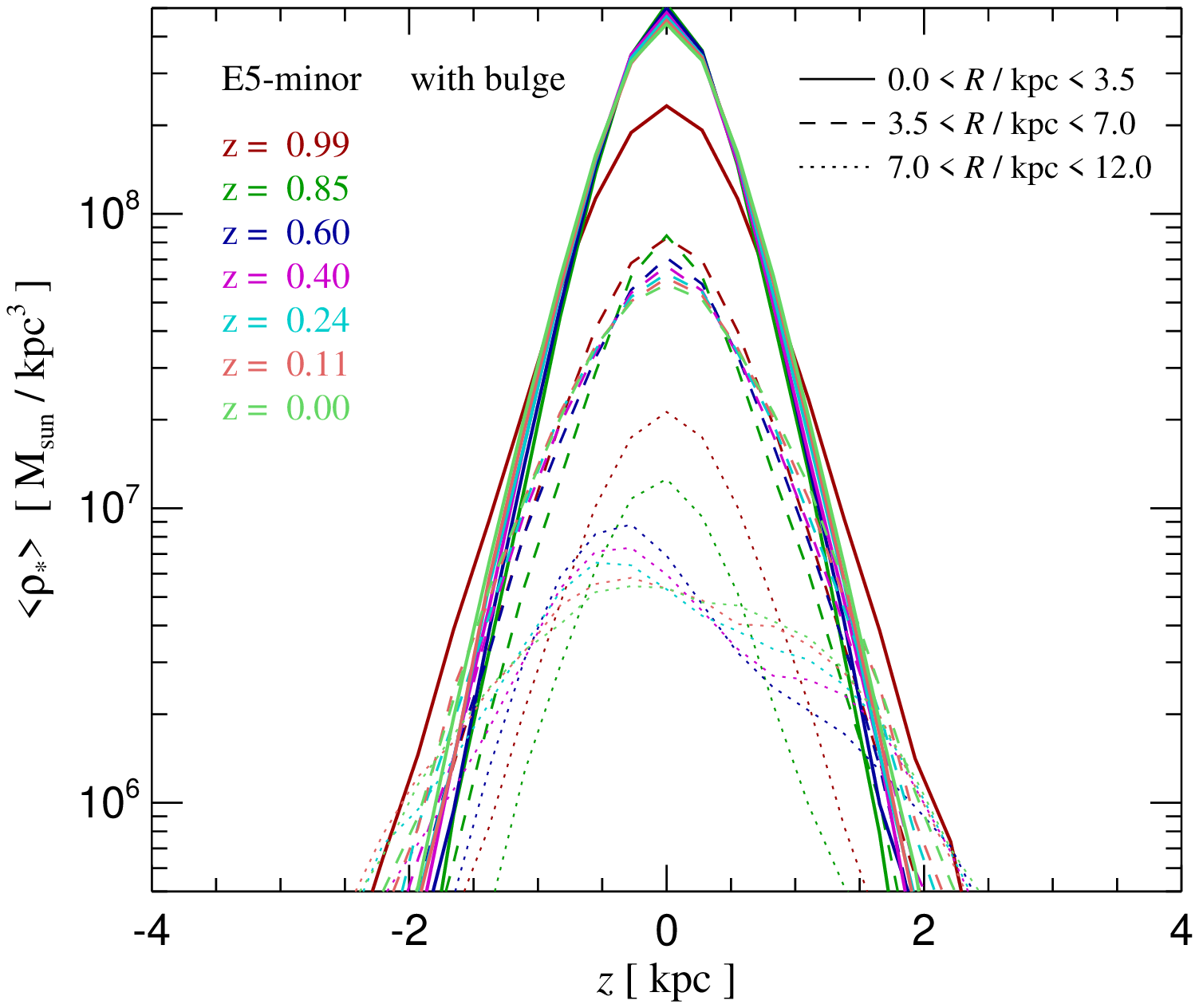}}%
\resizebox{4.5cm}{!}{\includegraphics{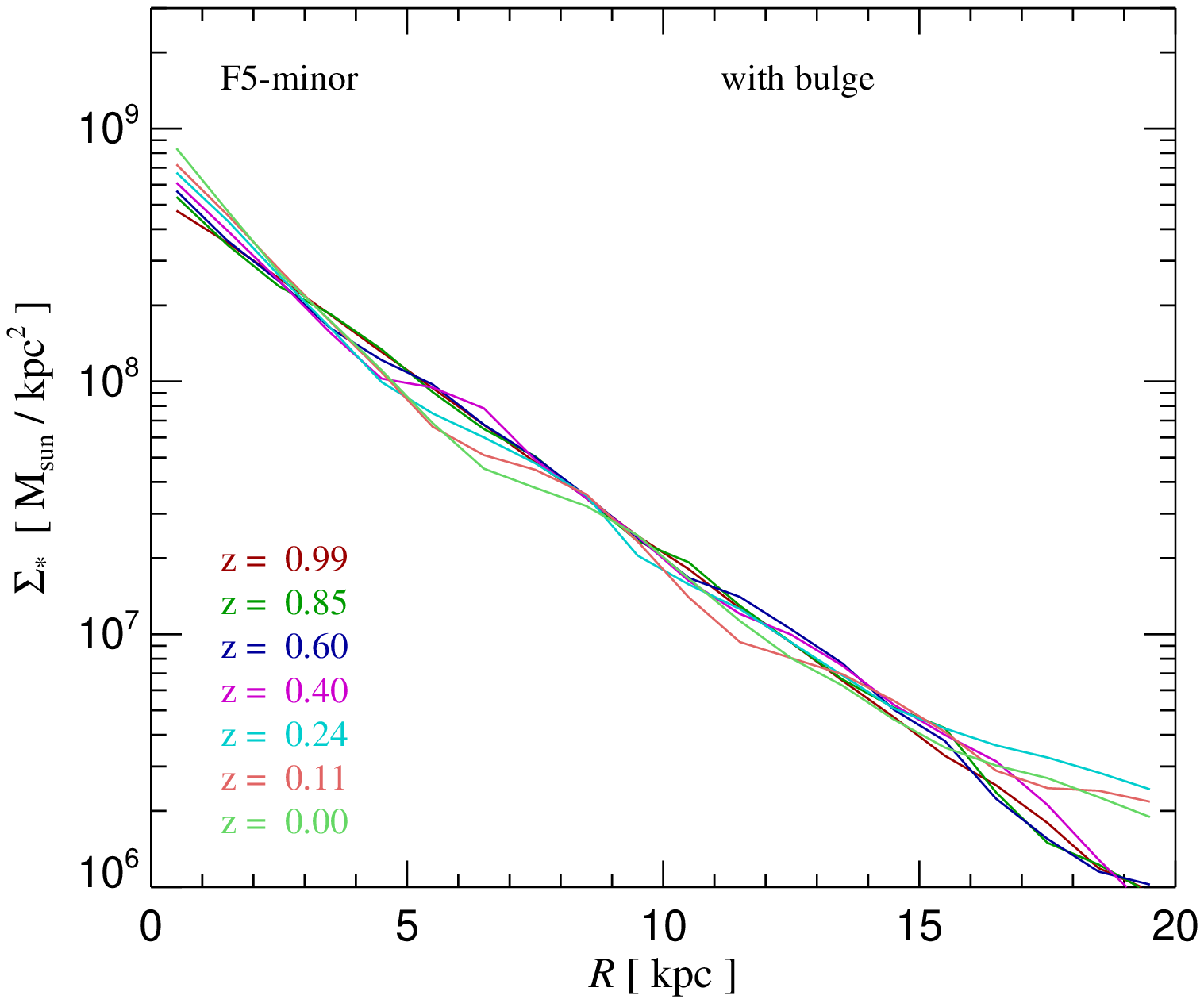}}%
\resizebox{4.5cm}{!}{\includegraphics{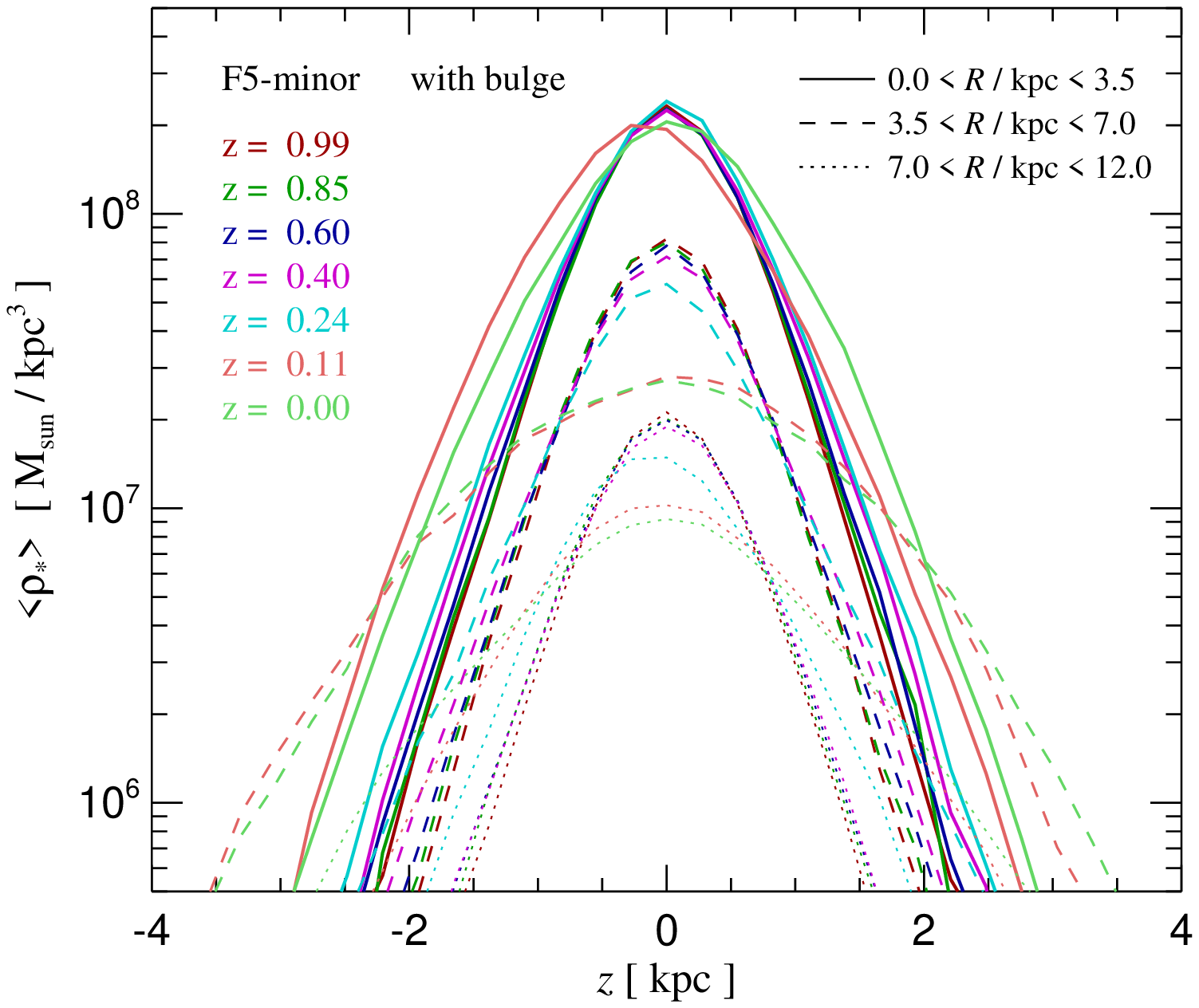}}\\%
\resizebox{4.5cm}{!}{\includegraphics{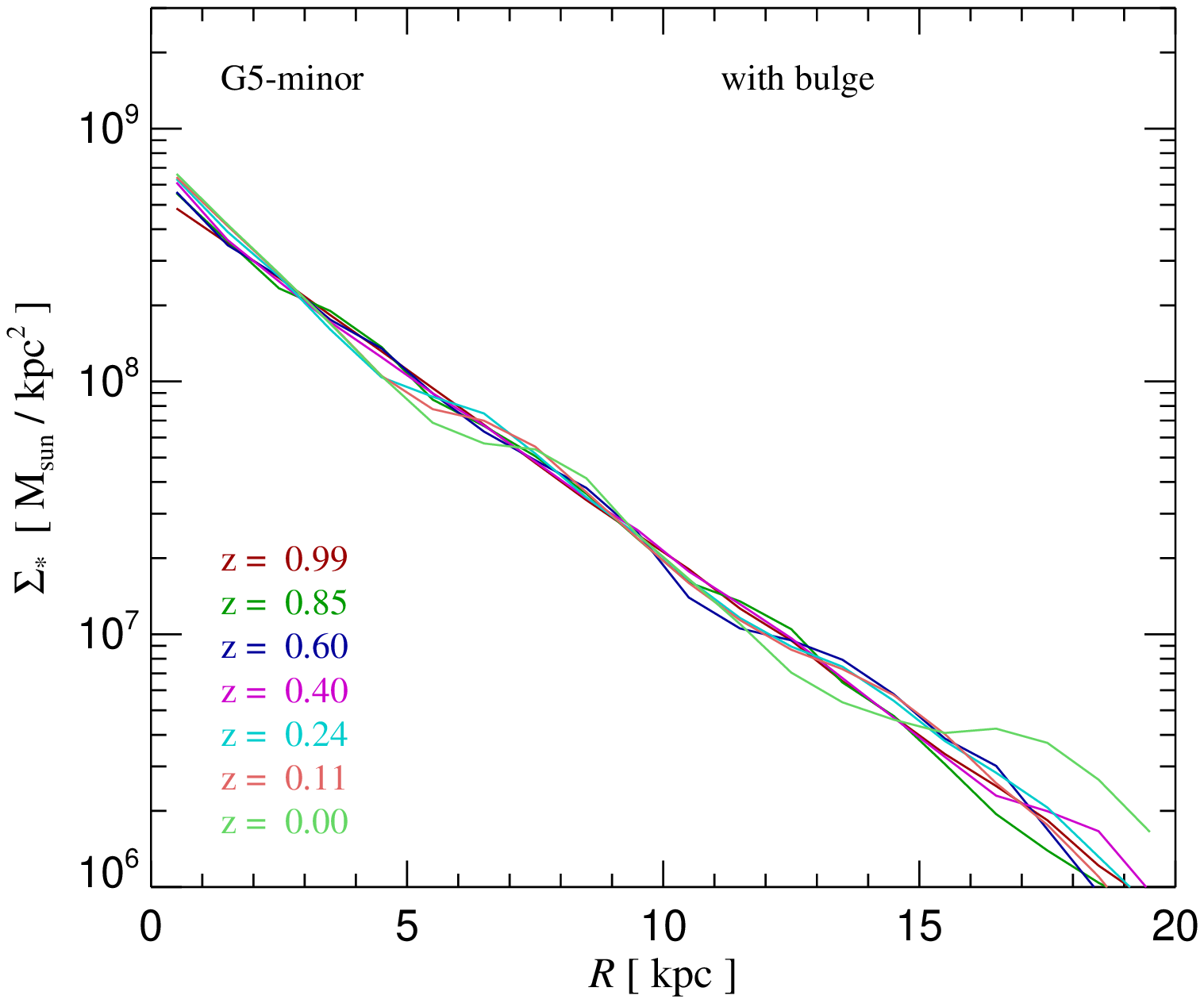}}%
\resizebox{4.5cm}{!}{\includegraphics{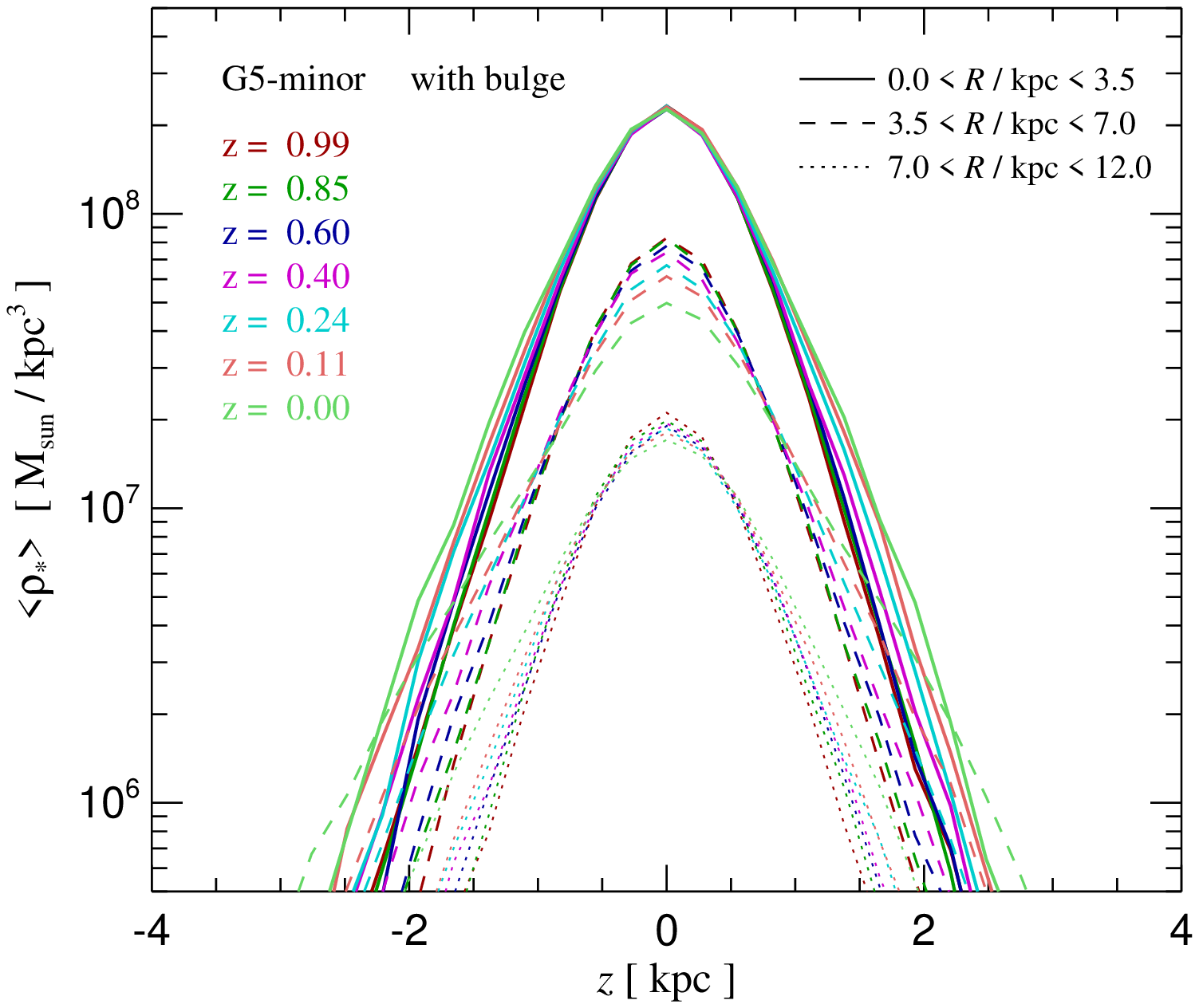}}%
\resizebox{4.5cm}{!}{\includegraphics{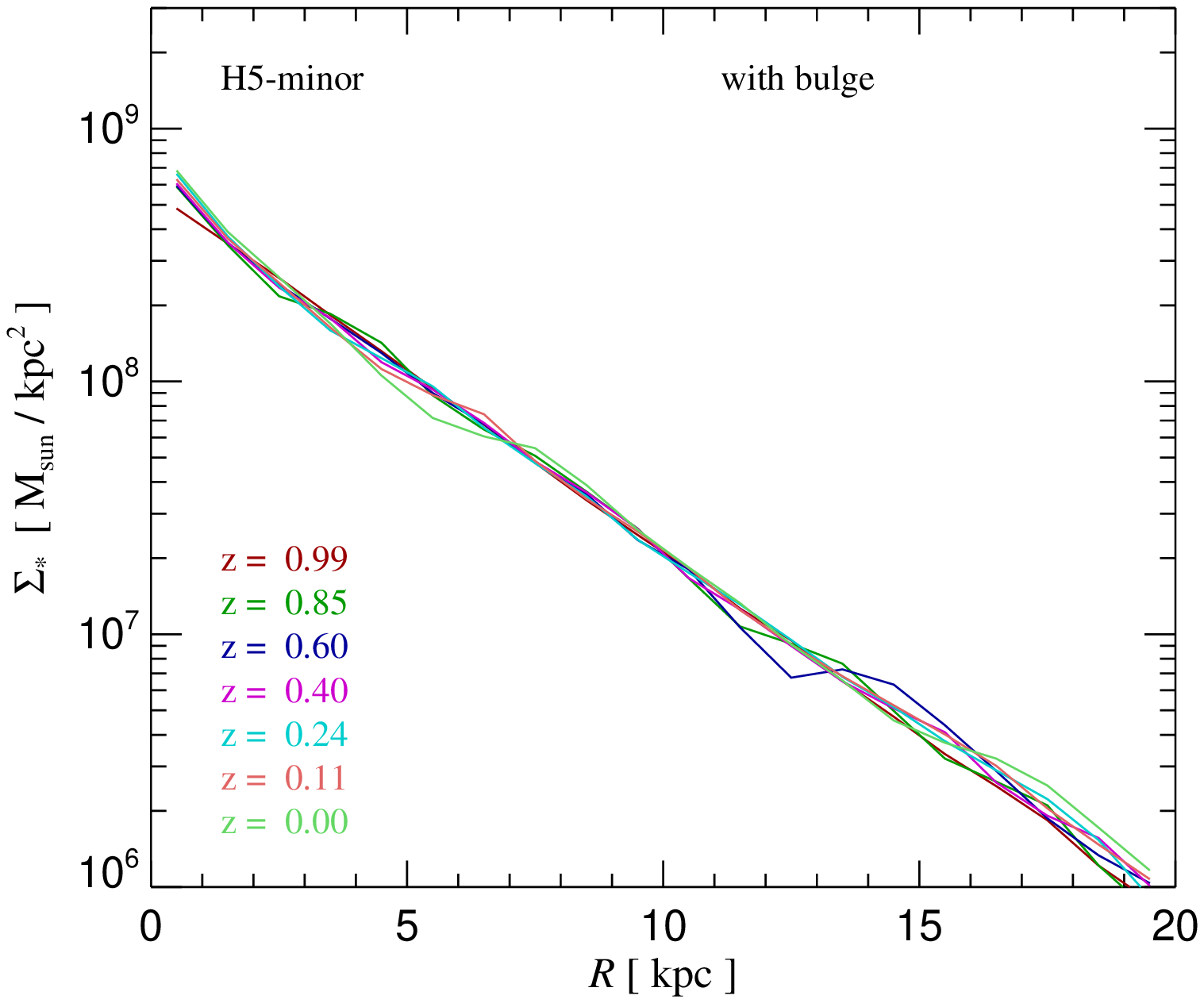}}%
\resizebox{4.5cm}{!}{\includegraphics{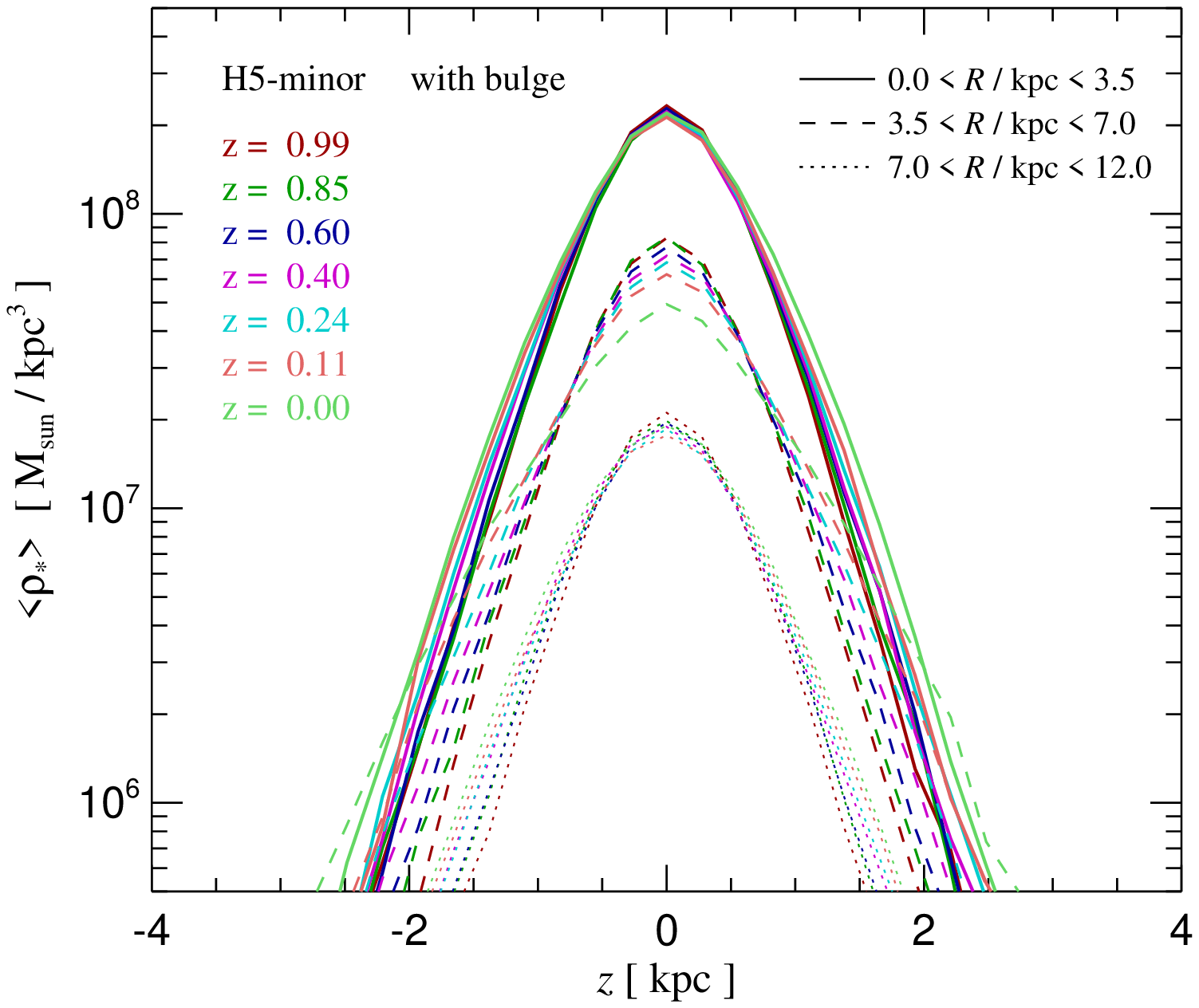}}%
\caption{Radial and vertical disk density profiles for our default
  disk+bulge simulations (series \#3) as a function of time. As in
  the corresponding Fig.~\ref{fig:default_radialprofiles} for the
  pure disk runs, we show results
  for all of our eight Aquarius halos, in each case with a pair of
  panels where the surface density profile is shown on the left, and the
  vertical density profiles on the right. Especially in the much more
  modest 
  evolution of the vertical
  structure it is evident that these disks evolve comparatively
  little; i.e.~the bulge has largely stabilized the disks against
  strong bar formation.
\label{fig:radialprofiles_withbulge}
}
\end{center}
\end{figure*}

\begin{figure*}
\begin{center}
\setlength{\unitlength}{1cm}
\resizebox{4.5cm}{!}{\includegraphics{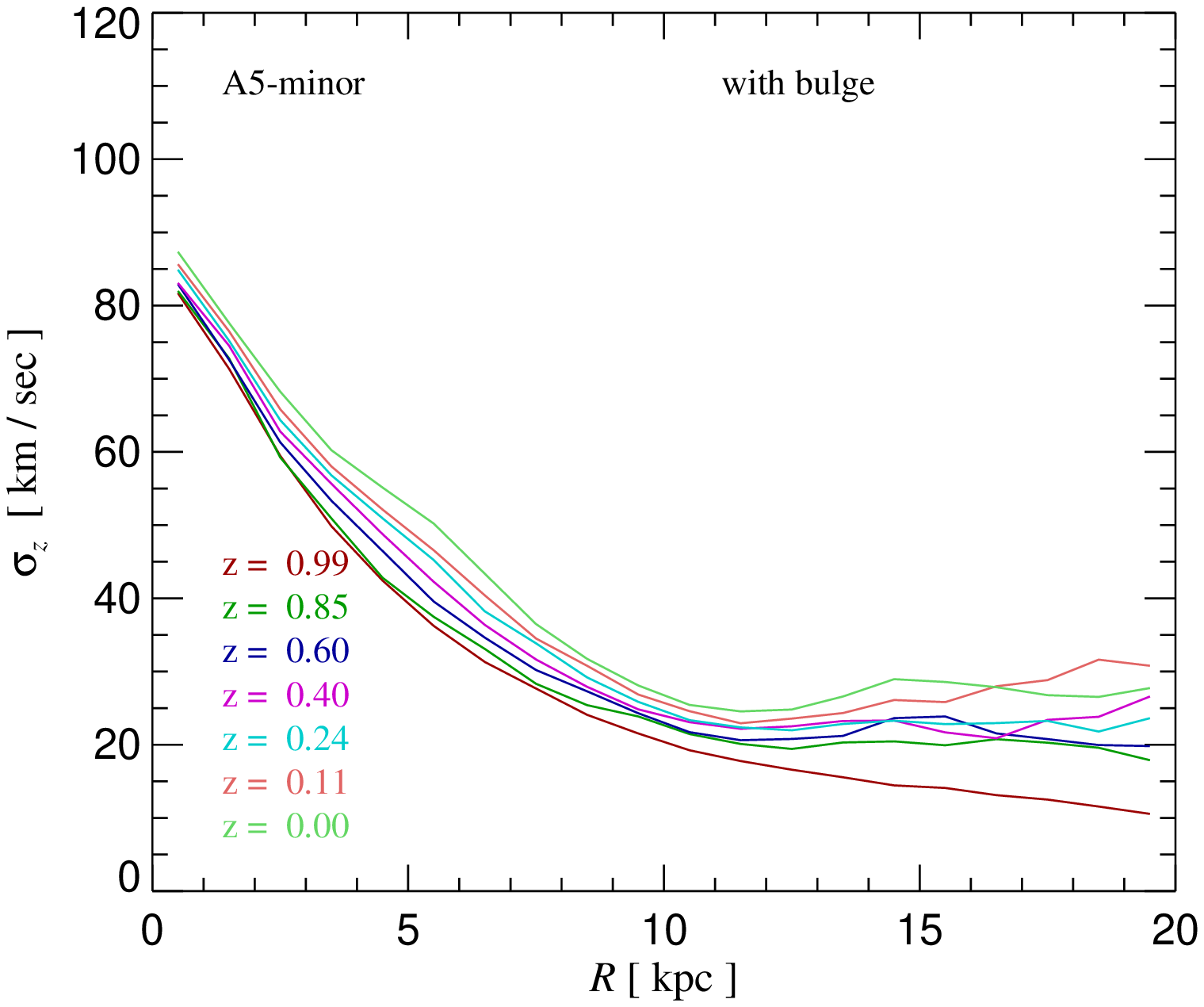}}%
\resizebox{4.5cm}{!}{\includegraphics{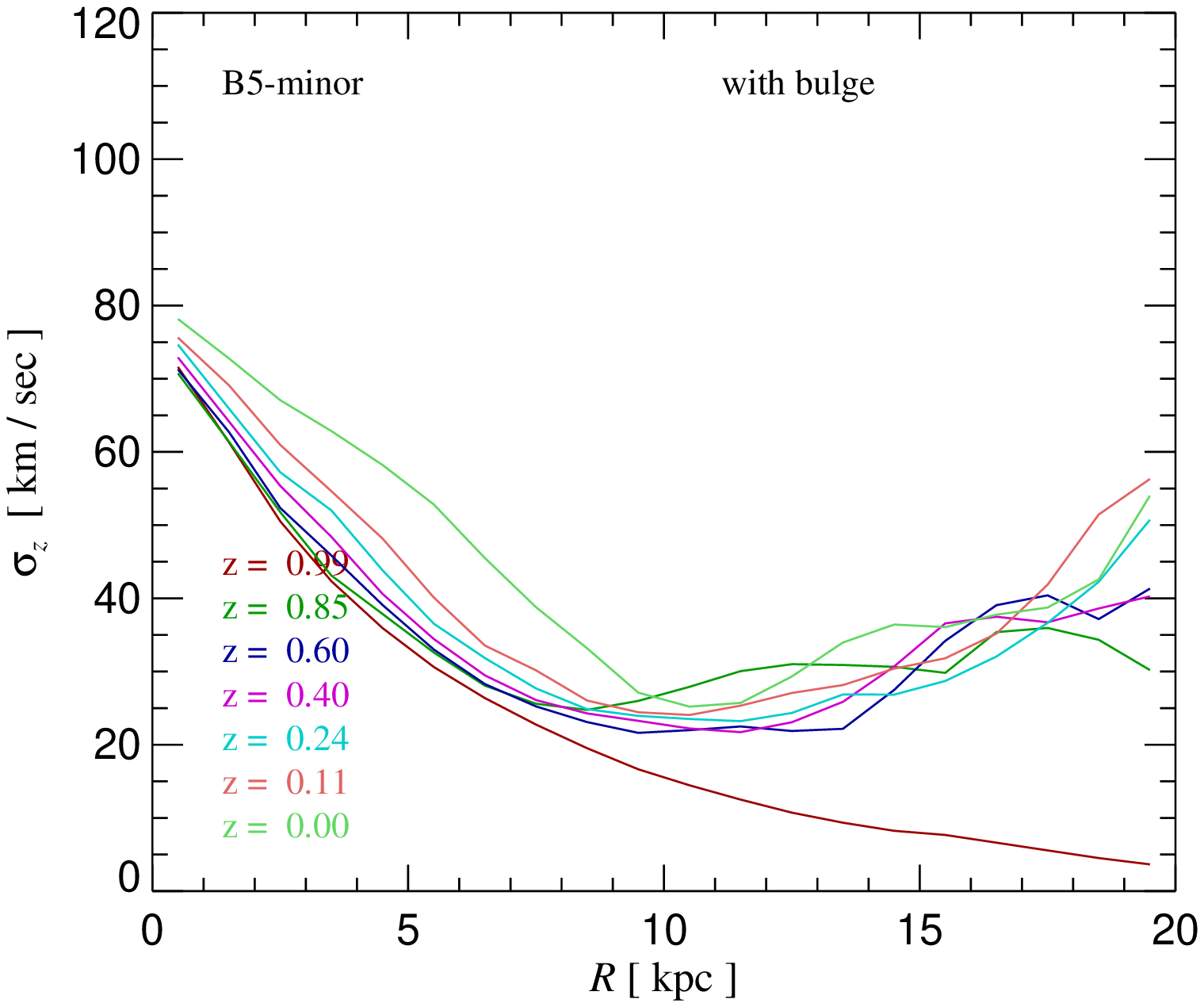}}%
\resizebox{4.5cm}{!}{\includegraphics{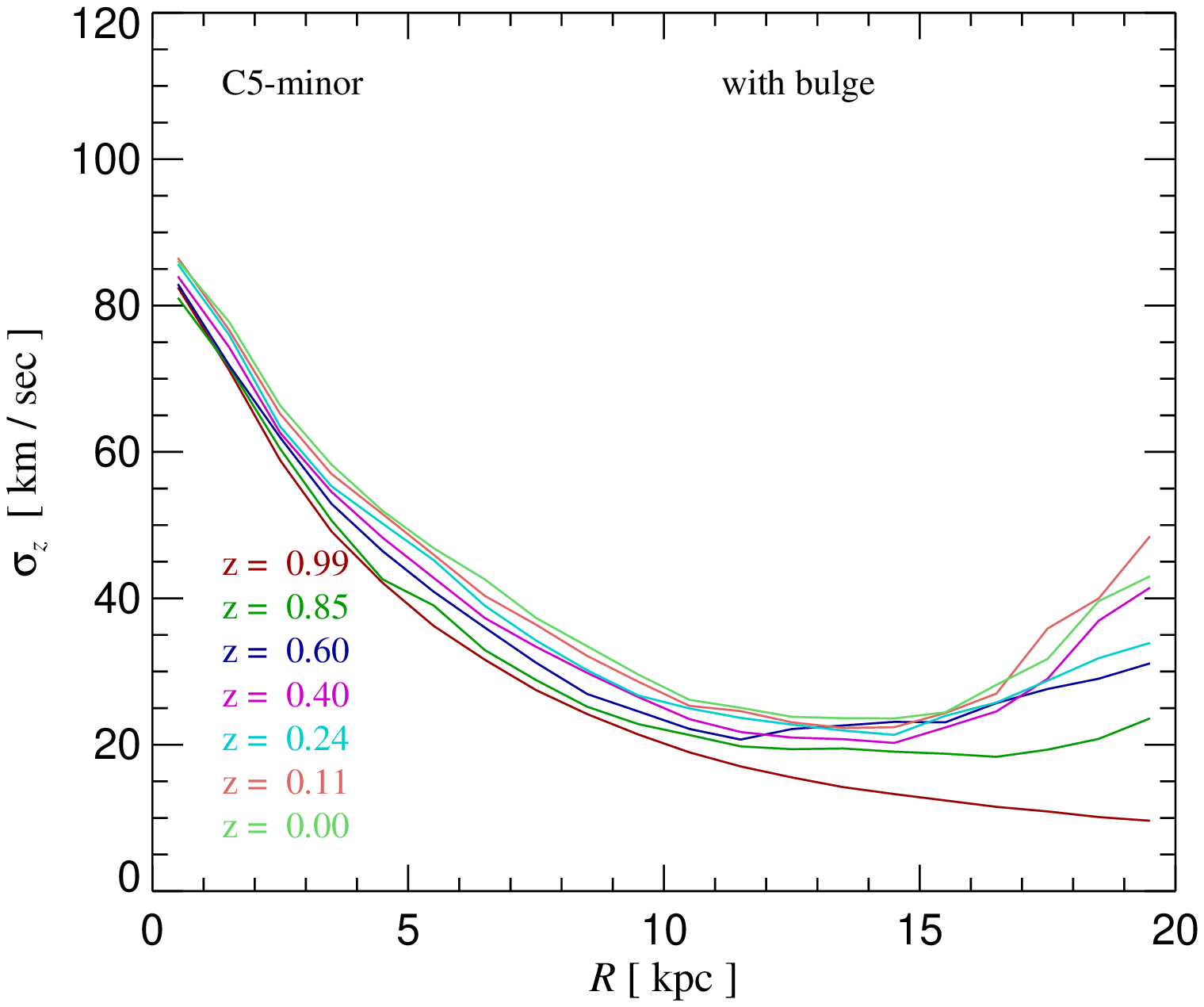}}%
\resizebox{4.5cm}{!}{\includegraphics{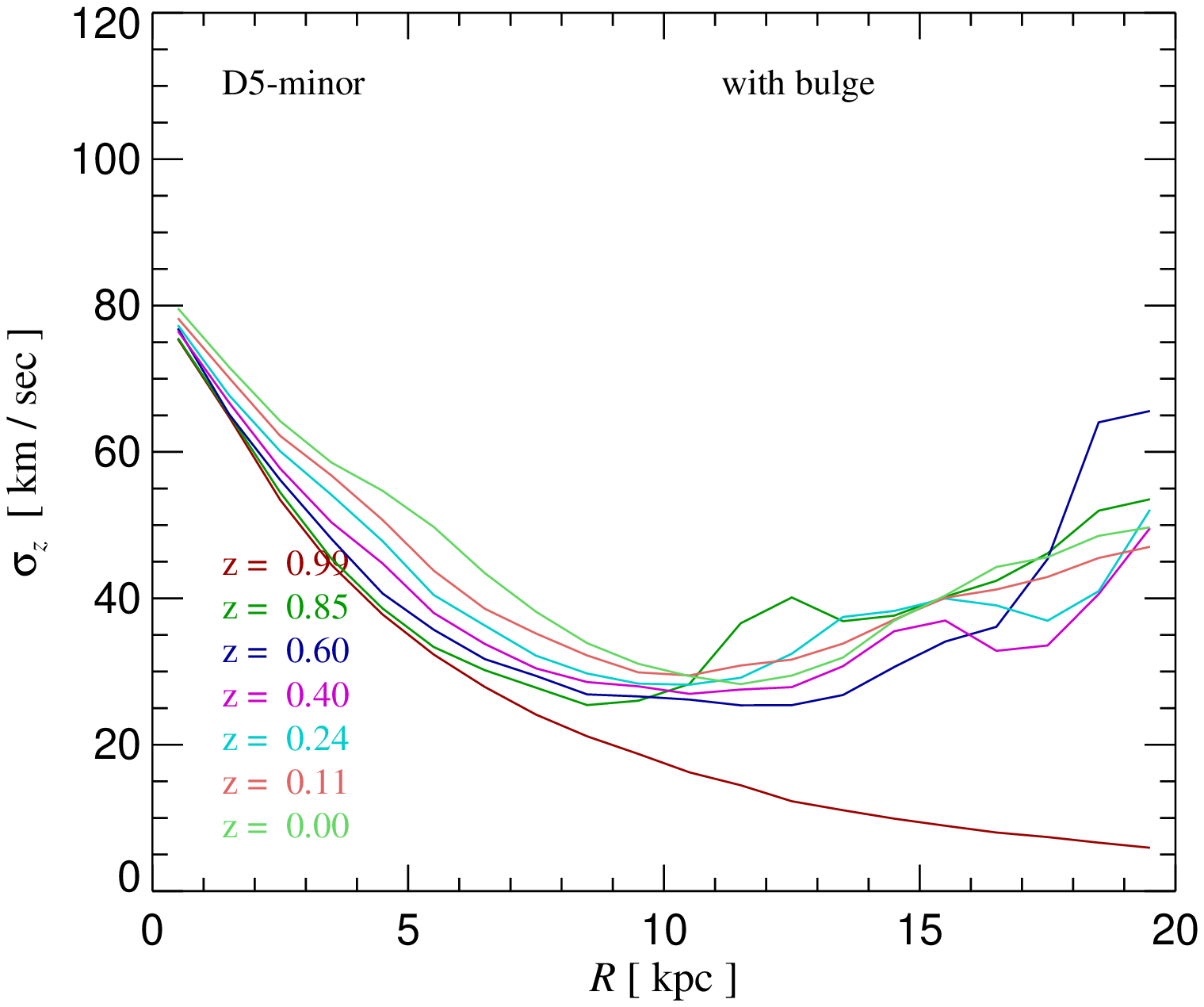}}\\%
\resizebox{4.5cm}{!}{\includegraphics{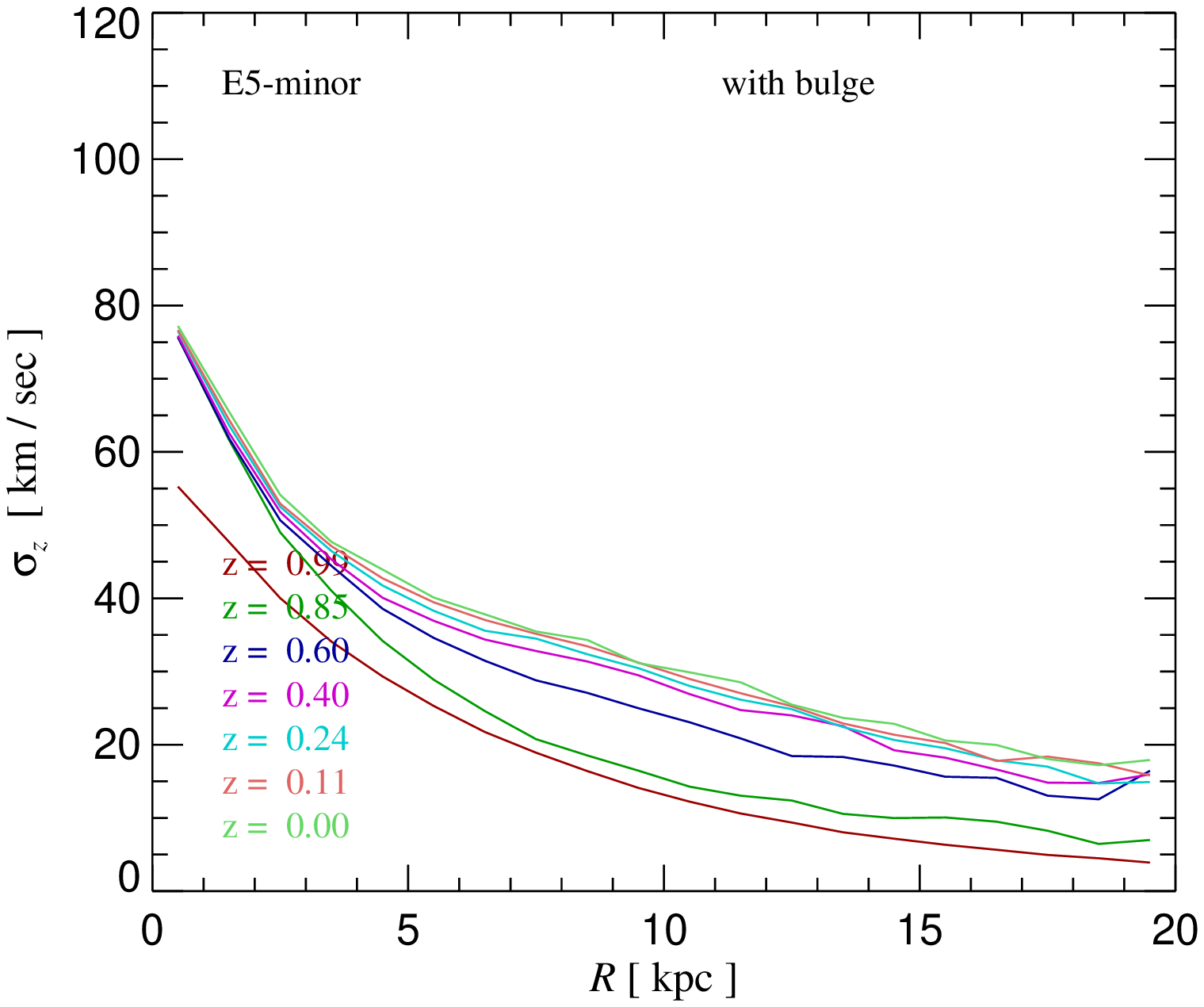}}%
\resizebox{4.5cm}{!}{\includegraphics{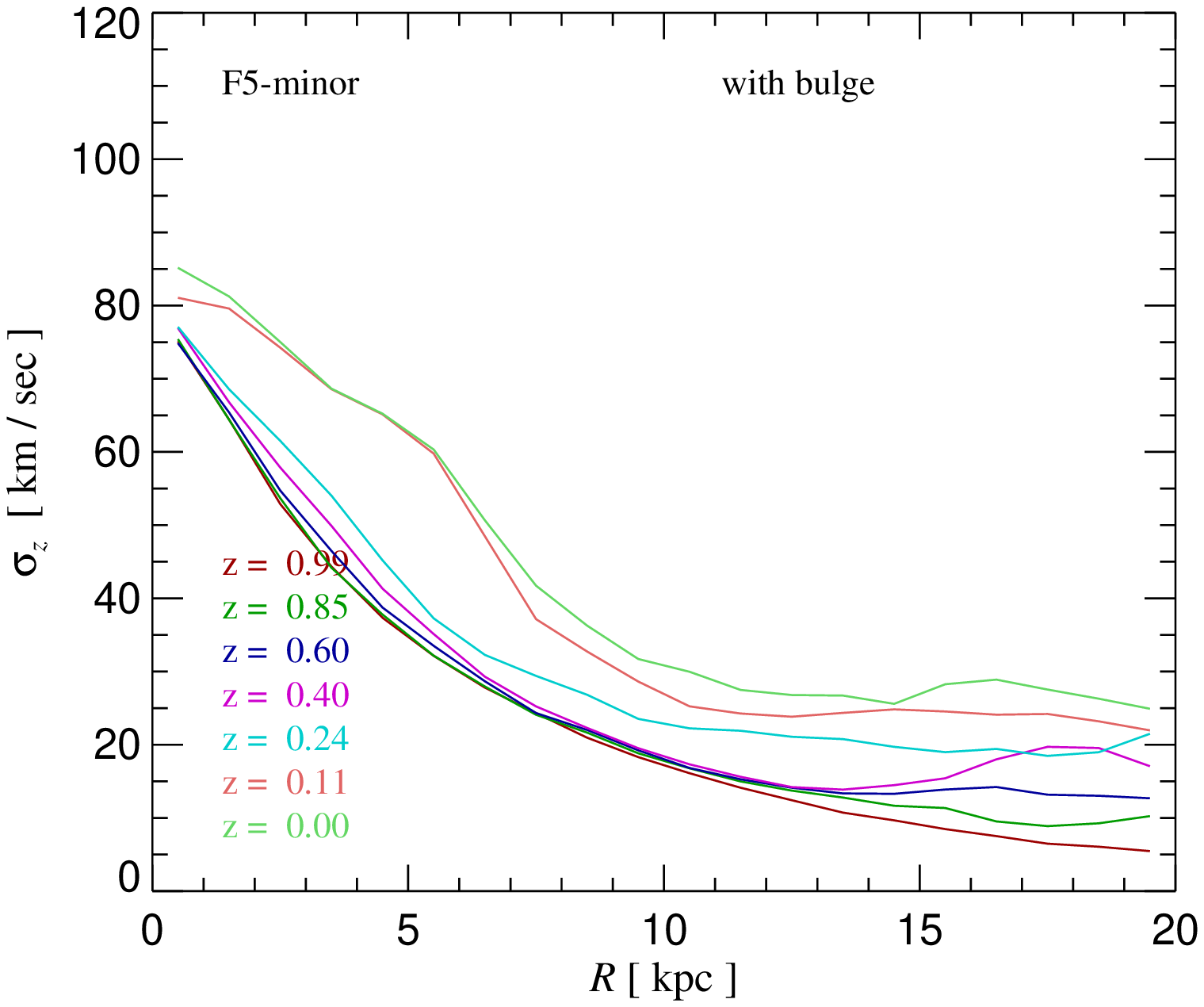}}%
\resizebox{4.5cm}{!}{\includegraphics{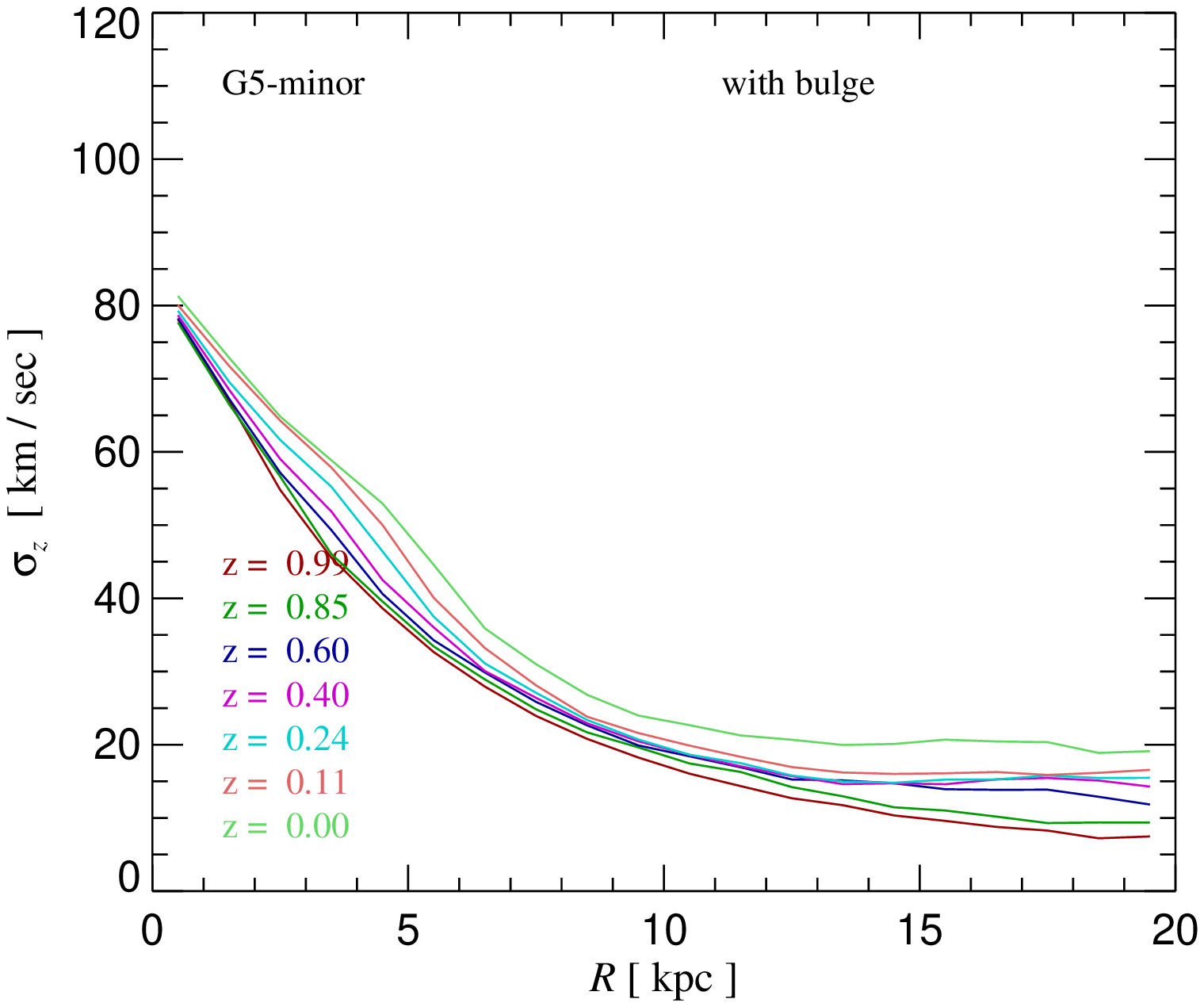}}%
\resizebox{4.5cm}{!}{\includegraphics{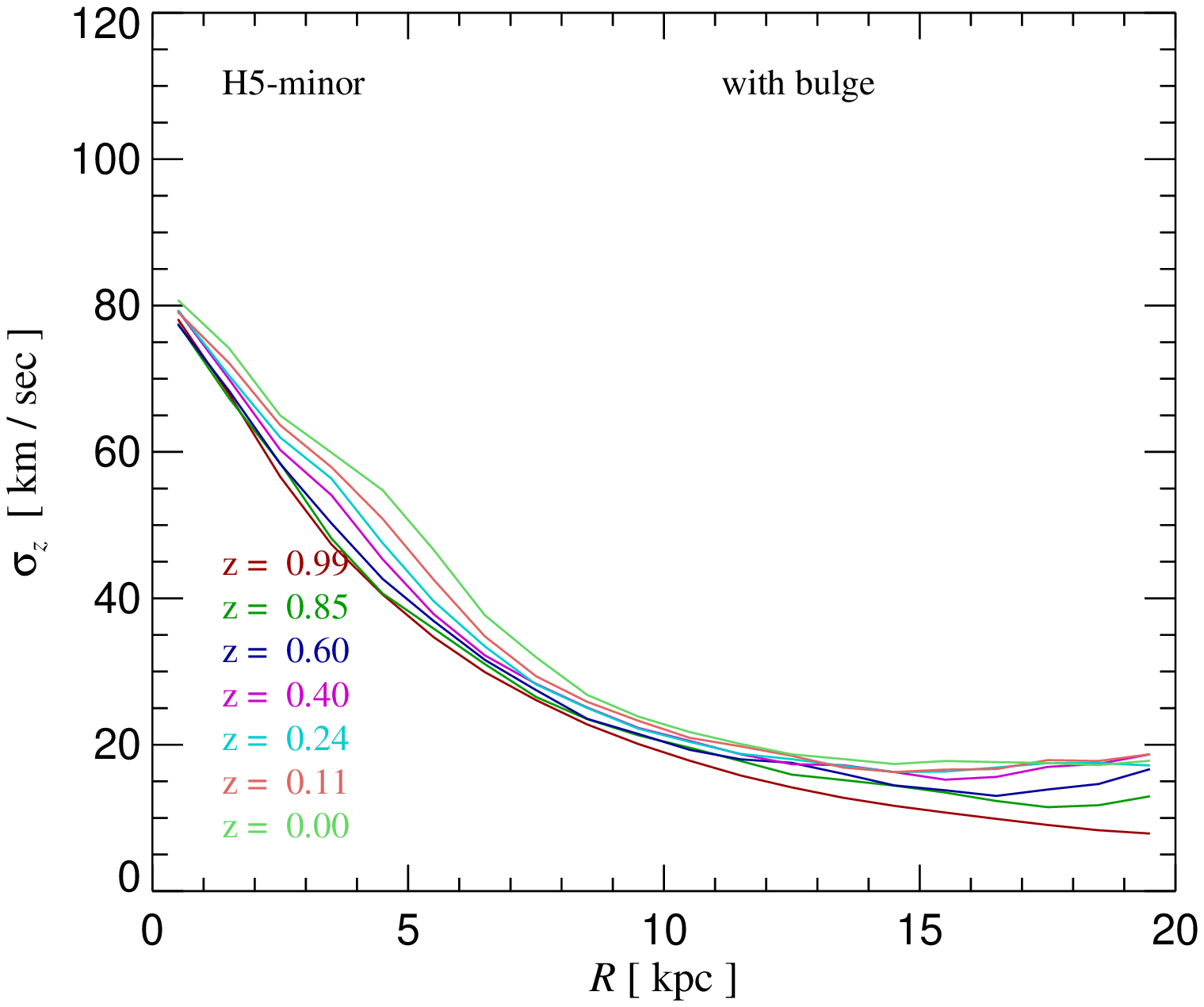}}\\%
\resizebox{4.5cm}{!}{\includegraphics{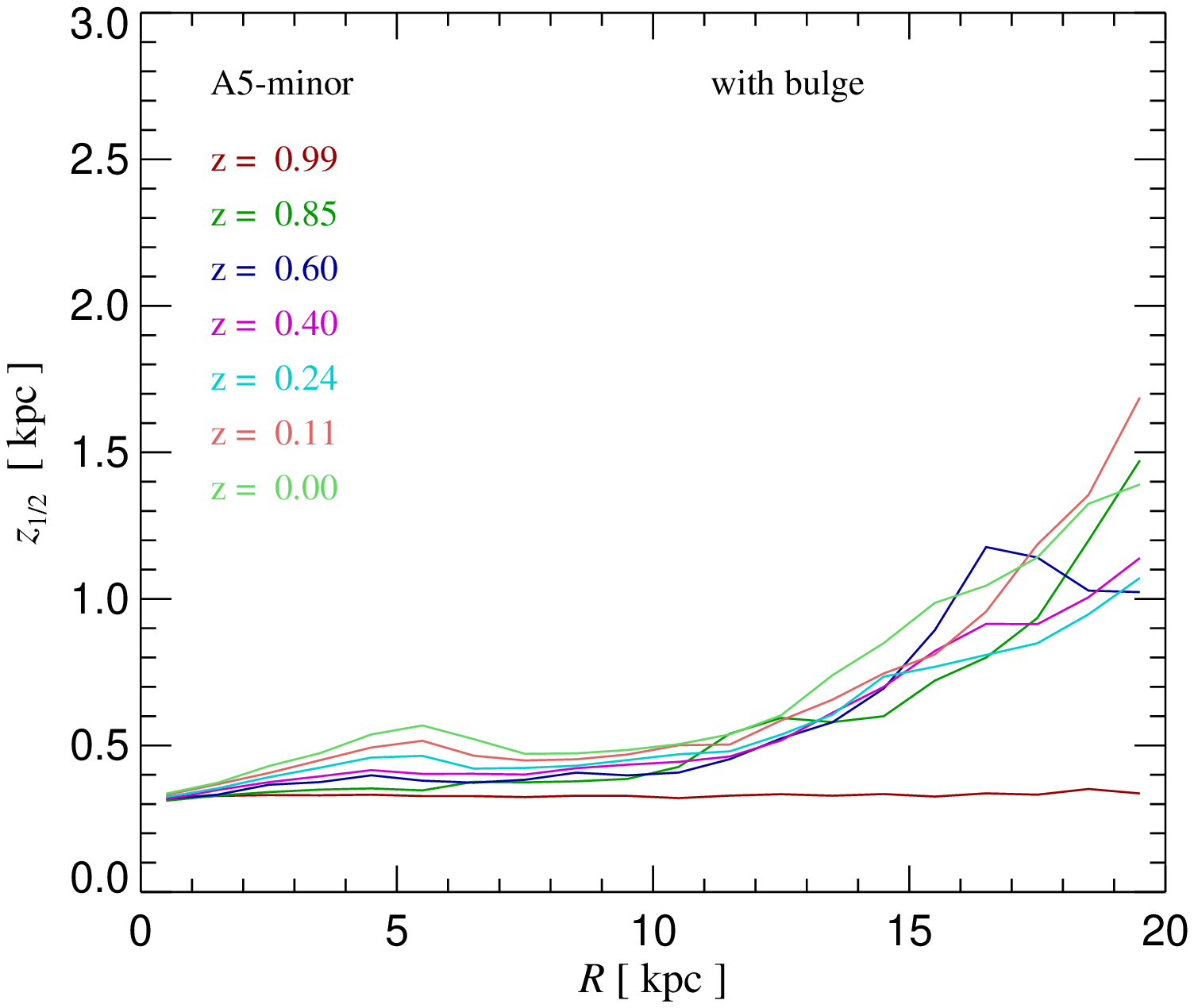}}%
\resizebox{4.5cm}{!}{\includegraphics{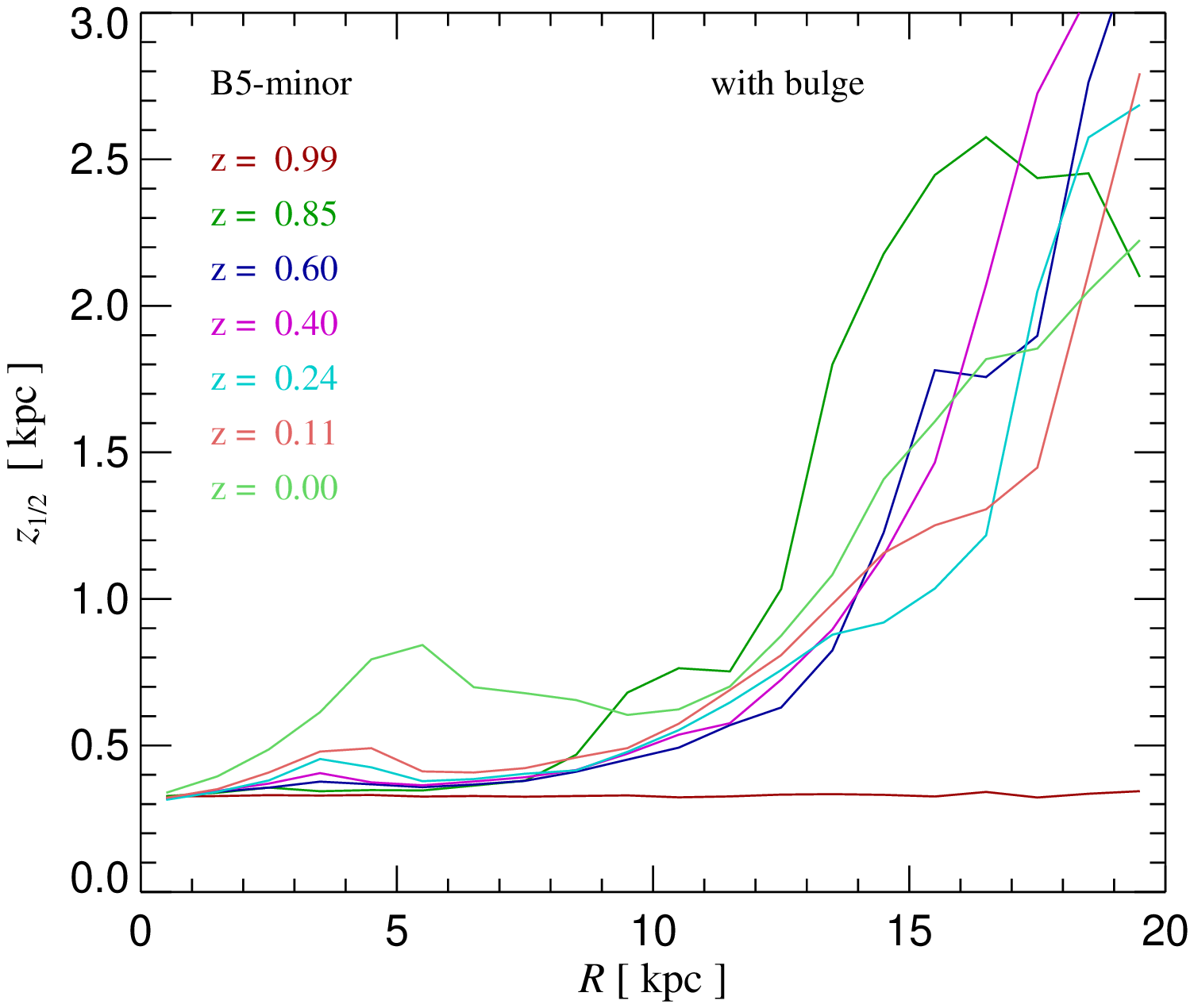}}%
\resizebox{4.5cm}{!}{\includegraphics{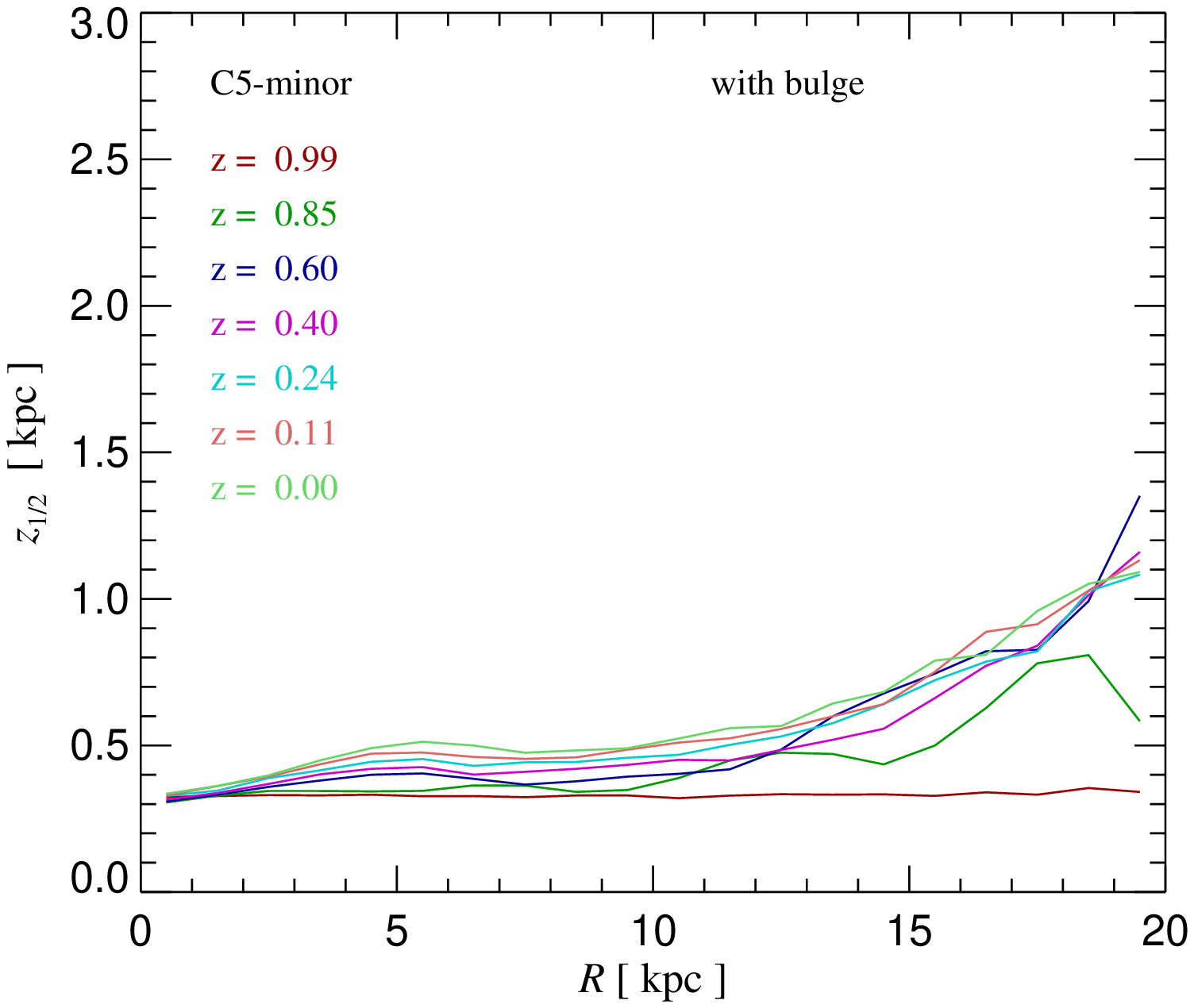}}%
\resizebox{4.5cm}{!}{\includegraphics{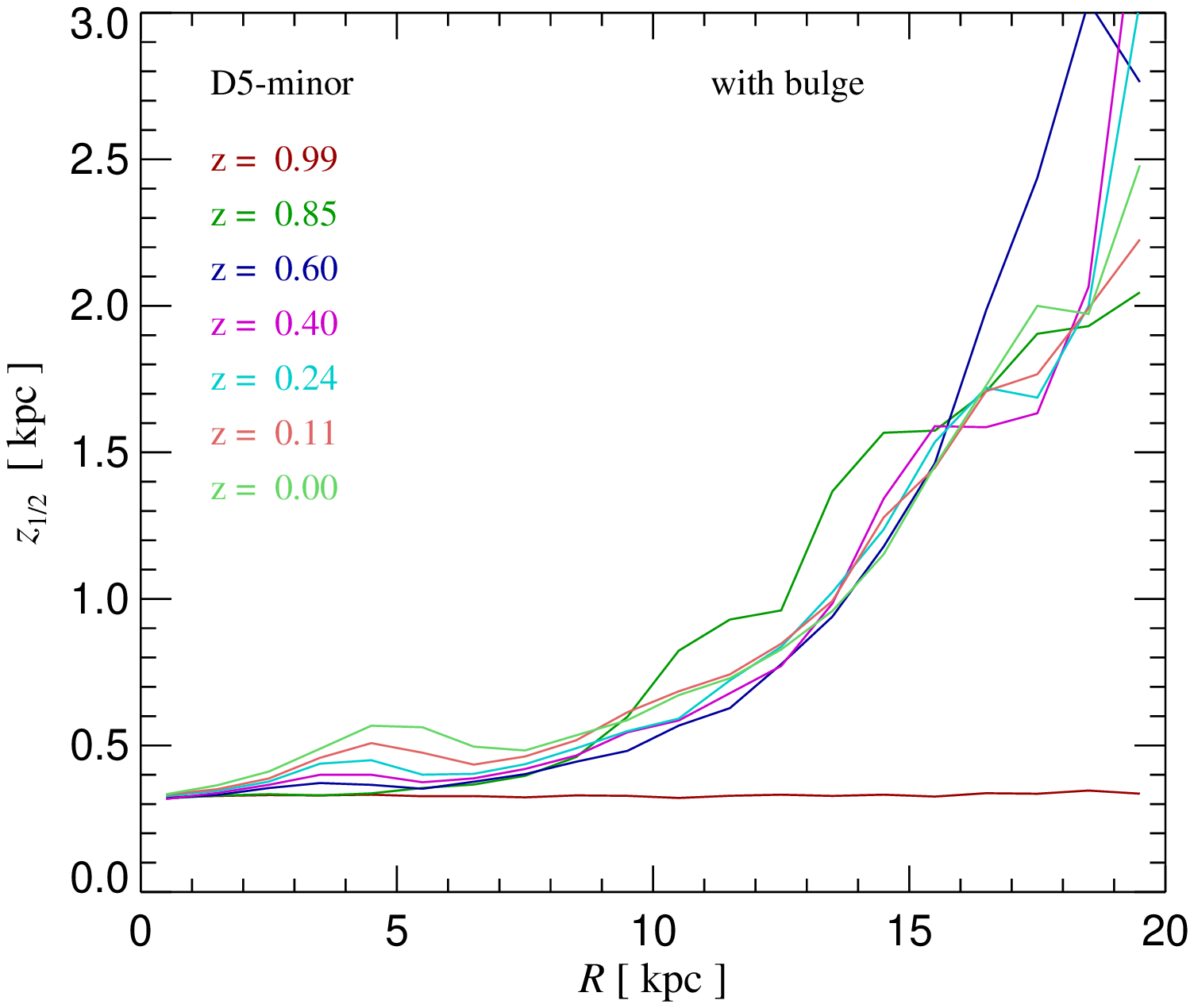}}\\%
\resizebox{4.5cm}{!}{\includegraphics{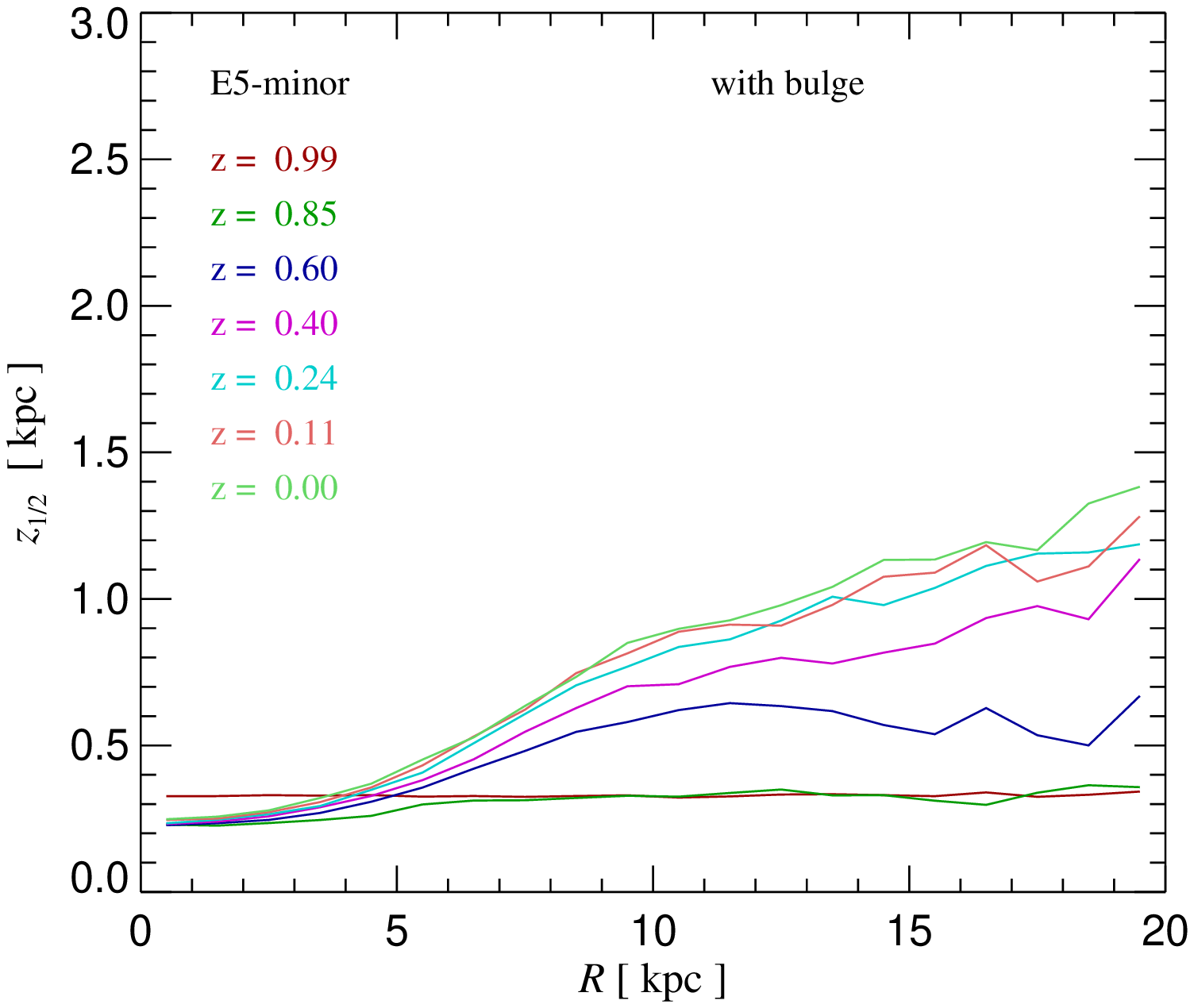}}%
\resizebox{4.5cm}{!}{\includegraphics{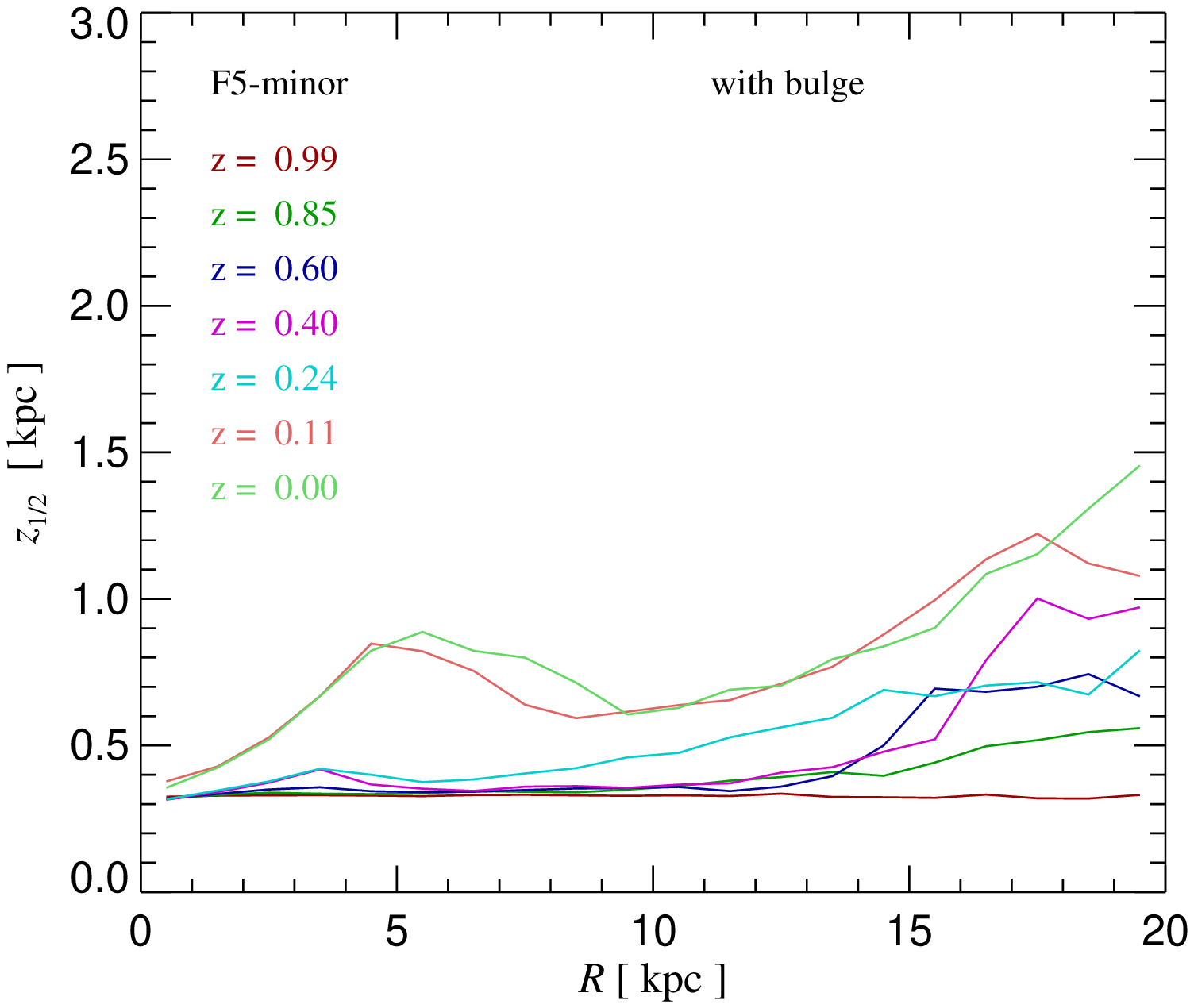}}%
\resizebox{4.5cm}{!}{\includegraphics{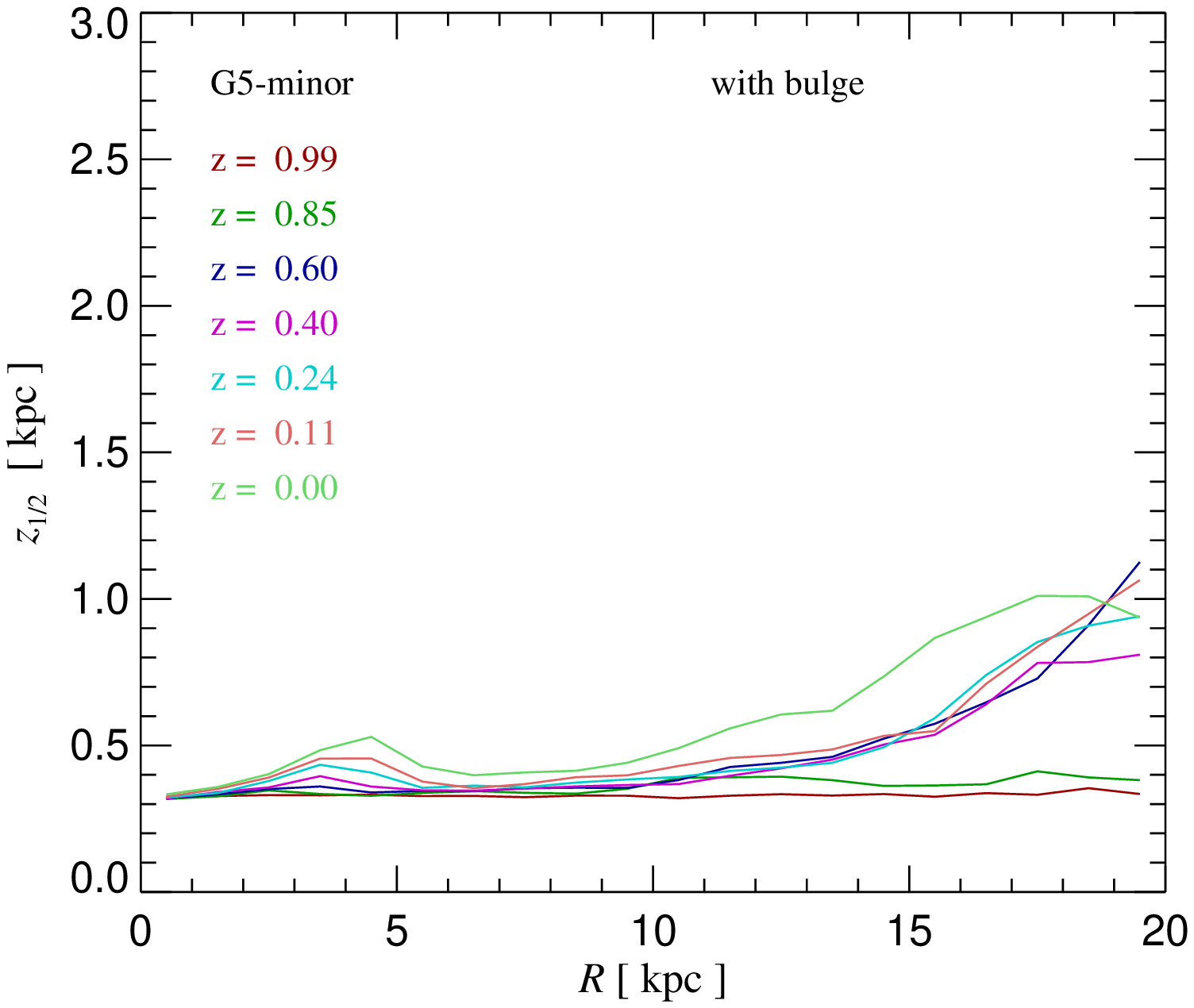}}%
\resizebox{4.5cm}{!}{\includegraphics{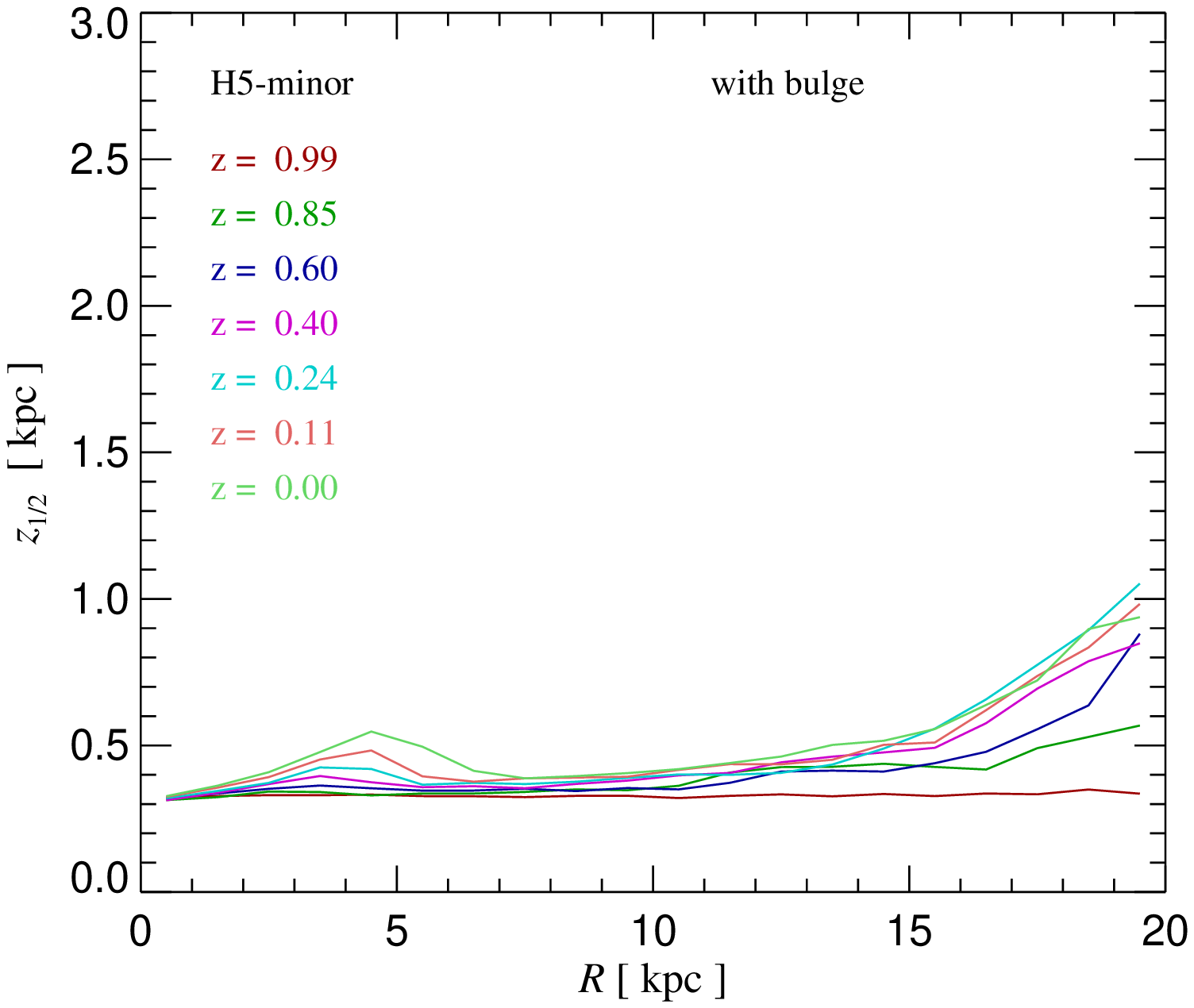}}\\%
\caption{
Time evolution of the vertical velocity dispersion profile and of
  the half-mass height profile of the disk particles in our default
  disk+bulge simulations (series \#3). These results correspond
  directly to the simulations for our pure disk simulations shown in
  Fig.~\ref{fig:default_heightprofiles}. Note that only disk star
  particles are included in the measurements.
\label{fig:dispersionprofile_bulges}
}
\end{center}
\end{figure*}

The equivalent simulations for inserting the disks along the major
orientation of the halos (series \#2) yield qualitatively very similar
results, and we therefore refrain from showing the corresponding
images. Figure~\ref{fig:haloaxes} gives instead an overview of the
time evolution of the orientation of the halos' minor, major and
intermediate axes, as defined based on the moment of inertia tensor of
the central dark matter distribution in a spherical aperture of size
$R_{\rm 200}/4$. We also include in the plots the evolving orientation
of the disk spin axis as a function of time (based on the minor
orientation runs of series \#1), as well as the dark matter spin
direction of the central region of the halo. The individual panels
are Mollweide projections of the corresponding direction angles, one
for each Aquarius halo. The origin of the projection has been shifted
such that the initial position of the disk orientation lies at the
center of the corresponding Mollweide map.

Interestingly, in several of the systems, the dark matter halo spin is
quite well aligned with the minor axis. This is in particular the case
for systems A, E, G, and H, and to a lesser degree in B. In halos C
and D, the dark matter spin is reasonably stable in orientation but
offset from the minor axis, while in F it wanders all over the
place. Studies of cosmological halos have long found a preference of
the dark matter halo spin to line up with the minor axis
\citep[e.g.][]{Hayashi2007}, a trend also seen here. Interestingly,
there is growing observational evidence from alignment studies of SDSS
galaxies \citep{Zhang2014} that favor a picture where disk spins line
up with the dark matter halo angular momentum of the inner regions of
halos, suggesting that the minor orientation arises naturally and is
actually preferred. Moreover, \citet{Aumer2013b} found in
  their idealized baryonic simulations of disk formation that disks
  indeed are most stable if the spin of the disk is aligned with the
  minor axis of the halo.

What is also evident from Figure~\ref{fig:haloaxes} is that the disk
spins of several of the models wander away significantly from their
original orientation, especially in those cases where the minor axis
shows a change of orientation as well. Of course, the change
  of spin orientation arises from gravitational torques,
  similar to how a spinning top precesses in an external field
  \citep{Landau1969}, but the detailed evolution of the live stellar
  disks is more complicated to understand due to the large-scale
  gravitational torques acting on the galaxy, as well as due to the
  coupling of the disk and the triaxial halo.  In cases where the
minor axis moves little, such as E and H, the disk spin exhibits a
fairly stable orientation \citep[see
  also][]{Debattista2013}. A clearer view of the size of the changes
in orientation is given in Figure~\ref{fig:disktilt}, where the angle
between the current disk orientation and its initial orientation is
shown as a function of time, from redshift $z=1$ to $z=0$. The left
panel shows our results when the disk is initially aligned with the
minor orientation, the right panel is for the major orientation. Some
of the models show rather substantial reorientations of the disks
reaching up to 65 degrees in the minor cases, and even larger angles
beyond 110 degrees for the major orientation. The average tumbling
angle (shown as a dashed line) for the eight systems is $35$ degrees
for the minor orientation, and $60$ degrees for the major
orientation. The substantially smaller average tilt angle for the
minor orientation suggests that this orientation typically offers
better long-term directional stability than the major
orientation, as expected.  We note that our results for A-D
show a large resemblance to those of \cite{DeBuhr2012}, for example,
we find the same characteristic evolution pattern for halos A and
C. This is reassuring, given the independent and at a technical level
quite different methodology to introduce and simulate the live
disks. However, there are also some quantitative differences, and for
a subset of the systems we tend to find somewhat smaller angles than
\cite{DeBuhr2012}.

\subsection{Radial and vertical structure, and its evolution}

As pointed out by \citet{DeBuhr2012}, it is perhaps not too surprising
that these systems show such strong tendencies to form bars. Whereas
their rotation curve structure, shown in
Figure~\ref{fig:default_rotcurve}, in principle suggests that the
disks are not exceeding the rotation curve contribution of the dark
matter anywhere (apart from halo G for a small region) and are thus
far away from being maximal disks, the simple criterion of
\citet{Efstathiou1982} for stability against bar formation,
\begin{equation}
Q_{\rm bar} \equiv \frac{v_{\rm max}}{(G M_d / R_d)^{1/2}} > 1.1
\label{EqnELN}
\end{equation}
is violated for all the models. Here $v_{\rm max}$ is the maximum
rotation curve velocity, and $M_d$ and $R_d$ refer to mass and
scale-length of the exponential stellar disk. In fact, the values for
$Q_{\rm bar}$ after the disks have been inserted are 0.99, 0.79, 1.0,
0.88, 0.87, 0.82, 0.85, and 0.92, for A to H, respectively.

It should be noted however that the simple criterion of
  Equation~(\ref{EqnELN}) is not particularly reliable in general disk
  systems. In fact, \citet{Athanassoula2008} has given a number of
  counter-examples that violate this criterion for bar
  instability. For example, if a disk is supported by a high radial
  velocity dispersion it can evade forming a bar even if the criterion
  by \citet{Efstathiou1982} is violated. Similarly,
  \citet{AthanassoulaMisiriotis2002} have shown examples for bar
  growth in massive, quite concentrated halos where this would have
  been unexpected according to the criterion. However, our disk
  systems were all initialized as isotropic rotators, and hence form a
  more restricted class of models for which Equation~(\ref{EqnELN})
  can still be useful to assess the relative susceptibility to bar
  formation.

In this context it is also interesting to look at Toomre's stability
parameter for axisymmetric stability of stellar disks
\citep{Toomre1964}, 
\begin{equation}
Q_{\rm Toomre} \equiv
\frac{\sigma_R\,\kappa}{3.36\, G \Sigma} \, > \, 1,
\end{equation}
where $\sigma_R$ is the radial velocity dispersion, $\Sigma$ the
surface density, and 
\begin{equation}
  \kappa^2 = \frac{3}{R} \frac{\partial\Phi}{\partial R} +
  \frac{\partial^2 \Phi}{\partial R^2}
\end{equation}
is the epicycle frequency.
The value of $Q_{\rm Toomre}$ is shown as a function of radius in
Figure~\ref{fig:default_toomre}. Interestingly, most of the models are
Toomre stable, with B and F being marginal cases, but the light halo E
is clearly predicted to be unstable against axisymmetric
instabilities. And indeed, inspecting the stellar images at $z=0.85$
in Fig.~\ref{fig:defaultruns_faceon} one can clearly see ring-like
spherical features that are absent in this form in the other models,
providing evidence that such instabilities have occurred in the early
evolution of the system.

We find further signs for this special evolution of halo E in the
evolution of the structural properties of the systems, which we
examine next. In Figure~\ref{fig:default_radialprofiles} we show the
radial and vertical density profiles, at a set of different times. The
exponential surface density profile measured for the face-on
orientation of the disks is quite robust and more or less retains its
initial shape, despite the rather dramatic bar formation events
occurring in these simulations. In contrast, the vertical density
profiles (three families of curves are shown, corresponding to
different radial ranges, as labelled) show the damaging impact of the
forming bars more clearly. In particular, a relatively sudden
transition to a new vertical equilibrium with a thicker profile is
apparent in most of the models. One interesting difference with
\citet{DeBuhr2012} is that our models A-D show substantially less
broadening in the outer parts of the disks. This is presumably a
reflection of our more accurate approach to initialize the velocities
of the initial disk models.

Further support for this is provided by the evolution of kinematic
quantities, for example those shown in
Figure~\ref{fig:default_heightprofiles}. The eight panels on top in
the figure give the evolution of the profiles of the vertical velocity
dispersion $\sigma_z^2$, while the bottom eight measure the disk
height in terms of the median $z_{1/2}$ of the absolute value of the
vertical $z$-coordinates of the star particles relative to the disk
plane. Or in other words, half of the stellar mass has a height above
the central disk plane less than $z_{1/2}$. Comparing again to
\citet{DeBuhr2012}, we see that our $\sigma_z(r)$ profile at $z=0.85$
is much closer to the initial profile than in their case, suggesting
that our disk models are in better dynamical equilibrium
initially. This equilibrium is however anyway destroyed relatively
quickly by the onset of bar formation. Another notable difference are
the much smaller values for $z_{1/2}$ we find in the outer parts of
the disks, corroborating the observation that our disk models appear
to be in better equilibrium.

\subsection{Bar strength and vertical heating}

The formation of the bar can also be studied more quantitatively, for
example through measuring a bar strength indicator, or by looking at
the vertical heating of the stellar disk. As a simple global measure
of the total amount of vertical heating we can use the quantity
\begin{equation}
\zeta = \frac{\left<v_z^2\right>}{\left<v_z^2\right>_0}, 
\end{equation}
which simply is equal to the total kinetic energy in vertical motion
relative to the initial value of this quantity at the time the disk
starts to evolve live.

As a characterization of the bar strength, we adopt a simple measure
for the $m=2$ Fourier mode of the disk, as is often done to
quantify bars. For definiteness and ease of comparison, we measure the
bar strength similarly as~\cite{DeBuhr2012}, by first determining
the quantities
\begin{equation}
a_2^{(b)} = \sum_{i \in b} m_i \cos(2\phi_i)
\end{equation}
\begin{equation}
b_2^{(b)} = \sum_{i \in b} m_i  \sin(2\phi_i)
\end{equation}
for a set of 30 radial bins between $R=0$ and $R=2\,R_d = 6\,{\rm
  kpc}$. Here $\phi_i$ refers to the azimuthal angle of each disk
star, and the sums extend only over the particles in a radial bin $b$.
Defining $c_2 = \sqrt{a_2^2 + b_2^2}$, we then calculate a bar strength
parameter as
\begin{equation}
A_2 = \frac{\sum_b R_b\, c_2^{(b)}}{\sum_b' R_b' M_b'}
\end{equation}
where $M_b$ is the mass falling into bin $b$, and $R_b$ is the bin radius.

In Figure~\ref{fig:default_barstrength}, we show our results for the
time evolution of the $A_2$ parameter in the left panel, and in the
right panel we give the time evolution of the vertical heating
parameter $\zeta$. Consistent with the evolution of the visual
morphology, all the systems show a rapidly growing bar signal in their
early evolution, with all the models except E converging to a
similarly high bar strength of $A_2 \simeq 0.6$ at the
end. 

Interestingly, the relative vertical heating parameter shows only a
small and slow growth in the beginning, but then jumps up rapidly by a
large factor, followed by a stabilization at a new high level.  In
contrast, the bar strength indicator becomes high already
significantly earlier. Presumably this is a combination of two
different effects. One is that some disk asymmetries quickly develop
in the early disk evolution simply because of the non-sphericity of
the halo potential, which for example manifests itself in pronounced
spiral patters in the disks. The other is that during the initial
phase of bar formation the density contrast of the bar grows without
yet leading to a notable change in the vertical structure. Only later,
once the bar ``collapses'' or buckles, a sudden transition to a new
equilibrium occurs, and this is associated with some degree of violent
relaxation and significant vertical heating
\citep[e.g.][]{Martinez-Valpuesta2006, Berentzen2006A}.

\section{Models with bulges and lighter disks} \label{sec:bulges}

The strong bars formed in the models considered in the previous
section raise the question under which conditions disk galaxies could
survive in the Aquarius halos and maintain a nice, disky morphology
all the way to $z=0$. It appears clear from the preceding results
that the bar criterion of \citet{Efstathiou1982} should be taken as an
important first guide. Increasing the value of $Q_{\rm bar}$ in a
given dark matter halo can be achieved first of all by making the disk
lighter, and/or by making it larger. However, we note that there is
only limited room for such changes if one wants to keep the disk
parameters close to observational inferences, such as the stellar
mass--halo mass relationship derived from abundance matching
\citep{Guo2010, Moster2010}, or direct constraints on the
size--stellar mass relationship \citep{Shen2003}. Another problematic
aspect of simply adopting a very light disk is that this will greatly
reduce the roundening effects of the dark matter halo due to the
growth of a baryonic mass distribution in the centre. Such a rounding
is however essential to keep the destructive effects of a highly
triaxial dark matter potential on disk stability and morphology at
bay.

Another approach to improve the stability of the disk against bar
formation is to add a central stellar bulge in addition to the
disk. Such a bulge increases the circular velocity of the spheroidal
component of the system, i.e.~it contributes to the numerator of the
bar stability criterion. At the same time, growing a central spherical
bulge in a dark matter halo is an effective way to rounden the dark
matter potential. Of course, on the other hand, adding a central bulge
is limited by the observed bulge-to-disk ratios, and is not a
promising option for explaining pure disk galaxies. 

To test these options, we have considered a few additional sets of
runs. Our default standard bulge models (series \#3) have the same
stellar mass as our default disk models, but one third of the stellar
mass is put into a bulge, with two thirds remaining in the disk. Such
bulge-to-disk ratios are about the smallest ones that the present
generation of hydrodynamical simulations of disk formation can achieve
\citep[e.g.][]{Marinacci2014}.

In addition, we have also run models were we omitted this bulge, which
is exploring the lighter disk option at some level.  Here the disk
mass was reduced to 2/3 of the value in our default models (series
\#4). This is complemented by a further set of runs in which we
swapped the masses of disk and bulge of our standard bulge models,
ending up with 1/3 in the disk and 2/3 in the bulge (series \#5). We would
expect these models to be extraordinarily stable against bar formation.

\begin{figure*}
\begin{center}
\resizebox{12cm}{!}{\includegraphics{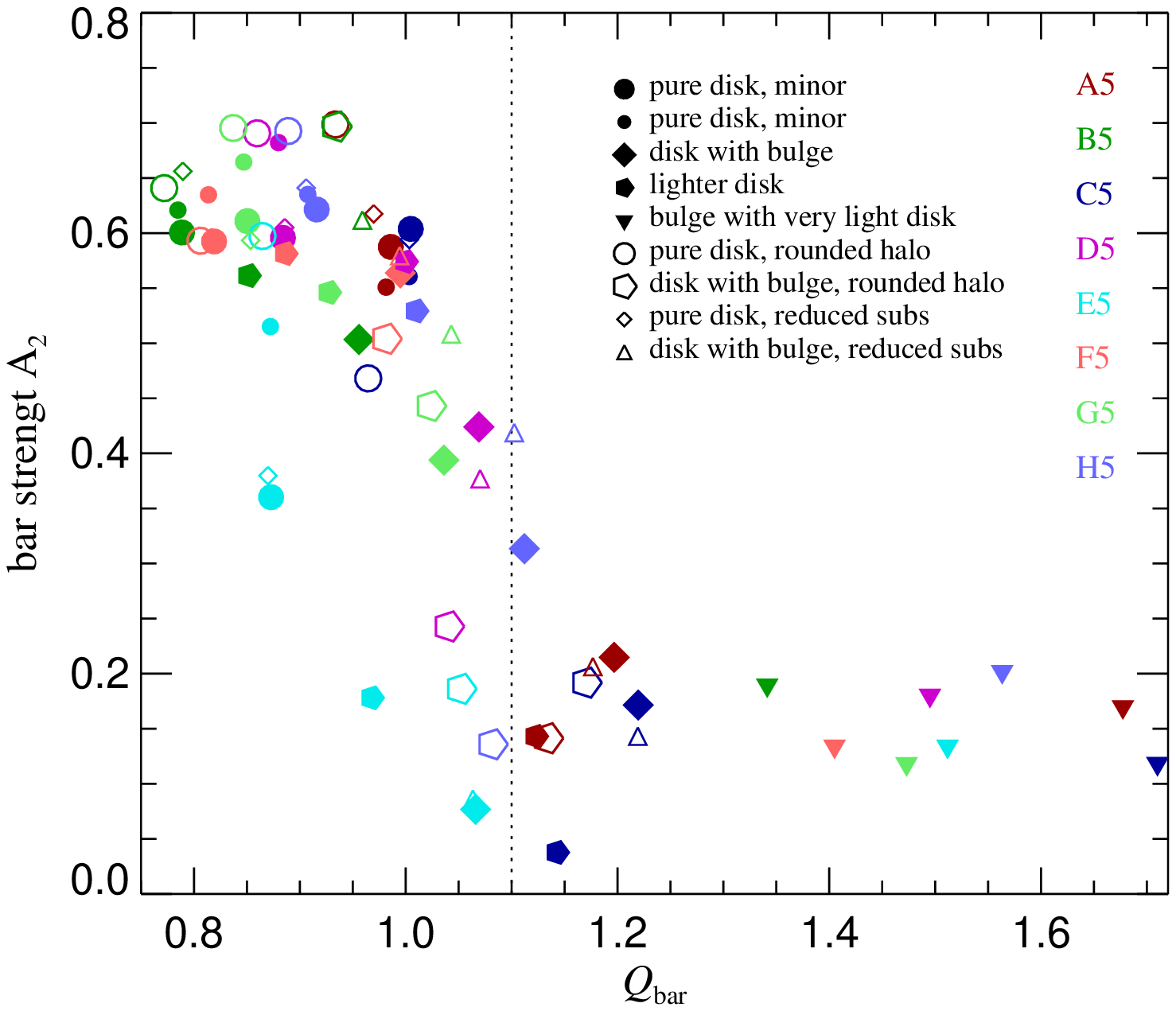}}%
\caption{Compilation of the measured bar strength parameter $A_2$ at
  $z=0$ in our simulation set versus the $Q_{\rm bar}$ parameter of
  \citet{Efstathiou1982}.  We here collect results for all of our runs
  (except for the late insertion and reorientation ones). Different
  colours are used for the different Aquarius halos, and different
  symbols for the different simulation series, as indicated in the
  legend (the number in the symbol key refers to series number in
  Table~\ref{tab:sims}). It is clear that the criterion $Q_{\rm bar}
  \ge 1.1$ formulated by  \citet{Efstathiou1982} for indicating
  stability against bar formation provides a rough 
    guide for our restricted class of simulations, even though
it has been shown to fail in more general situations \citep{Athanassoula2008}.
 Parameters such as minor or major disk
  orientation, or the residual asphericity of the dark matter halo,
  appear to be only of secondary importance.
  \label{fig:bar_vs_Q}}
\end{center}
\end{figure*}

\begin{figure*}
\begin{center}
\setlength{\unitlength}{1cm}
\resizebox{8cm}{!}{\includegraphics{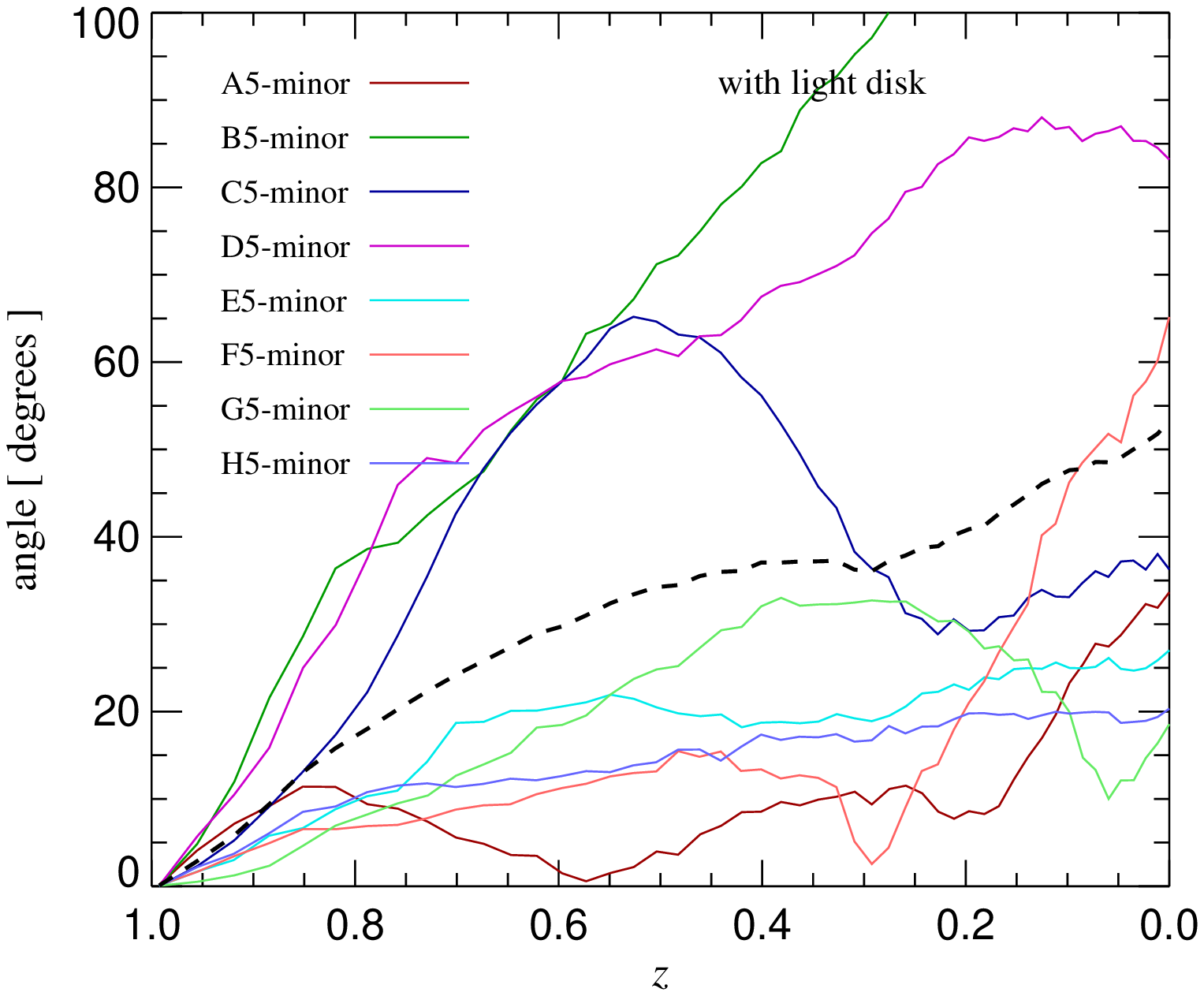}}%
\resizebox{8cm}{!}{\includegraphics{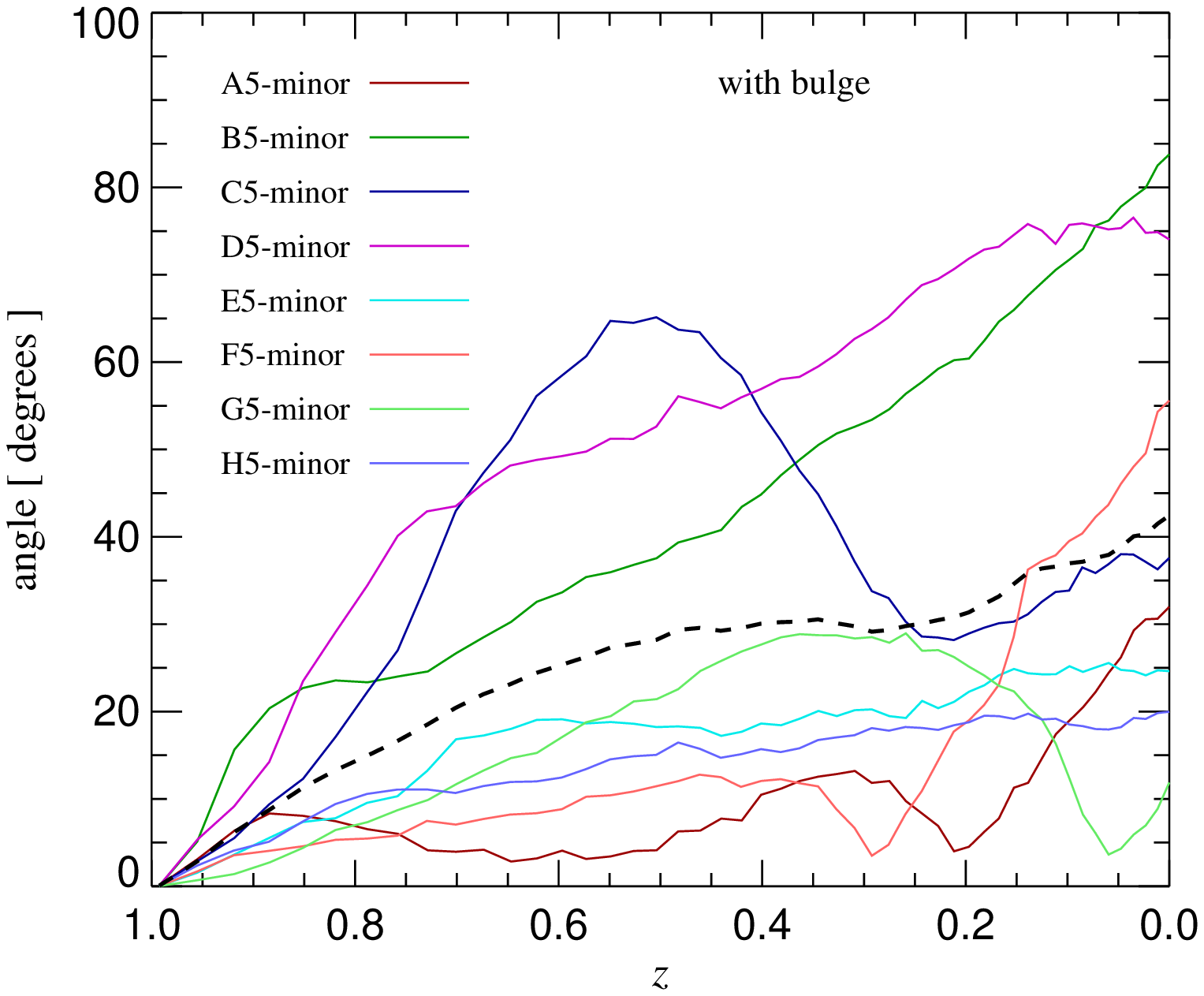}}\\
\resizebox{8cm}{!}{\includegraphics{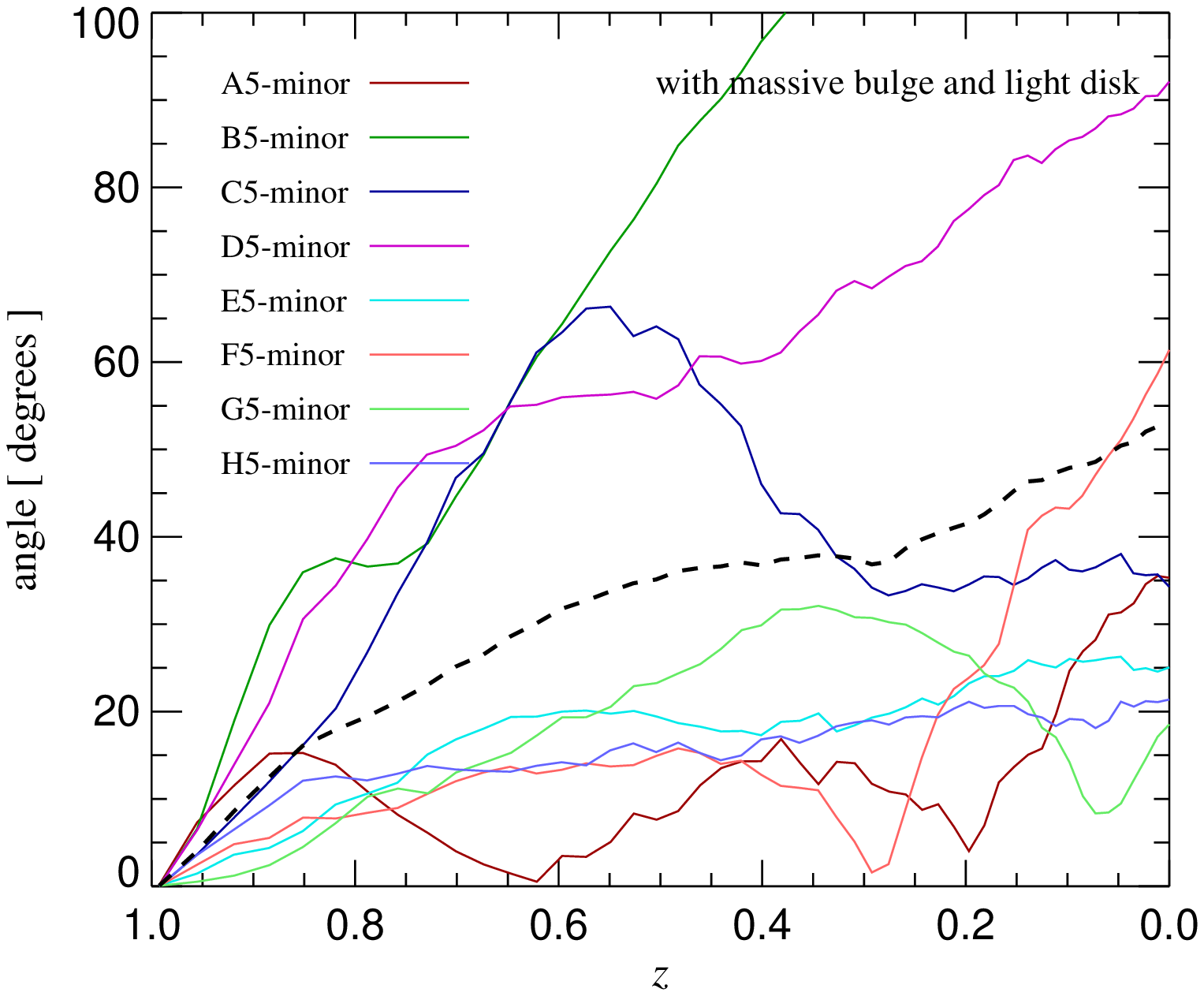}}%
\resizebox{8cm}{!}{\includegraphics{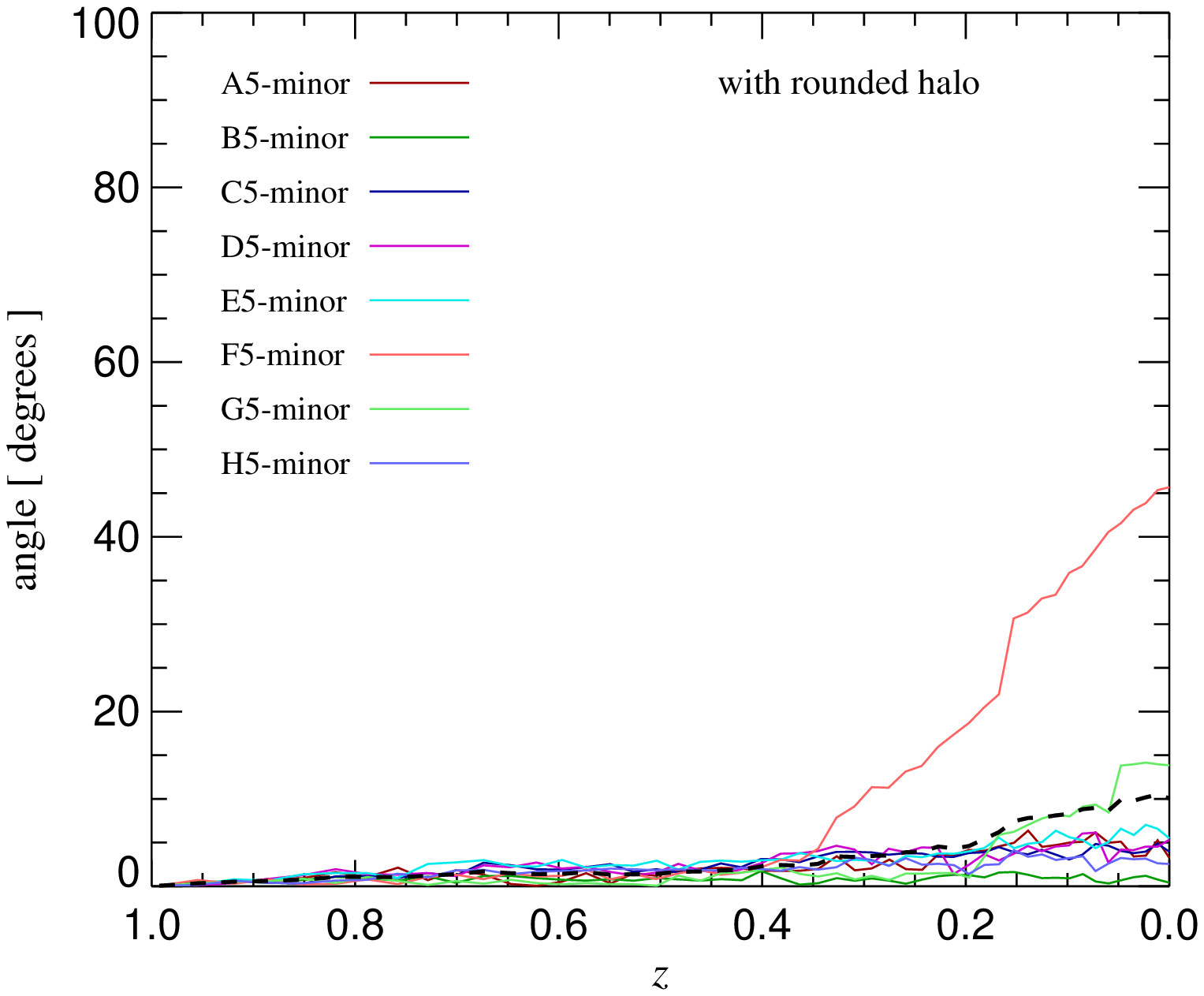}}\\
\caption{Angle between the current disk spin axis and the initial disk
  orientation as a function of time, for different types of
  simulations. The top left panel shows results for our simulation
  series \#4, where a lighter disk is used compared with our default
  disk runs. The top right panel shows our default disk+bulge models
  instead (series \#3), whereas the bottom left gives the tilt angle
  evolution for our simulations with massive bulges and a light disk
  (series \#5). Finally the bottom right is for our default pure disk
  simulations, but this time the dark matter halo has been
  artificially rounded at $z=1.3$ when the disk was inserted (series
  \#6). Interestingly, the disk structural properties appear to play
  only a minor role for the disk tumbling (compare also to
  Fig.~\ref{fig:disktilt}). The latter is clearly governed by the dark
  matter halo and can only be eliminated by substantial roundening
  that is far stronger than the effects achieved by the halo's
  response to the growing baryonic component.
\label{fig:disktilt_othermodels}}
\end{center}
\end{figure*}

Finally, to test the influence of the residual triaxiality of the dark
matter halos on the disk stability, we have also run two sets of
simulations where we artificially sphericalized the dark matter halos
at the instant of disk insertion, i.e.~at $z=1.3$ (series \#6 and
\#7). To this end, we simply took all dark matter particles in the FOF
halo at that time and rotated them randomly around the halo
center. Also, their velocities were isotropized by turning them
randomly in the rest frame of the halo. In this way, the density
structure and potential energy of the halo was approximately
maintained, with any deviation from non-equilibrium decaying away
during the disk growth phase. When the disk goes live at $z=1.0$ it
then does so in an essentially spherical halo, but the mass growth of
the halo, including the accretion of newly infalling substructures
on to the halo, stays unaffected.

We begin our review of the results of these runs with images of the
disks in the runs with the small bulges. These are shown in face-on
and edge-on orientations in Figures~\ref{fig:bulgeruns_faceon} and
\ref{fig:bulgeruns_edgeon}, respectively. Note that the bulge stars
are not included in these images. Compared to the corresponding images
for the simulations with pure disk, the disks are clearly much less
effected by bulge formation, although most systems still do form bars,
albeit of weaker strength and at later times. The edge-on projects of
the disk stars correspondingly show much less evidence for central
bars.

The rotation curves of the models with bulges, as well as the
corresponding profiles of the Toomre stability parameter are shown in
Figure~\ref{fig:rotcurve_toomre_withbulges}. None of the models is any
more dominated by baryons at any radius, and as expected, the
resistance against axisymmetric instabilities is increased
substantially.  Note however that halo E is still predicted to be
Toomre unstable. This is confirmed by visual inspection of simulation
E at $z=0.85$ in Fig.~\ref{fig:bulgeruns_faceon}, which reveals
conspicuous axisymmetric rings of stars.

The weaker bar strengths in the bulge-runs also become evident in
Figure~\ref{fig:barstrength_withbulge}, which shows the time evolution
of the bar strength parameter $A_2$ and the vertical heating parameter
$\xi$ of the disks of our eight standard bulge models. Comparing to
the corresponding results for the pure disk models
(Fig.~\ref{fig:default_barstrength}) clearly shows a much weaker trend
towards making strong bars. Most notably the early evolution is very
different, where all the models show a small value of $A_2$ as a
result of deviations from axisymmetry due to the aspherical dark halo
potential, but no strong bulge signal is present yet. Only after a few
Gyr, a subset of the models starts to grow a significant bar.

It is also interesting to examine differences in the evolution of the
structural properties of the models with bulges. This is shown in
Figure~\ref{fig:radialprofiles_withbulge} in terms of the radial and
vertical density profiles, and in
Figure~\ref{fig:dispersionprofile_bulges} in terms of the vertical
velocity dispersion and half mass height as a function of radius. We
see a very substantial improvement in the structural stability of the
models with respect to the runs with pure disks. This is reflecting
both, the stronger stability against bar formation, and the slightly
rounder dark matter halo potential due to the more concentrated
distribution of the baryons.

Getting almost completely rid of the bars is achieved in our models that
adopt a massive bulge and a light disk (series \#5). Here the disk
images show a remarkable degree of stability.  For conciseness, we
refrain from showing their uneventful time evolution. Instead, we
collect all our simulations in a single plot of the bar strength
versus the initial $Q_{\rm bar}$-parameter of \citet{Efstathiou1982},
allowing us to assess how well this venerable criterion works in the
context of full cosmological CDM models. Figure~\ref{fig:bar_vs_Q}
compiles our different runs, with the bar strength $A_2$ measured at
$z=0$, and the $Q_{\rm bar}$-parameter evaluated at $z=1.0$ when the
disk goes live.  Remarkably, the threshold value $1.1$ originally
introduced by \citet{Efstathiou1982} for the dividing line between
bar-unstable and stable models still serves as a surprisingly robust
indicator, even in the light of all sorts of other complicating
factors. In particular, we note that the different symbols in the
figure show a broad range of simulation models, including runs with
and without bulge, with minor or major axis orientation, with ordinary
or artificially rounded dark matter halos, etc. Irrespective of these
factors, it appears that the strength of the disk self-gravity
relative to the supporting spheroidal potential is by far the most
decisive parameter for governing stability against the formation of
strong bars. We note that this therefore cannot be ignored in the
interpretation of mass models derived for the Milky Way. For example,
the rotation curve decomposition derived by \citet{Bovy2013} suggests
that the corresponding live galaxy model should be violently bar
unstable according to our results.

In light of these differences, it is now interesting to consider the
stability of these models with respect to the spatial orientation of
the disks. In Figure~\ref{fig:disktilt_othermodels} we show results
for several of our modified runs, including the ones with lighter
disks, the ones with a bulge, the ones with a very massive bulge, and
the ones for a rounded dark matter halo. The latter are for the pure
disk case (series \#6), but the results for the bulge case (series
\#7) look essentially identical.

Comparing with the corresponding results in Fig.~\ref{fig:disktilt}
for the pure disk case, it is evident that the amount of disk tumbling
is fairly independent of the structural properties of the galaxies. In
particular, it does not matter much whether a bar is present or
not. Apparently, the reorientation of the disk is primarily controlled
by the tumbling of the dark matter halo and the torques it exerts on
the disk, and this is only marginally affected by the growth of the
baryonic disk/bulge system. Only when the halos are artificially
rounded and any figure rotation of the inner dark matter halo is
stopped by construction, the disk orientation remains
stationary. Except for halo F -- its disk turns even in this case by a
substantial angle, starting at $z\simeq 0.35$. The same characteristic
turning motion of F-disks is also seen in the standard runs at this
time, indicating that this is caused by the fly-by of a massive
substructure that interacts with the disk at this time and torques it
substantially.

The results above suggest that disk tumbling of significant size is
virtually inevitable in CDM halos. We expect typical tumbling rates of
about 40 degrees from $z=1$ to $z=0$, or about $\sim 6-7$ degrees per
Gigayear on average. Some systems may have up to 2-3 times that, while
others stay below it by a similar factor. Occasionally, disks may also
be brought into a turning motion by a close encounters with a
substructure. Importantly, our results show that disks can survive such
reorientations largely unaffected, i.e.~they are not in apparent
conflict with the observed abundance of thin stellar disks.

\section{Impact of substructures} \label{sec:subs}

Cold dark matter subhalos contain a large amount of substructures,
raising the question whether they may interfere with the stability of
cold stellar disks and induce substantial heating. We note however
that it is well established that substructures populate primarily the
outer parts of dark matter halos \citep[e.g.][]{Ghigna1998,
  Diemand2004, Springel2008}, leaving the inner halo relatively
smooth. Also, the subhalo mass function is skewed very slightly to
being dominated by the most massive subhalo systems. Those are
expected to dominate the heating \citep{Springel2008}, but their
number is small.

\begin{figure}
\begin{center}
\resizebox{8cm}{!}{\includegraphics{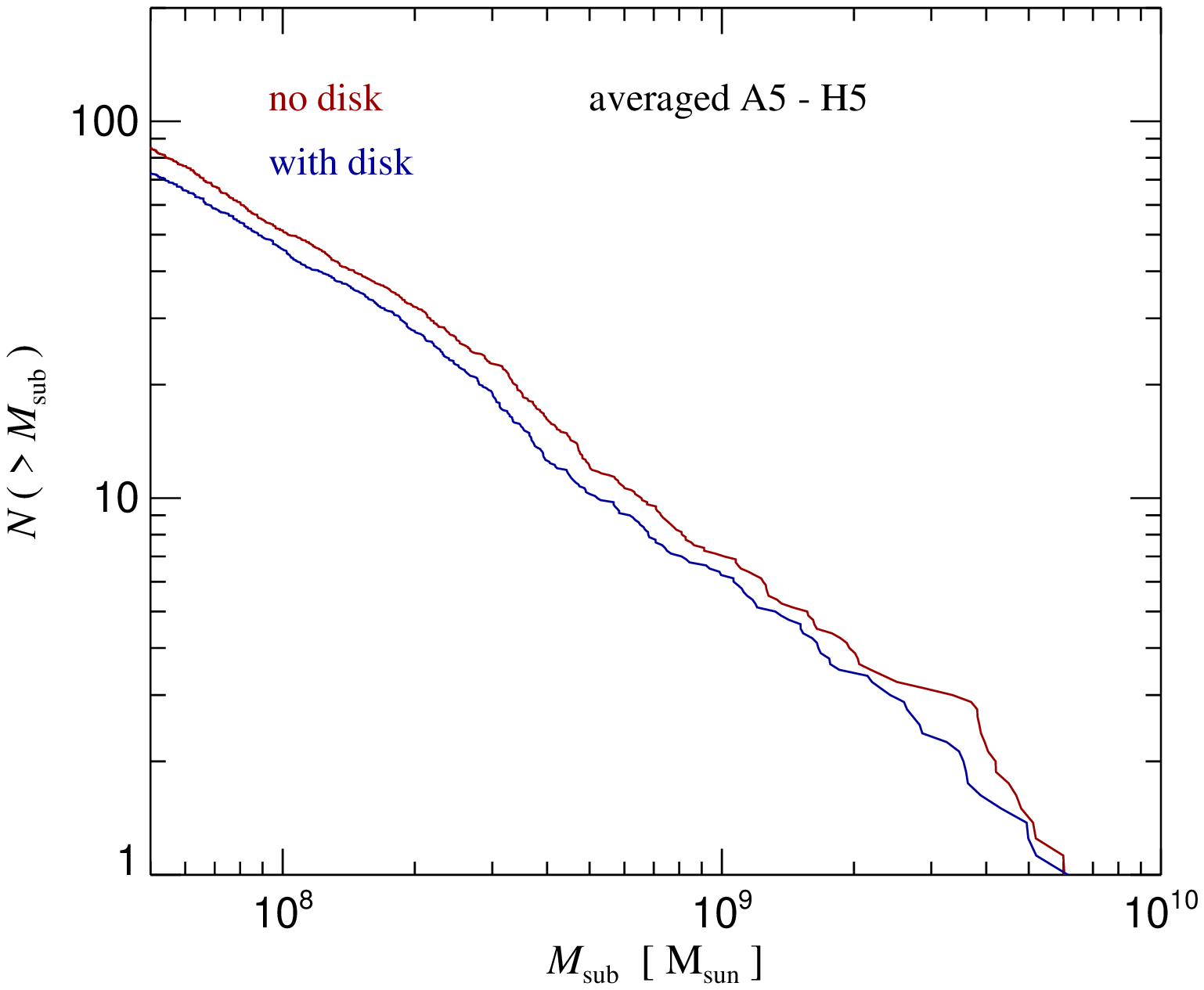}}%
\caption{Change of the cumulative abundance of substructures as a
  function of mass due to presence of a stellar disk. We count all
  substructures identified by {\small SUBFIND} within $200\,{\rm kpc}$ of the 
halo centers, and averaged over our 8 Aquarius halos. The ``no disk''
runs refer to results at $z=0$ of pure dark matter only runs of the
halos where no stellar component is inserted, while  the ``with disk''
results are for our default disk systems of simulation series \#1.   
\label{fig:subabunancemass}}
\end{center}
\end{figure}

\begin{figure}
\begin{center}
\resizebox{8cm}{!}{\includegraphics{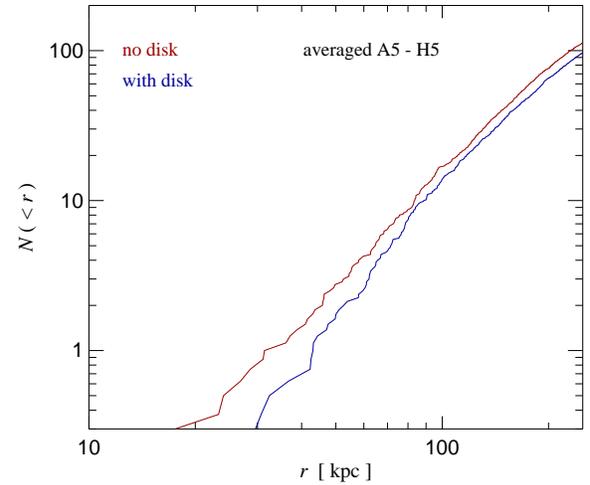}}%
\caption{Change of the cumulative abundance of substructure as a
  function of radius due to presence of a stellar disk. We count all
  substructures identified by {\small SUBFIND} with a mass larger than
  $6\times 10^{7}\,{\rm M}_\odot$, and average over our 8 primary
  galaxy models.
The comparison corresponds to that shown in
Fig.~\ref{fig:subabunancemass}, but considers the radial distribution
instead of the mass distribution.
\label{fig:subabunanceradial}}
\end{center}
\end{figure}

\begin{figure}
\begin{center}
\resizebox{8cm}{!}{\includegraphics{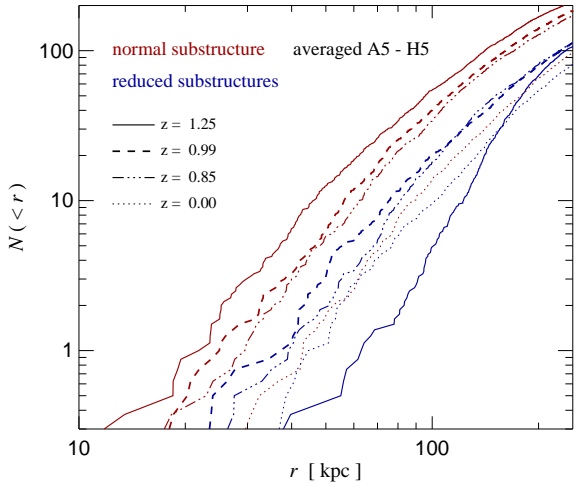}}%
\caption{Illustration of the impact of our ``substructure cleaning''
  procedure where at the disk insertion redshift of $z=1.3$ all
  particles in bound substructures of the central FOF group are
  spherically redistributed. The plot shows the cumulative radial
  abundance of substructures with mass above $6\times 10^{7}\,{\rm
    M}_\odot$ at different times, comparing runs where the
  substructure reduction was carried out with normal simulations
  where this was not done. At $z=1.25$, the reduction of subhalos in
  the inner parts of the target halo is still very strong, but soon
  the subhalo population is replenished by the accretion of additional mass
  and substructures. This reduces the substructure reduction effect at late
  times substantially, but in the range $1.0 < z < 0.85$ it is
  more or less constant as a function of radius and amounts to a
  suppression of around a factor of 2.
  \label{fig:reducedsubs}}
\end{center}
\end{figure}

\begin{figure}
\begin{center}
\resizebox{8cm}{!}{\includegraphics{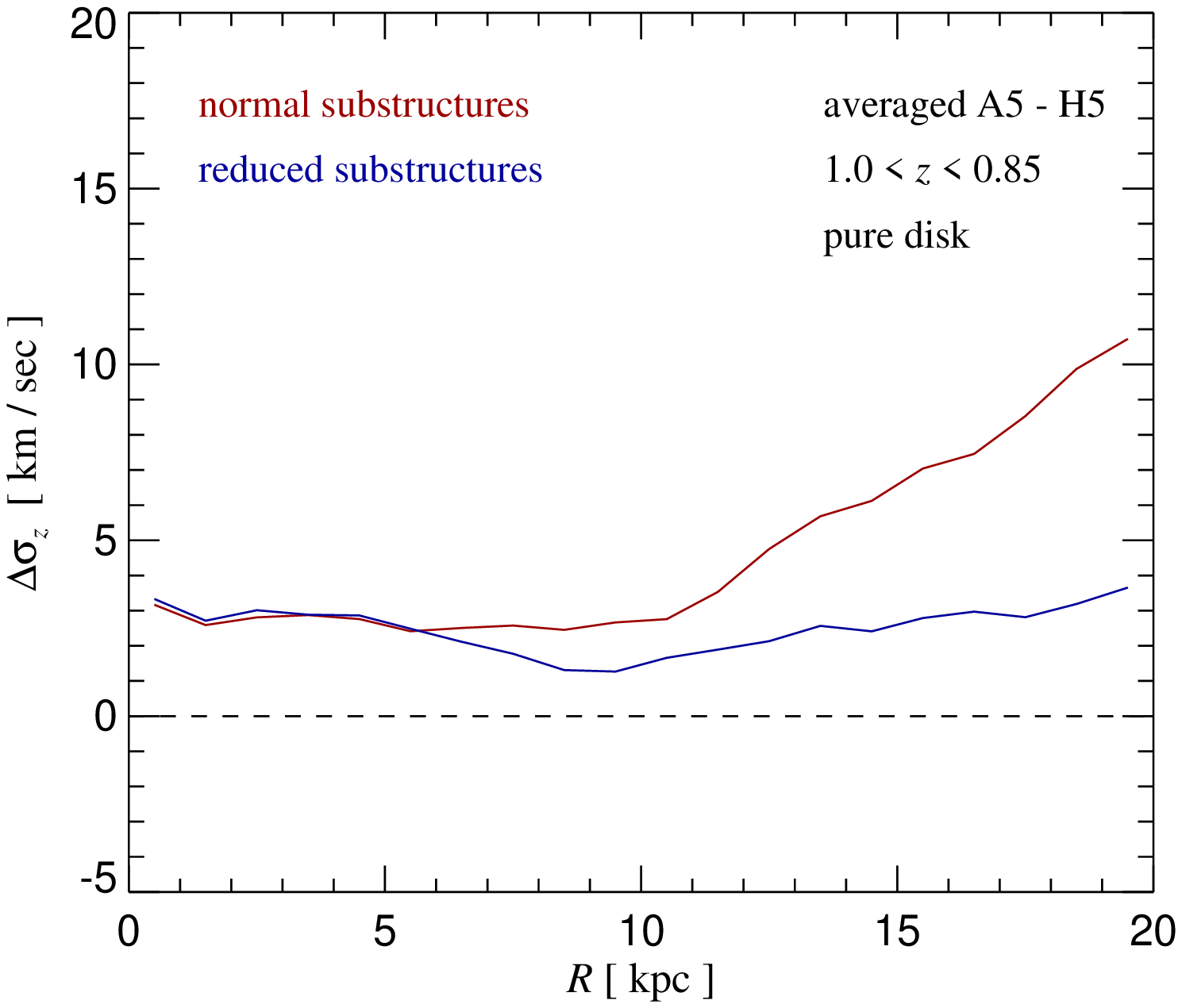}}%
\caption{Differential heating of the stellar disk in runs with normal
  or reduced substructure abundance. We here show radial profiles of
  the increase of the vertical velocity dispersion of disk stars
  between times $z=1$ and $z=0.85$ in our runs with pure disks. The
  reference simulations correspond to series \#1, while in the
  comparison runs of series \#8 the substructure has been smoothed out
  at the disk insertion time ($z_{\rm insert}=1.3$). In the runs with
  reduced substructure abundance, we find a substantially reduced
  heating rate in the outer parts of the disk, whereas in the inner
  parts within two disk scale lengths no difference is detected.
  \label{fig:reducedsubs_heating}}
\end{center}
\end{figure}

We here use our models to check whether subhalos contribute
significantly to the disk heating, and whether the disk in turn plays
a significant role in reducing substructure abundance when they pass
through pericenter and experience gravitational tidal shocks from the
disk or the enhanced central cusp. There is a body of previous work on
this subject \citep[e.g.][]{Kazantzidis2008, Kazantzidis2009,
  Purcell2009, D'Onghia2010}, largely based on much simpler toy
simulations than studied here. Our analysis is so far the most
elaborate attempt to study this in the correct cosmological setting,
and in particular takes the expected system-to-system variation into
account.

We consider first the subhalo abundance in runs without any disk
(i.e.~these are {\em dark-matter only} runs of A to H at $z=0$), and
compare it to the one found at $z=0$ in our default runs with disk. In
Figure~\ref{fig:subabunancemass}, we show the cumulative abundance of
substructures as a function of mass in both types of simulations. To
emphasize the mean difference in a clear way we show the averaged
abundance over all eight systems we simulated; we note that orienting
the disk along minor or major axis makes no difference here. There is
a $\sim 30\%$ reduction of substructure abundance across all mass
scales in the runs with the disk. This can be understood as an effect
of accelerated substructure depletion due to the gravitational shocks
the substructures experience as they pass through the disk. This
enhanced destruction rate shows up particularly strongly in the halo
center, as evidenced by Figure~\ref{fig:subabunanceradial}, which
gives the cumulative average abundance of substructures with mass
larger than $6\times 10^7\,{\rm M}_\odot$ as a function of radius. In
the inner parts of the halo, there is about a factor of two reduction
of the subhalo abundance. These results are in good agreement with
the analysis of \cite{D'Onghia2010}, who carried out orbit
integrations for the subhalos found in the dark matter only
simulations of Aquarius and estimated their evaporation rate
analytically by summing up the impact of gravitational shocks
\citep{Ostriker1972} experienced during disk passages.

\begin{figure*}
\begin{center}
\setlength{\unitlength}{1cm}%
\resizebox{16cm}{!}{\includegraphics{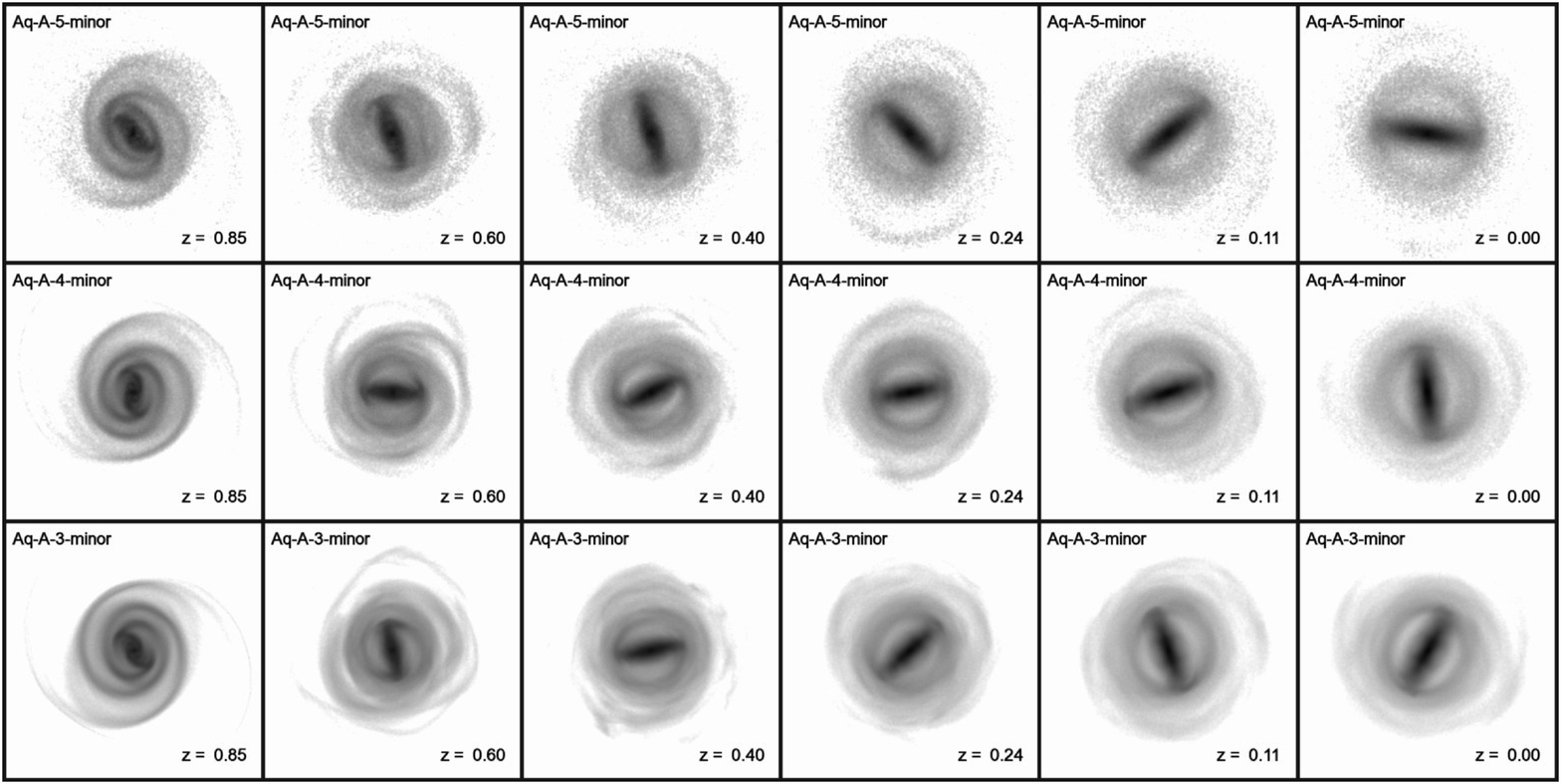}}\vspace*{0.5cm}\\
\resizebox{16cm}{!}{\includegraphics{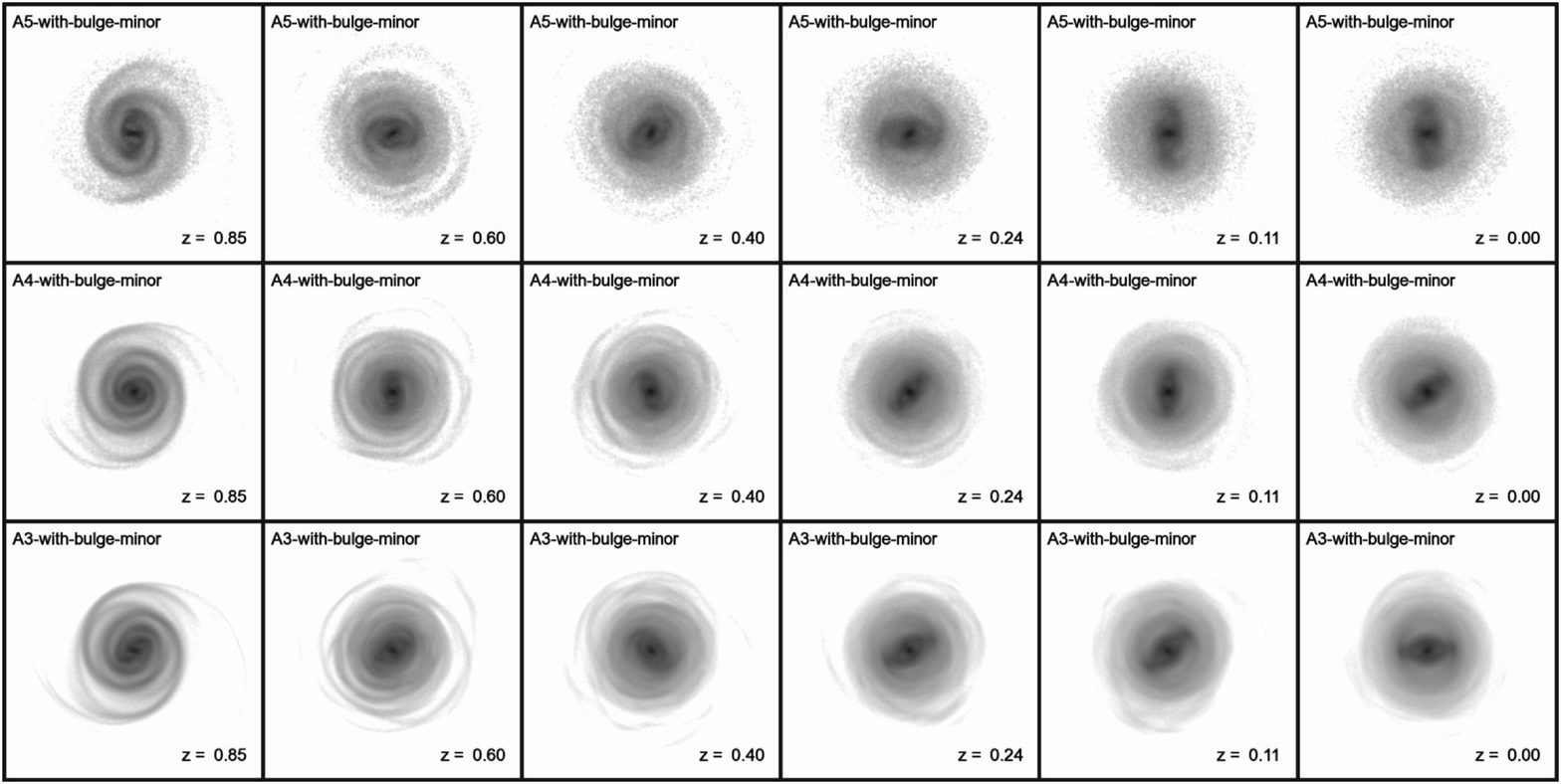}}
\caption{Resolution dependence of disk morphology in our default runs
  with pure disks (series \#1) and disks+bulges (series \#3). The top
  three rows of images compare the face-on projections of
the stellar disks at different times of our simulations of A5, A4,
and A3 in the runs with pure disks. The corresponding simulations
where a third of the stellar mass is moved to a central bulge are shown
in the bottom three rows. It is reassuring that the runs agree rather
well, apart from the phase angle of the bar.
\label{fig:resolution}
}
\end{center}
\end{figure*}

In order to explicitly test for the influence of substructures on disk
heating and disk stability, we have carried out a series of runs where
we varied our ``rounding halo'' experiment above. Instead of
sphericalizing all dark matter particles in the halo at $z=1.3$, we
have done so only for the dark matter particles bound in
substructures. Combined, these subhalos amount to a few percent of the
mass of the halo, so by redistributing the substructure particles in a
spherical fashion, the halo is made smooth without affecting its
dynamical equilibrium much. We note however that a large fraction of
the substructures found in a halo at low redshift will be accreted at
$z<1.3$; they are unaffected, and so this cleaning of substructure
will only temporarily make the halo smooth. This is seen explicitly
in Figure~\ref{fig:reducedsubs}, where we show the abundance of
substructures as a function or radius at different times, comparing an
ordinary disk run with a run where the substructure cleaning has been
carried out at $z=1.3$. While at $z=1.25$ the inner halo is still largely
devoid of subhalos in the cleaned run, in the redshift range $1< z <
0.85$, the abundance is suppressed on average by a factor of 2,
whereas towards $z=0$ it is down by only $\sim 20\%$.

We now use the subhalo cleaning run to look at differences of the disk
heating rates between $z=1.0$ and $z=0.85$. At this time, the disks
are still largely intact even in the runs that form strong bulges, and
here the substructure suppression in the runs with the subhalo
cleaning is still substantial and fairly uniform across radius and in
time.  In Figure~\ref{fig:reducedsubs_heating} we show the difference
in the vertical velocity dispersion in the standard runs and the runs
with subhalo cleaning. We clearly see evidence for an enhanced heating
rate of the disks in their outer parts, amounting to several ${\rm
  km\, s^{-1}}$ over the course of 1 Gyr. However, within two disk
scale lengths, there is virtually no detectable difference in the disk
heating rate, suggesting that substructure heating is negligible for
the bulk of the disk's stellar mass. It may however play an important
role in contributing to flaring of the stellar disk in the outer
parts.

\section{Resolution dependence} \label{sec:resolution}

So far, all our results have been based on `level-5' Aquarius
simulations, with a resolution of $2\times 10^5$ particles in the
disk, a dark matter halo resolved by about $10^6$ particles, and a
gravitational softening length of $680\,{\rm pc}$. Some effects of
galactic dynamics can depend strongly on numerical resolution
\citep[e.g.][]{Weinberg2002}, so it is advisable to check the
robustness and numerical convergence of our primary results in the
relevant regime.

To this end we have repeated our primary A-5 simulations at two higher
resolution levels, where the particle number is increased by factors
of 8 and 64, respectively, in both the dark matter and stellar
components. The gravitational softening lengths are reduced by factors
of 2 and 4, respectively. In our highest resolution simulation A-3,
this means that the stellar disk is represented with 12.8 million
particles, the stellar bulge (if present) with 6.4 million, and the
dark matter halo with about 50 million particles.

In Figure~\ref{fig:resolution} we show a visual comparison of the
stellar disk evolution seen in the three resolution levels. The top
three rows compare pure disk models at the resolution levels 5, 4, and
3, whereas the bottom three rows give the same comparison for our
standard models with bulges. Overall, the disk morphologies are very
similar in both resolution sequences.  The position angles of the bar
motion do not line up exactly, but such phase differences are to be
expected. However, the overall morphological evolution is clearly very
similar, which suggests good convergence of the structural evolution
of the galaxies. 

\begin{figure}
\begin{center}
\setlength{\unitlength}{1cm}
\resizebox{8cm}{!}{\includegraphics{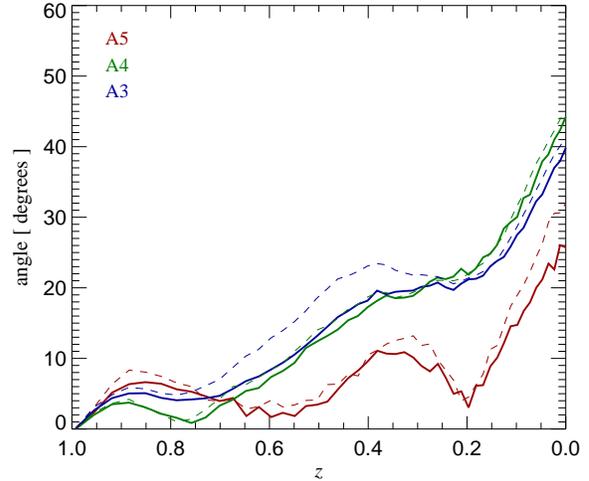}}
\caption{Disk tilt angle measured for simulations at different
  numerical resolutions, ranging from Aquarius level-5 to level-3. Two
  types of runs are shown, the A-halo with a pure disk (solid lines,
  from series \#1), and the A-halo with a disk+bulge (dashed lines,
  from series \#3). The tilt angles of the disks do not line up
  quantitatively in detail for the different resolutions, but the
  evolution qualitatively still agrees reasonably well, without any
  indication of systematic trends with resolution. The tilt angle
  evolution for a given resolution is almost identical for runs with
  or without a bulge, showing that it is governed almost exclusively
  by the dark matter halo, so that the difference we find between A3,
  A4, and A5 likely originate in small differences of the dynamical
  states of their dark matter halos when the disk is inserted.
\label{fig:tilt_convergence}
}
\end{center}
\end{figure}

More quantitatively, we show in Figure~\ref{fig:tilt_convergence} a
convergence study of the tilt angle evolution in the two resolution
sequences. Especially the two high resolution runs line up remarkably
well. The lower resolution run shows a small offset in its evolution
in comparison, but since there is no systematic trend with resolution
we interpret this small difference in angle as a chance effect. Other
quantities we examined are similarly stable with respect to
resolution. We hence believe that already the level-5 resolution
provides robust results for the quantities studied in this work.

\section{Discussion and conclusions} \label{sec:discussion}

In this study, we have analyzed the stability of disk galaxies inserted
into high-resolution zoom-simulations of the formation of Milky
Way-sized dark matter halos, with initial conditions taken from the
Aquarius project. We have refined a methodology previously used by
\citet{DeBuhr2012}, most notably by employing a more sophisticated
approach to determine the initial velocity distribution of the star
particles. For the latter we made use of the iterative method realized in
the {\small GALIC} code \citep{Yurin2014}, which is capable of computing
high-quality stationary solutions in general dark matter
halos. For the dark matter potential, we directly used the
distribution of dark matter particles found in the Aquarius halos,
without using any approximations besides imposing axisymmetry on the
force field.

Using this improved methodology, we have extended the analysis of the
Aquarius halos to a larger halo sample (using eight systems, A to H),
and to structural variants that also include systems with central
bulges. We have also investigated a number of toy simulations where
the dark matter halos were artificially rounded or dark matter
substructures were erased, in order to highlight the impact of
residual triaxiality and of substructures on the dynamical evolution of
the disks.

Our main findings can be summarized as follows:
\begin{enumerate}
\item The presence/absence of a stellar bulge, as well as the
  presence/absence of a stellar bar, do not significantly affect the tumbling of
  disk galaxies. The turning motion of disks appears to be primarily
  driven by the triaxiality of the halo and its figure rotation; in
  rare cases encounters with massive substructures can also initiate
  substantial disk tilt.
\item Disks initially oriented along the dark matter halo's minor axis
  show on average better directional stability than disks oriented
  along the major axis.
\item We predict that an average tumbling angle of about $40$ degrees
  over 7.6 Gyr between $z=1$ and $z=0$ should be quite typical for disk
  galaxies, corresponding to $5-6$ degrees per Gyr. Importantly, thin
  disks can survive such tumbling rates in a largely unaffected way.
\item There is a significant depletion of dark matter substructures
  due to the presence of a massive disk. If part of the disk mass is
  put into a bulge instead, the effect is very slightly reduced,
  suggesting that gravitational shocking at the disk is indeed more
  important than the enhancement of the central core density and the
  associated increase of pericenter at the halo cusp.
\item Dark matter substructure can significantly contribute to
  disk heating in the outer parts of disks, while this appears
  largely negligible in the inner regions within $\simeq 2$ disk scale lengths.
\item Our quantitative numerical results are unaffected by numerical
  resolution, as evidence by our resolution tests that cover a factor
  of 64 in mass resolution, and a factor of 4 in gravitational
  softening length.
\end{enumerate}

Overall, our results suggest that the survival of thin stellar disks
is in principle not a problem in CDM halos. In particular, the
triaxiality and high substructure abundance in CDM halos do not
preclude the survival of thin stellar disks, even though we would
expect them to tumble slowly with time. What is arguably more
difficult to understand is how massive, cold stellar disks can survive
strong bar formation. Cuspy cold dark matter halos actually help here,
whereas the low central dark matter densities often inferred
observationally for the inner parts of galaxies \citep[even for the
Milky Way, see][]{Bovy2013} make it challenging to understand how such
observed systems manage to sport only a small or no bar. This remains
an interesting topic for further study.

The results obtained here demonstrate the power of our techniques to
introduce stellar systems into cosmologically consistent, growing dark
matter halos. This facilitates research in galactic dynamics at very
high resolution within a realistic and complex cosmological
environment, something that promises to be a worthwhile avenue for
future work.

\section*{ACKNOWLEDGMENTS}
 
We thank the anonymous referee for constructive criticism and
  valuable suggestions that helped to improve the paper.
D.Y.~acknowledges support by the DFG Research Centre SFB-881 `The
Milky Way System' through project A6.  This work has also been
supported by the European Research Council under ERC-StG grant
EXAGAL-308037 and by the Klaus Tschira Foundation.

\bibliography{paper}

\end{document}